\documentclass[12pt]{oxarticle}
\pdfoutput=1

\usepackage{graphicx, float, array, xspace, amscd, amsmath, amsthm, amssymb, latexsym, relsize, bbm}
\usepackage[letterpaper, left=1.7in, right=0.3in, top=0.8in, bottom=1.2in]{geometry}
\usepackage[bbgreekl]{mathbbol}
\usepackage{amsfonts, tikz-cd}
\usepackage[T1]{fontenc}

\usepackage{tikz}
\usetikzlibrary{backgrounds}
\usepackage[framemethod=tikz]{mdframed}
\AtBeginEnvironment{mdframed}{%
\tikzset{every picture/.style={}}%
}
\mdfsetup{roundcorner=.5ex}
{\begin{mdframed}[backgroundcolor=white]}%
{\end{mdframed}}

\DeclareSymbolFontAlphabet{\mathbb}{AMSb}
\DeclareSymbolFontAlphabet{\mathbbl}{bbold}

\usepackage[sort&compress, comma, square, numbers]{natbib}
\usepackage[all,cmtip]{xy}
\usepackage{color}
\definecolor{MyDarkBlue}{rgb}{0.15,0.25,0.45}
\usepackage{braket}      
\usepackage[linktocpage=true]{hyperref}
\hypersetup{colorlinks=true, citecolor=blue, linkcolor=blue, urlcolor=blue, pdfauthor={}, pdftitle={},breaklinks=true,pageanchor=false, hypertexnames=false}

\numberwithin{equation}{section}

\let\SS=\S 

\newcommand{\contr}{\,\lrcorner\,}

\renewcommand{\#}{^{\sharp}}

\newcommand{\Het}{\hbox{\sf Het}}

\newcommand{\ol}[1]{{\overline{#1}}}

\newcommand{\LC}{\text{\tiny LC}}

\newcommand{\Hu}{{\text{\tiny H}}}
\newcommand{\Bi}{{\text{\tiny B}}}

\newcommand{\Ch}{\text{\tiny CH}}
\newcommand{\eb}{{{\overline{\eta}}}}

\renewcommand{\sb}{{\overline{\sigma}}}

\newcommand{\rb}{{\overline{ r}}}

\newcommand{\Ob}{{\overline{ \Omega}}}

\newcommand{\w}{{\,\wedge\,}}
\newcommand{\wt}{\widetilde}

\newcommand{\fD}{{\mathfrak{D}}}
\newcommand{\fDb}{{\overline\fD}}

\newcommand{\half}{\frac{1}{2}}

\def\CS{{\text{CS}}}

\newcommand{\ab}{{\overline\alpha}}
\newcommand{\bb}{{\overline\beta}}

\newcommand{\A}{\cA}
\newcommand{\M}{\ccM}
\newcommand{\X}{X}
\newcommand{\B}{\ccB}
\newcommand{\Z}{\cZ}
\newcommand{\Zb}{{\overline \cZ}}

\renewcommand{\aa}{\mathfrak{a}}
\newcommand{\ccb}{\mathfrak{b}}

\renewcommand{\a}{\alpha}
\renewcommand{\b}{\beta}
\newcommand{\g}{\gamma}\newcommand{\G}{\Gamma}
\renewcommand{\d}{\delta}\newcommand{\D}{\Delta}
\newcommand{\e}{\epsilon}\newcommand{\ve}{\varepsilon}

\renewcommand{\th}{\theta}\newcommand{\Th}{\Theta}

\renewcommand{\k}{\kappa}
\renewcommand{\l}{\lambda}\renewcommand{\L}{\Lambda}
\newcommand{\m}{\mu}
\newcommand{\n}{\nu}
\newcommand{\x}{\xi}

\renewcommand{\r}{\rho}
\newcommand{\s}{\sigma}\renewcommand{\S}{\Sigma}
\renewcommand{\t}{\tau}

\newcommand{\ph}{\phi}\newcommand{\Ph}{\Phi}

\newcommand{\Ps}{\Psi}
\renewcommand{\o}{\omega}\renewcommand{\O}{\Omega}

\DeclareFontFamily{OT1}{pzc}{}
\DeclareFontShape{OT1}{pzc}{m}{it}{<-> s * [1.200] pzcmi7t}{}
\DeclareMathAlphabet{\mathpzc}{OT1}{pzc}{m}{it}

\newcommand{\cA}{\mathcal{A}}
\newcommand{\ccB}{\mathpzc B}

\newcommand{\ccD}{\mathpzc D}

\newcommand{\cF}{\mathcal{F}}
\newcommand{\cG}{\mathcal{G}}
\newcommand{\cH}{\mathcal{H}}

\newcommand{\cL}{\mathcal{L}}
\newcommand{\cM}{\mathcal{M}}\newcommand{\ccM}{\mathpzc M}

\newcommand{\cO}{\mathcal{O}}

\newcommand{\cR}{\mathcal{R}}

\newcommand{\cT}{\mathcal{T}}\newcommand{\ccT}{\mathpzc T}
\newcommand{\cU}{\ccU}\newcommand{\ccU}{\mathpzc U}

\newcommand{\ccX}{\mathpzc X}
\newcommand{\ccY}{\mathpzc Y}
\newcommand{\cZ}{\mathcal{Z}}\newcommand{\ccZ}{\mathpzc Z}

\newcommand{\ccZb}{{\overline \ccZ}}

\DeclareFontFamily{U}{bbold}{}
\DeclareFontShape{U}{bbold}{m}{n}
 {  <-5.5> s*[1.05] bbold5
    <5.5-6.5> s*[1.05] bbold6
    <6.5-7.5> s*[1.05] bbold7
    <7.5-8.5> s*[1.05] bbold8
    <8.5-9.5> s*[1.05] bbold9
    <9.5-11.5> s*[1.05] bbold10
    <11.5-16> s*[1.05] bbold12
    <16-> s*[1.05] bbold17
 }{}

\newcommand{\IA}{\mathbbl{A}}
\newcommand{\IB}{\mathbbl{B}}
\newcommand{\IC}{\mathbbl{C}}
\newcommand{\ID}{\mathbbl{D}}\newcommand{\Id}{\mathbbl{d}}
\newcommand{\IE}{\mathbbl{E}}
\newcommand{\IF}{\mathbbl{F}}
\newcommand{\Ig}{\mathbbl{g}}
\newcommand{\IH}{\mathbbl{H}}

\renewcommand{\IJ}{\mathbbl{J}}

\newcommand{\IL}{\mathbbl{L}}

\newcommand{\IQ}{\mathbbl{Q}}\newcommand{\Iq}{\mathbbl{q}}
\newcommand{\IR}{\mathbbl{R}}
\newcommand{\IS}{\mathbbl{S}}\newcommand{\Is}{\mathbbl{s}}

\newcommand{\IU}{\mathbbl{U}}

\newcommand{\IX}{\mathbbl{X}}
\newcommand{\IY}{\mathbbl{Y}}
\newcommand{\IZ}{\mathbbl{Z}}

\newcommand{\IDb}{{\overline{\mathbbl{D}}}}

\newcommand{\ITheta}{\mathbbl{\Theta}}

\newcommand{\Iomega}{\mathbbl{\hskip1pt\bbomega}}

\newcommand{\inbar}{\vrule height 0.43em width 0.048em}
\newcommand{\Idel}{\hbox{$\partial\hspace{-0.38em\inbar\hspace{0.3em}}$}}
\newcommand{\Idelb}{{\hbox{$\bar\partial\hspace{-0.38em\inbar\hspace{0.3em}}$}}}

\newcommand{\deth}{\eth} 
\newcommand{\dethb}{\overline{\eth}}

\font\eightrm=cmr8 at 8pt
\font\csc=cmcsc10

\newcommand{\beq}{\begin{equation}}
\newcommand{\eeq}{\end{equation}}
\newcommand{\beqnn}{\begin{equation*}}
\newcommand{\eeqnn}{\end{equation*}}
\newcommand{\bea}{\begin{eqnarray}}
\newcommand{\eea}{\end{eqnarray}}
\newcommand{\bean}{\begin{eqnarray*}}
\newcommand{\eean}{\end{eqnarray*}}

\newcommand{\sref}[1]{\SS\ref{#1}}

\newcommand{\pd}[2]{\frac{\partial #1}{\partial #2}}

\newcommand{\ii}{\text{i}}

\newcommand{\place}[3]{\vbox to0pt{\kern-\parskip\kern-7pt
                             \kern-#2truein\hbox{\kern#1truein #3}
                             \vss}\nointerlineskip}
\newcommand{\smallfrac}[2]{\frac{\scriptstyle #1}{\scriptstyle #2}}

\DeclareFontFamily{U}{wncy}{}
\DeclareFontShape{U}{wncy}{m}{n}{<->wncyr10}{}
\DeclareSymbolFont{mcy}{U}{wncy}{m}{n}
\DeclareMathSymbol{\sha}{\mathord}{mcy}{"58}

\newcommand{\capt}[3]{\parbox{#1}{\renewcommand{\baselinestretch}{1.0}
                                                           \caption{\label{#2}\small\it #3}}}

\newcommand{\del}{{\partial}}
\newcommand{\delb}{{\overline{\partial}}}

\newcommand{\lb}{{\overline\lambda}}
\newcommand{\nb}{{\overline\n}}
\newcommand{\mb}{{\overline\m}}

\newcommand{\Db}{{\overline D}}
\newcommand{\Dbar}{{\overline D}}

\newcommand{\Ione}{\mathbbl{1}}

\renewcommand{\aa}{\mathfrak{a}}
\newcommand{\mfb}{\mathfrak{b}}

\newcommand{\End}{{\text{End}\,}}
\newcommand{\EndE}{{\text{End}\,E}}

\newcommand{\dd}{{\text{d}}}

\newcommand{\K}{K\"ahler\xspace}

\renewcommand{\H}{\text{H}}

\def\ker{{\rm ker ~}}

\newcommand{\vol}{\mbox{\,vol}}
\newcommand{\tr}{\text{Tr}\hskip2pt}

\newcommand{\tb}{{\overline{\tau}}}

\newcommand{\ap}{{\a^{\backprime}\,}}
\renewcommand{\sb}{{\overline{\sigma}}}

\renewcommand{\rb}{{\overline{\rho}}}
\renewcommand{\=}{\;=\;}

\makeatletter
\g@addto@macro\bfseries{\boldmath}
\makeatother

\newcommand{\citeHHP}{\cite{Herbst:2008jq}\xspace}
\newcommand{\citeE}{\cite{McOrist:2021dnd}\xspace}
\newcommand{\citeES}{\cite{McOrist:2024zdz}\xspace}
\newcommand{\citeM}{\cite{Candelas:2016usb}\xspace}

\newcommand{\citeSG}{\cite{McOrist:2019mxh}\xspace}

\newcommand{\citeUG}{\cite{Candelas:2018lib}\xspace}
\newcommand{\citeUGSG}{\cite{Candelas:2018lib, McOrist:2019mxh}\xspace}

\newcommand{\citeUS}{\cite{Candelas:2018lib, Candelas:2016usb,McOrist:2016cfl}\xspace}

\newcommand{\citeOS}{\cite{delaOssa:2014cia}\xspace}

\newcommand{\citeS}{\cite{Ashmore:2018ybe}\xspace}
\newcommand{\citeW}{\cite{Witten:1985bz}\xspace}

\renewcommand{\baselinestretch}{1.1}
\proofmodefalse
\begin{document}
\pagestyle{empty}      
\ifproofmode\underline{\underline{\Large Working notes. Not for circulation.}}\else{}\fi

\begin{center}
\null\vskip0.2in
{\Huge The moduli of the universal geometry of heterotic moduli \\[0.5in]}
{\csc   Jock McOrist$^{\sharp 1}$,  Martin Sticka$^{\sharp 2}$ and Eirik Eik Svanes$^{\dagger 3}$\\[0.5in]}

{\it 
$^\sharp$ Department of Mathematics\hphantom{$^2$}\\
School of Science and Technology\\
University of New England\\
Armidale, 2351, Australia\\[3ex]
}

{\it 
$^\dagger$ Department of Mathematics and Physics\hphantom{$^1$}\\
Faculty of Science and Technology\\
University of Stavanger\\
N-4036, Stavanger, Norway\\[3ex]
}

\footnotetext[1]{{\tt jmcorist@une.edu.au}}
\footnotetext[2]{{\tt msticka@myune.edu.au}}
\footnotetext[3]{{\tt eirik.e.svanes@uis.no}}
\vspace{1cm}
{\bf Abstract\\[-8pt]}
\end{center}

We study the moduli of the universal geometry of $d=4$ $N=1$ heterotic vacua. Universal geometry refers to a family of heterotic vacua fibered over the moduli space. The universal geometry mimics aspects of the original heterotic vacua, in particular holomorphic data such as F-terms, as well as the Green-Schwarz Bianchi identity.  Here we study  first order deformations of the universal geometry and find this provides a shortcut to computing second order deformations of the original problem. The equations governing the moduli of the universal geometry are remarkably similar to the equations of the underlying heterotic theory and we find a fascinating double extension structure that mirrors the original heterotic problem.  As an application we find first order universal deformations determine second order deformations of the original heterotic theory. This gives a  shortcut to determining results that are otherwise algebraically unwieldy. The role of the D-terms is closely related to the existence of flat connections on the moduli space. Finally, we re-derive some of these results by direct differentiation - this direct approach requires significantly more calculation.

\vskip150pt

\newgeometry{left=1.5in, right=0.5in, top=0.75in, bottom=0.8in}
%
\newpage

{\baselineskip=10pt\tableofcontents}
\restoregeometry
\setcounter{page}{1}
\pagestyle{plain}
\renewcommand{\baselinestretch}{1.3}
\null\vskip-10pt
\pagenumbering{arabic}
 
\section{Introduction}

Heterotic geometry describes the moduli space of a heterotic string vacuum, focusing on the geometry at large volume, where it can be approximated by a spacetime geometry, defining a non-linear sigma model. This sigma model flows under the renormalization group to a super-conformal field theory (SCFT), defining the string background perturbatively in terms of the string coupling constant $ g_s $. The geometry also receives corrections proportional to $ \alpha' $, which are linked to loop corrections in the non-linear sigma model. These corrections are determined at first order in terms of geometric conditions on the manifold; at second order they are known as far as variations of the fermionic fields go but not yet translated into conditions on the geometry.

The moduli space of Calabi-Yau manifolds exhibits intriguing properties, particularly when quantum corrections, including both perturbative and non-perturbative $\ap$-corrections, are taken into account. These corrections give rise to mirror symmetry and special geometry. Special geometry can be described in several ways; one perspective involves flat connections on the moduli space, enabling the parallel transport of states in the associated SCFT. A natural question arises: does this framework generalize to the study of heterotic vacua?\footnote{The moduli of Calabi-Yau manifolds can be explored within the context of the standard embedding of heteortic vacua, which unlike type II string theory, preserves $N=1$, $d=4$ spacetime supersymmetry. Smooth deformations of the bundle away from the standard embedding at a generic point in moduli space does not break any supersymmetry and from the point of view of a  four-dimensional field theorist, is a perfectly benign movement in moduli space and nothing dramatic is expected to happen e.g. \cite{McOrist:2010ae,Melnikov:2012hk}. This leads to the question of whether special geometry persists away from the standard embedding.}

In \citeUG, a fibration of the heterotic vacuum space over its moduli space, termed universal heterotic geometry, was introduced as a generalisation of the mathematical concept of a universal bundle. This fibration is defined by connections on the moduli space, and our goal here is to compute the associated curvatures. While direct computation is possible, it is cumbersome. A key insight of \citeUG is that the universal bundle mimics many of the holomorphic properties of the original heterotic vacuum. That is, remarkably both the F-terms and the Green-Schwarz Bianchi identity hold on the total space of the fibration. These relations encode significant information about the first-order deformations of the heterotic problem.

In this paper, we study the deformations of universal heterotic geometry and show that first-order deformations yield, in an elegant manner, second-order deformations of the heterotic vacuum. Notably, many second-order derivatives vanish, leading to the conclusion that numerous components of the curvature of the connections on the moduli space also vanish. We verify these findings through direct calculations. A critical aspect of our analysis involves requiring that first-order deformations preserve D-terms fiberwise. This suggests a profound connection between D-terms and special geometry, which warrants further exploration.

To set the stage, we briefly review the geometry of a heterotic vacuum, correct to first order in $\alpha'$, and key features of its moduli space.

\subsection{Heterotic geometry}
The large volume approximation in which we work considers a spacetime background of the form $ \mathbb{R}^{1,3} \times X $, where $ X $ is a complex three-dimensional manifold with a vanishing first Chern class, and $ E $ is a holomorphic vector bundle over $ X $ with a connection $ A $ satisfying the Hermitian Yang-Mills equations. The background also involves the Kalb-Ramond field $ B $ and its field strength $ H $, both of which must satisfy field equations and anomaly cancellation conditions. This defines the heterotic structure, denoted by $ \text{Het} $, which comprises the geometry of $ X $, the bundle $ E $, and the background $ H $.

The moduli space of heterotic string theory at large radius, accurate to first order in $\ap$, is computed from a dimensional reduction of the supergravity theory. Due to supersymmetry, this moduli space must have a \K structure, and the relationships between $ H $, the connection on $ E $, and the Hermitian form $ \omega $ on $ X $ must satisfy the Green-Schwarz anomaly cancellation and $ N=1 $ supersymmetry in four dimensions.

While these conditions are cited in many places in the literature, its a useful setting for us to follow \citeE in which these conditions are phrased in terms of F-terms and D-terms. We define a bundle $Q$ given topologically as
\beq
Q \= (\ccT_\X{}^{1,0})^* \oplus {\rm End} E \oplus \ccT_\X^{1,0}~,
\eeq
and there is a $\Dbar$-operator and its adjoint $\Dbar^\dag$ that act on sections of this bundle. In \citeE, we demonstrate the kernel of this operator corresponds to holomorphic deformations of set of equations we term F-terms:
\beq
N_J \= 0~, \quad F^{0,2} \= 0~, \quad H \=  \dd^c \o \= \frac{1}{3!} J^m J^n J^p (\dd \o)_{mnp}~, 
\label{eq:F-terms}
\eeq
Let $ x^m $ denote the real coordinates of $ \X $, and $(x^\mu, x^{\bar{\nu}})$ the holomorphic coordinates. The vector-valued form $ J^m \equiv J^m{}_n \, \dd x^n $ is a 1-form derived from the complex structure. In the following, we will generally omit the wedge product symbol ``$\wedge$'' between forms, except where its omission might cause ambiguity.  The anomaly relation yields a modified Bianchi identity for $H$. 
\beq
   \dd H \=\! -\frac{\ap}{4} \tr F\w F + \frac{\ap}{4} \tr R^\H\w R^\H ~.
\label{eq:Anomaly0}\eeq
The Bianchi identity is required so that $\Dbar^2 = 0$ defines a cohomology, and so it is sometimes viewed in close conjunction with the F-terms. Of course, in the ten-dimensional theory it is a constraint. 

Deformations of equations we call D-terms corresponds to the adjoint $\Dbar^\dag$ which is computed with respect to the moduli space metric \cite{Candelas:2016usb, McOrist:2019mxh}, see also \cite{Garcia-Fernandez:2020awc, Ashmore:2019rkx}. These equations are
\beq\label{eq:D-terms}
\o^2 F \= 0~, \qquad \dd \o^2 \= 0~, \qquad e^{\Phi} \= e^{\Phi_0} = g_s~,
\eeq
where the dilaton is a constant so that the non-linear sigma model is a guaranteed to be a relevant description of the background. If the dilaton is not constant, which amounts to $H=\cO(1)$ (not subleading in $\ap$), then while we might find solutions of the mathematical problem defined by \eqref{eq:F-terms}-\eqref{eq:D-terms}, the resulting geometry does not necessarily have anything to do with the string theory without some string non-perturbative arguments such as duality. For example, it means the string coupling varies over the internal manifold and so in regions where $g_s$ is large we expect string loop corrections and non-perturbative $g_s$ effects to be relevant.  It is a double edged sword because at $H=\cO(1)$,  the non-linear sigma model on a Riemann sphere is itself strongly coupled in its coupling constant $\ap$, meaning we would need to correct both the solution and the equations of motion to all orders in $\ap$. 

Returning now to the problem of interest: $H=\cO(\ap)$ with a geometry satisfying \eqref{eq:F-terms}--\eqref{eq:D-terms}. The deformations of this collection of equations, for a fixed topology,  correspond to the points of the moduli space $\M$, which is itself a complex manifold. This has real coordinates~$y^a$ and complex coordinates  
$(y^\a,y^\bb)$. Said differently, given a point $y\in \M$ we get a solution to the heterotic equations of motion and we call that solution a heterotic structure. Our goal is to understand how the space of vacua are fibered over the moduli space. The role of singularities is clearly important to the moduli space. For our work here however, we assume we are working in the neighbourhood of a smooth point. Studying global issues,  singularities and other novel features such as jumps in dimension remains as future work.  

\subsection{The universal bundle}
For that we come to a universal bundle. In general terms, a universal bundle is a fibration. Its  base is the moduli space of some geometric objects such as vector bundles or complex structures, satisfying certain equations and the fibres are the geometric objects themselves. For our purposes, these data derive from heterotic string theory and the universal bundle $\ccU$ encodes the heterotic structures as a fibres over each point in this moduli space. The bundle is `universal' in the sense that it systematically encodes the structure (e.g., vector bundles, connections, or complex structures) across the entire moduli space. It provides a framework for computing quantum corrections to the leading order heterotic structures.

This was studied at length in \citeUG. A summary of the key aspects are as follows:
\begin{itemize}

   \item Extended versions of $B$ and $H$ are introduced, denoted $\IB$ and $\IH$, which satisfy a supersymmetric relation and a Bianchi identity. The mixed components of $\IH$ yield key relations for the moduli space.

   \item The geometry is extended by introducing a connection $c_a{}^m$, which helps split the tangent space of the fibration $X \to \M$ into horizontal and vertical subspaces. The splitting of tangent spaces is crucial for defining how vectors and tensors extend across the larger space.

   \item A covariant derivative structure is developed, decomposing the standard de Rham operator into components acting on both the fiber $X$ and the base $\M$, taking into account changes in the complex structure and parameter variations.

   \item Two metrics, one on the manifold $X$ and the other on the moduli space $\M$, combine into an extended metric that includes cross terms involving the connection $c_a{}^m$. This connection also influences the variation of the complex structure on $X$, reflecting how moduli variations manifest geometrically.
   
\end{itemize}

This universal heterotic geometry framework extends the moduli space and field theory in a way that integrates both gauge and gravitational degrees of freedom.

At a more technical level, we have the diagram for the universal bundle $\ccU$:
\beq\label{eq:HeteroticFamily}
\begin{tikzcd}
\Het \arrow[r]  & \ccU \arrow[d]\\
& \M
\end{tikzcd}~,
\eeq
where $\Het = [\omega,J,A]$ denotes the geometric data solving \eqref{eq:F-terms}--\eqref{eq:D-terms} given some fixed topology. Note: the connection on the tangent bundle $\ccT_\X$ is fixed in terms of the other fields to be the Hull connection. 

It is important to take into account the spaces are curved and that there are gauge symmetries to be respect. We account for both of these simultaneously by considering covariant derivatives. As a warm-up, consider gauge transformation of Yang-Mills connection~$A$ 
\beq
A\to\; ^\Ph\!A \= \Ph A \Ph^{-1} - \dd\Ph\,\Ph^{-1}~.
\label{eq:Atransf}\eeq
The connection $A$ depends on the manifold $X$ and moduli space $\M$. That is, $A$ depends on parameters and  $\Phi$ necessarily does so too.  A deformation  $\d A$  is defined by a parameter space covariant derivative $D_a A$. This is cooked up to transform homogeneously under gauge transformations and requires the introduction of a connection 
$A\# {\=} {A\#}_a\, \dd y^a$ transforming under the same group as $A$ but with legs on the moduli space. Then $A\#$ defines for us a choice of covariant derivative:
\beq\label{eq:covariantderivative1}
D_a A \= \del_a A - \dd_A A\#_a~,\quad \dd_A A\#_a \= \dd A\#_a + [A,\,A\#_a]~.
\eeq
Deformations of $A\#$ correspond to small gauge transformations of $D_a A$  and a natural choice of gauge is holomorphic gauge in which holomorphic deformations of $\A = A^{0,1}$ correspond to holomorphic tangent vectors of $\M$ \citeSG.

It is a conceptual leap but $A$ and $A\#_a$ are to be unified into a single connection for a bigger gauge bundle
\beq
\IA\= A + A\#_a \dd y^a~.
\notag\eeq
 It has an associated field strength
\beq
\IF \= \Id \IA + \IA^2~,~~~\text{with}~~~\Id\= \dd + \dd y^a \del_a~.
\notag\eeq
In some sense we can view $\IF$ as the extension of $F$ to the universal bundle. It naturally encodes the curvature of the Yang-Millls bundle $F$, covariant derivatives $\IF_{am} = \fD_a A_m$ and an additional curvature $\IF_{ab}$ on the moduli space. The last curvature is of interest -- its the curvature of the connection $A\#$ restricted to the moduli space. These curvatures are closely related to special geometry in the context of CY manifolds. The curvature tells us how fields or geometric objects (such as complex structure and hermitian structure) change as one moves around in moduli space. This is the genesis of universal geometry. Democracy dictates that we shouldn't stop at just the symmetries of the Yang-Mills bundle $E$, but include other gauge symmetries: diffeomorphisms of $X$ and the gerbe symmetries of the B-field. This gives a universal geometry for heterotic vacua.

This gives us a nice mathematical structure. It encodes otherwise difficult to derive algebraic relations and these were crucial in deriving  a \K potential for the moduli space metric in \citeM. The motivation for this article is to address two closely related questions: is there a meaning to the curvature of the moduli space connection $\IF_{ab}$ and how is the universal geometry sensitive to the second order deformations?  Both of these are closely related to studying the space of deformations of the universal bundle structure.

\hypersetup{pageanchor=false}
\subsection{Deformations of the universal bundle}
In this article, we focus on understanding the space of deformations of the universal bundle,  with two seperate motivations in mind. First, in separate work \cite{MPSap2}, we explore the $\ap^2$ corrections to the Hull-Strominger system, which are treated as small deformations. By studying how the universal bundle adapts to such a deformation, we may learn something about the nature of the $\ap$--corrections. Second, the study of second-order deformations in the Hull-Strominger system plays a crucial role in defining the curvature on the moduli space, which is integral to any notion of special geometry. Our findings reveal that investigating second-order deformations in the Hull-Strominger system is closely tied to understanding first-order deformations of the universal bundle. Imposing the condition that the holomorphic structure of the universal bundle is preserved leads to intriguing constraints on second-order derivatives.

Let's explain what we mean by holomorphic structure. The moduli space $\M$ is associated to the space of deformations of $\X$, $E\to\X$ preserving the supersymmetry conditions. It is complex \K with a natural \K metric and \K potential. The supersymmetry conditions can be expressed as a combination of F-terms and D-terms and the Bianchi identity \citeE. The F-term conditions and Bianchi identity are
\beq
N_J \= 0~, \quad F^{0,2} \= 0~, \quad H \=  \dd^c \o~, \quad   \dd H \=\! -\frac{\ap}{4} \tr F^2 + \frac{\ap}{4} \tr R^2~.
\eeq
Part of the motivation for calling these F-terms is they can be derived from a superpotential functional \cite{LopesCardoso:2003dvb, Gurrieri:2004dt, Ashmore:2018ybe}. What we show in \citeUGSG is that these equations are satisfied on the universal geometry
\beq
N_\IJ \= 0~, \quad \IF^{0,2} \= 0~, \quad \IH \=  \Id^c \Iomega~, \quad   \Id \IH \=\! -\frac{\ap}{4} \tr \IF^2 + \frac{\ap}{4} \tr \IR^2~.
\eeq
The equations governing the universal geometry mimic that of the original heterotic vacuum. 
This holomorphic structure provides valuable insights into the deformation space of the universal bundle. On one hand, we can approach the problem using the framework developed in \citeOS, where the deformations of the F-term equations are linked to sections of a complex of specific vector bundles. The kernel of the operator in this complex corresponds to deformations of the F-terms and the Bianchi identity. On the other hand, the deformation space of the universal bundle can be studied more directly, which proves to be a fruitful approach. It offers deeper understanding of the original problem, particularly by illuminating the behaviour of second-order covariant derivatives. Similar to the first-order case explored in \citeUG, the holomorphy of the universal bundle is closely tied to the proper fixing of small gauge transformations, appropriately generalised as discussed in \citeSG. We find that for the universal bundle to exhibit a well-defined deformation theory, certain second-order derivatives must vanish. This requirement naturally extends the concept of holomorphic gauge to second-order covariant derivatives. We find a double extension structure for the universal geometry moduli space, whose structure mimics that of the original heterotic problem. Consequently, the moduli of the heterotic theory and its universal geometry are captured by similar equations. As a consistency check, we compute these second-order covariant derivatives directly by varying the original supersymmetry equations. The results align with our expectations, confirming that the extension of gauge fixing to the second-order problem is indeed a natural and necessary generalisation.

Returning the perturbation problem of the universal geometry. The first order deformation problem of the universal geometry tells us about the second order covariant derivatives of the fields of the original heterotic theory. Many of these vanish. The commutators of these derivatives are related to the curvatures on the moduli space such as $\IF_{ab}$. Provided we demand the D-terms we conjecture this gives a flat connection on the universal family. This is intriguingly similar to the work in \cite{Herbst:2008jq}, in which parallel transport of D-branes over the \K moduli space of Calabi-Yau manifolds defines a flat connection on the moduli space provided the D-terms hold. This is suggestive the universal bundle is a good structure to study heterotic special geometry.

\subsection{Outline of the article}
\vskip-10pt
In the body of this paper, a detailed discussion of the points outlined above is given.

In section 2, we review the first order deformation problem of heterotic theories. This is a two-fold calculation: computing the variations of the supersymmetry variations and then fixing the small gauge transformations. This calculation is described in detail in \citeSG and this section summsrises the salient points, improving some of the discussion as appropriate. The fixing of holomorphic gauge is required for an embedding for the heterotic moduli problem into a holomorphic universal  bundle. This theme will be repeated at second order.

In section 3, we describe some of the relations between second order deformation theory, connections on the moduli space and their curvatures. We demonstrate with the baby example of a fixed manifold $\X$ and consider deformations of just the vector bundle $E$. In this situation we can differentiate the first order results directly and show that fixing holomorphic gauge at first order, sets some but not all of the second order derivatives of the connection $A$ to zero. This is enough to show that the connection on the moduli space $\M$ is holomorphic but it is not enough to demonstrate the entire field strength vanishes.

In section 4, we embed the toy example discussed in the previous section into the universal bundle. We demonstrate demanding the universal bundle remain holomorphic under deformations requires the gauge fixing constructed directly in the previous section; the universal calculation is significantly cheaper than the direct calculation. The main lesson is that the universal bundle forces the curvature $\IF_{ab}$ of the connection on $\M$ to vanish.  We extend these results to the Atiyah-type problem in which the complex structure of $\ccU$ is allowed to vary. We then study the space of deformations both directly and via the kernel of an operator $\ID$ acting on sections of a bundle topologically given by $\IQ = \ccT_\IX^{1,0} \oplus {\rm End} \ccU$. 

In section 5, we take lessons learnt to the full problem: the space of deformations of the universal bundle. Working around a smooth point in moduli space, demanding an embedding into the universal bundle forces many second order deformations to vanish. This results in either  vanishing of commutators of deformations or vast simplifications. 

In section 6, we study a double extension structure on the universal bundle, inspired by the work of \citeOS. A long exact sequence determines a cohomology for the moduli of universal geometry. These moduli are controlled by similar equations to the original heterotic problem. 

In section 7, we demonstrate the direct calculation of second order derivatives of the full heterotic theory and proposed gauge fixing. Our results here reproduce the results in the universal calculation but a significantly more difficult and lengthy. Furthermore, their geometric significance is not clear until on embeds in the universal space. 

In section 8, we resolve an outstanding issue in \citeM: we show that the $(0,2)$ component of the deformation of the hermitian form decouples from the moduli space metric when differentiated from the heterotic \K potential 
$$
K \=\! -  \log \frac{4}{3} \int \o^3 - \log \ii \int (\Omega \bar \Omega)~.
$$
\subsection{Some notation and terminology}
\label{sref:Notation}
\vskip-10pt
It is useful to summarise some notation and terminology that we will introduce later.
\begin{itemize}
\item The fibration is depicted as
\beq\notag
\begin{tikzcd}
X \arrow[r]  & \IX \arrow[d]\\
& \M
\end{tikzcd}~, 
\qquad
\begin{tikzcd}
\Het \arrow[r]  & \ccU \arrow[d]\\
& \M
\end{tikzcd}~.
\eeq
 \item Tangibility $[p,q]$ means the form has $p$ legs along the moduli space $\M$ and $q$ legs along the fibre $X$. Our convention is that legs along the moduli space are written first.  Holomorphic or complex type is denoted by superscript e.g. $\o^{p,q}$ is a $(p,q)$-form.
\smallskip  
\end{itemize}

\begin{table}[H]
\begin{center}
\setlength{\extrarowheight}{3pt}
\begin{tabular}{|m{6cm}|c|c|}
\hline
\hfil Coordinates &~~Real indices~~ & ~~Complex indices~~\\[3pt]
\hline\hline
~Total space $\IX$\hfill $u^P$~ & $P,Q,R,S$ & \\[3pt]
\hline
~Moduli space  $\M$\hfill $y^a$~ & $a,b,c,d$ & $\a,\b, \g$ \\[3pt]
\hline
~Heterotic manifold $\X$\hfill $x^m$~ & $m,n,k,l$ & $\m,\n,\k,\l$ \\[3pt]
\hline
\end{tabular}
\capt{6.0in}{tab:coords}{The coordinates and indices for the total space, fiber and base of the fibration $\IX$.}
\end{center}
\end{table}
\begin{itemize}
 \item The family of universal bundles will be denoted  $\{ \ccU_Y\}_{Y\in \wt \M}$, in which the deformation parameters $Y^A$ are coordinates for the  moduli space $\wt\M$ of universal bundles. 
 
 \item The product structure has a curvature $S_{ab}{}^m$ with Ehresmann connection $c_a{}^m$. It being integrable is equivalent to $S_{ab}{}^m=0$. 
 
\item The vector bundle $\ccU$ on $\IX$ has gauge group $G$, the same as the heterotic theory. Its connection is $\IA = A + A^\sharp$. We use $\A \cong A^{0,1}$ and $\A\# = A^{\sharp\,0,1}$. The field strength $\IF = \Id \IA + \IA^2$ has $[2,0]$-component  $F^\sharp = \half \IF_{ab} \dd y^a \dd y^b$. Sometimes we just write $\IF_{ab}$ as a shorthand. 

\item The 3-form $\IH$ on $\IX$ has $[0,3]$ component $H$; $[1,2]$ component $\IH_a = \ccB_a$ and $[2,1]$-component $\IH_{ab}$. The $[3,0]$-component vanishes as $\M$ is \K.

\item The holotypical derivative comes from projection onto complex type $(\del_a \o)^{p,q}  \cong \fD_\a \o^{p,q}$. On real forms there's no difference $\fD_a \o = \del_a \o$. The $B$-field has a complicated gauge transformation property; deformations of it are gauge invariant and written $\ccB_a \d y^a$, where $\d y^a$ is a tangent vector to $\M$. Deformations of the hermitian form and $B$-field are combined into a complex object $\ccZ_a = \ccB_a + \ii \del_a \o$ with $\ccZb_a$ its complex conjugate.
 
\item We work with manifolds whose underlying geometry is Calabi-Yau in the sense that the manifold has vanishing first Chern class and to zeroth order in the $\ap$-expansion is Ricci-flat, \K. We include the first order $\ap$-correction which renders the metric non-\K. At a practical level this means we work with \K metric whenever there is an $\ap$-present, and the Hull connection, Levi-Civita and Bismut connection are all the same as the Chern connection.

\end{itemize}

\newpage
\section{First order deformations of \texorpdfstring{$N=1$}{N=1} \texorpdfstring{$d=4$}{d=4}  heterotic compactifications}

\subsection{A review of the background spacetime}
We consider a heterotic string compactification at large radius that is well approximated by $\ap$--corrected heterotic supergravity. We work to first order in $\ap$. The compactification is defined by a spacetime geometry 
$$
\IR^{3,1} \times \X~.
$$
We require  $N=1$ supersymmetry in $d=4$. To first order in $\ap$, the supersymmetry conditions in terms of the geometric data described in  \cite{Hull:1986kz,Strominger:1986uh}. The manifold is complex, which means it admits an integrable complex structure $J = J_m{}^n \dd x^m \otimes \del_n$, where $x^m$ are real coordinates on $\X$, equivalently a closed holomorphic volume form $\O$.

We shall henceforth omit the wedge symbol, writing for example $\dd x^{mn} \cong \dd x^m \w \dd x^n$, unless it is ambigious. The manifold has vanishing first Chern class $c_1(\X) = 0$, with a Riemannian metric $\dd s^2 = g_{mn} \dd x^m \otimes \dd x^n$,  compatible complex structure and so a hermitian 2-form $\o = \half \o_{mn} \dd x^{mn} = \ii g_{\m\nb} \dd x^{\m\nb}$. The manifold has an $SU(3)$-structure. There exists a vector bundle $V \to \X$, with a connection $A$, and a structure group being a subgroup of $E_8\times E_8$. The gauge connection is anti-hermitian meaning
$$
A \= \A - \A^\dagger~,
$$
where $\A$ is the $(0,1)$ component of $A$. The connection has a field strength $F= \dd A + A^2$. This datum satisfies the equations \eqref{eq:F-terms}--\eqref{eq:D-terms}.

\subsection{First order deformations review}	
 The deformations will always preserve $N=1$ supersymmetry. Therefore,  the moduli space  is a complex \K manifold. We denote $y^a = (y^\a, y^\bb)$ the corresponding complex  coordinates: small deformations of fields correspond to tangent vectors $\d y^a$. This is known as the Kuranishi map in the context of complex structure deformations \cite{Kuranishi:1965} and we extend it to deformations of heterotic structures $\Het$. \footnote{We thank Roberto Sisca for this summary.}

In \citeSG, our approach followed these steps:

\begin{enumerate}
\item Deform the F-term equations and Bianchi identity.
\item Analyse gauge symmetries (diffeomorphisms, rescaling of $\O$, gauge and gerbe transformations).
\item Define holomorphy by choosing a complex structure on the moduli space $\M$.
\item Fix the gauge (holomorphic gauge and a condition on $\d\O$).
\item Use the D-term equations to fix the exact and co-exact terms in the Hodge decomposition. 
\end{enumerate}
We now summarise the key results.

That $J^2 = -1$ implies that $\d J_\m{}^\n = \d J_\mb{}^\nb = 0$. We introduce projectors onto holomorphic and antiholomorphic coordinates respectively \cite{YanoBook}
$$
P \= \half (1 - \ii J)~, \qquad Q \= \half (1+\ii J)~.
$$
To connect with the notation of \citeUG, we write $\d J^\m = 2\ii \D^\m$ or $\d P^\m = - \d Q^\m = \D^\m$. 
We can write $\O$ in the following way
\beqnn
 \O \= \frac{1}{3!} \, f \, \e_{\m\n\r} \, \dd x^{\m\n\r} \quad \text{with} \quad |\O|^2 \= \frac{|f|^2}{\sqrt{g}} ~. 
\eeqnn
with the antisymmetric  symbol having $\e_{123} = 1$. Then,
\beq
\begin{split}
\label{eq:Fterms}
 (\dd\O = 0) \qquad &  ~ \delb \, \d P^\m \= 0 ~,\\[0.1cm]
& \nabla_\m^{\Ch} \, \d P^\m ~=~ \delb \, \d\log f \= \delb \, \big( \d\log{|\O|^2} + \d\log{\sqrt g} - \d\log{\overline f} \big) ~,\\[0.2cm]
 (F^{0,2} = 0) \qquad &\delb_{\cA}\d\cA ~=~ \d P^\m \, F_\m ~,\\[0.3cm]
 (H - \dd^c\o = 0) \qquad &\delb \, \d\ccZ^{0,2} \= 0 ~, \\
 &\delb \, \d\ccZ^{1,1 } + \del \, \d\ccZ^{0,2} \= 2\ii \, \D^\m (\del\o)_\m + \frac{\ap}{2} \tr{(\d\cA \, F)} ~,\\[0.2cm]
   (\dd \log{|\O|^2} = 0) \qquad & \dd \, \d\log{|\O|^2} \= 0 ~,\\[0.2cm]
 (\dd\o^2 = 0) \qquad 
  &\del^\dag \d\o^{1,1 } ~=~ \delb^\dag \d\o^{0,2} - \ii \, \delb \, \d\log{\sqrt g} ~,\\[0.2cm]
 (F\o^2 = 0) \qquad &\delb^\dag_{\cA} \d\cA - \del^\dag_{\cA^\dag} \d\cA^\dag  + \half \, ( \ccZ \contr F )\=0 ~.\\[0.3cm]
\end{split}\raisetag{4.5cm}\eeq

The quantity $\d\ccZ = \d y^a \ccZ_a$  --- where we expand in $\d y^a$, being tangent vectors to the moduli space at a point, and do so  with a prescience for later on --- is given by the complex combination
\beq
 \d\ccZ \= \d\ccB + \ii \, \d\o \quad \text{where} \quad \d\ccB \= \d B - \frac{\ap}{4} \big( \tr{(\d A \, A)} - \tr{(\d\Th \, \Th)} \big) ~.
\eeq
We can expand $\d \ccB = \d y^a \ccB_a$. 

The gauge symmetries, diffeomorphisms, gauge transformations, gerbe transformations have infinitesemal limits with parameters denoted by a vector $\ve^m = \ve^\m, \ve^\nb$, an endomorphism valued scalar $\phi$ and a gerbe $\mathfrak{b}$ (locally a 1-form, globally not). There is also a $\IC^*$ gauge symmetry over $\M$: $\O \to e^\l \O$, where $\l$ depends on parameters, constant over $\X$. The first order deformations transform  
\beq
\begin{split}
 (\ve) \qquad &\D^\m \sim \D^\m + \delb \ve^\m ~,\\[0.3cm]
 (\ve,\l) \qquad &\d\log f \sim \d\log f + \nabla^{\Ch}_\m\ve^\m + \l ~,\\[0.3cm]
 (\ve,\ph) \qquad &\d\cA \sim \d\cA + \ve^\m \, F_\m + \delb_\cA \phi ~,\\[0.2cm]
 (\ve,\ph,\mathfrak{b}) \qquad &\d\ccZ \sim \d\ccZ + \ve^m (H + \ii \, \dd \o)_m + \dd \left (\mathfrak{b} + \ii \, \ve^m \, \o_m\right) + \frac{\ap}{2} \tr\! \left( F \phi \right) ~.
\end{split}
\eeq
We expand a deformation in terms of holomorphic and antiholomorphic parameters using an appropriate covariant derivative, for example
\beq
 \d P^\m \= \d y^a \, \fD_a P^\m \= \d y^\a \, \fD_\a P^\m + \d y^{\ab} \, \fD_{\ab}P^\m ~.
\eeq
Note that $\fD_a P = \D_a^{0,1} - \D_a^{1,0}$, not $\D_b^{0,1} + \D_b^{1,0}$ so we work with $P$ and not $\D_a$.  Holomorphy means that certain combinations of antiholomorphic variations are exact, for example
\beq
 \D_{\ab}{}^{\m} \= \delb \k_{\ab}{}^\m ~, \quad \fD_{\ab}\cA \= \k_{\ab}{}^\m \, F_\m + \delb_{\cA}\Ph_{\ab} \quad \text{and so on...}
\eeq
This defines a complex structure on $\M$, because under a reparameterisations $\wt{y}^\a = \wt{y}^\a(y^\b, y^{\bb})$, which are not holomorphic, the combinations above acquire a non-exact bit. Holomorphic gauge is defined by killing these exact terms as far as possible. 
\beq\begin{split}\label{eq:holomorphicgauge}
 &\hspace{1cm}\D_\ab{}^\m \= 0 ~, \quad  \fD_\ab \cA \= 0 ~, \quad \fD_{\ab}\O \= 0 ~,\\[0.2cm]
 &\ccZb_\ab^{2,0 } \= 0 ~, \quad \ccZ_\ab^{1,1 } \= 0 ~, \quad \ccZ_\ab^{0,2} \= \ccZb_\ab^{0,2} \= 0 ~.
\end{split}\eeq
There is a residual freedom, which we use to impose that $\fD_\a\O^{(3,0)}$ is harmonic. This implies a condition on hermitian deformations from the $SU(3)$ structure relations as 
\beq
 \dd ~ \fD_\a\log{\sqrt g} \= \dd (\o \contr \fD_\a\o) \= 0 ~,
\eeq
that simplifies the variation of the balanced condition.

The physical degrees of freedom are
\beq\label{eq:DOF}
\fD_\a \A~,\quad \D_\a{}^\m~,\quad    \ccZ_\a^{1,1 }~,
\eeq
where note that $\fD_\a \o^{0,2} = \D_\a{}^\m \o_\m$ is non-vanishing and
 $$
 \ccZ_\a^{1,1 } \= 2\ii \fD_\a \o^{1,1 } = 2 \ccB_\a^{1,1 }~, \qquad \ccZb_\a^{0,2} {\=}2 \ccB_\a^{0,2} {\=}\! -2\ii \ccD_\a \o^{0,2}~.
 $$

 Now that the gauge symmetries have been fixed and the physical degrees of freedom isolated, go back to the equations of motion 
\beq\begin{split}\label{eq:ALLeomsgf}
 (\dd\O = 0) \qquad &\delb \, \D_\a{}^\m \= 0 ~, \qquad \nabla_\m^{\Ch} \, \D_\a{}^\m ~=~ 0 ~,\\[0.3cm]
(F^{0,2} = 0) \qquad &\delb_{\cA} \, \fD_\a\cA ~=~ \D_\a{}^\m \, F_\m ~,\\[0.1cm]
 (H - \dd^c\o = 0) \qquad &\delb \, \fD_\a\o^{1,1 } \= \D_\a{}^\m (\del\o)_\m - \frac{\ii\ap}{4} \Big( \tr{(\fD_\a\cA \, F)} - \tr{(\fD_\a\Th \, R)} \Big) ~,\\[0.1cm]
 (\dd\o^2 = 0) \qquad &\delb^\dag \, \fD_\a\o^{1,1 } ~=~ 0 ~, \qquad \del^\dag \, \fD_\a\o^{1,1 } ~=~ \delb^\dag \, \fD_\a\o^{0,2} ~,\\[0.3cm]
 (F\o^2 = 0) \qquad &\delb^\dag_{\cA} \, \fD_\a\cA + \half \ccZ_\a \contr F \= 0 ~.\\[0.2cm]
 \end{split}\eeq

We can identify a subset of these equations as F-terms:
\beq\label{eq:moduliEqnHol}
\begin{split}
\delb \D_\a{}^\m &\= 0~, \\[0.15cm]
\delb_\A \fD_\a \A &\= \D_\a{}^\m F_\m~, \\
 \delb \ccZ_\a^{1,1 } &\= 2\ii \, \D_\a{}^\m (\del\o)_\m +\frac{\ap}{2} \tr \big( \fD_\a \A \, F\big) - \ap \, \big( \nabla_\n \D_\a{}^\m + \ii \, \nabla^\m \, \fD_\a\o_\n^{0,1} \big) R^\n{}_\m ~,
\end{split}
\eeq
where $R^\m{}_\n = R_{\r\sb}{}^\m{}_\n \, \dd x^{\r\sb}$ is the Riemann tensor and the spin connection depends on the moduli of $\X$ \citeUG:
\beq\label{eq:spinmoduli}
 \fD_{\a}\Th{}_\mb{}^\n{}_{\s} \= \nabla_\s \, \D_{\a\mb}{}^\n + \ii \, \nabla^{\n} \, \fD_\a\o_{\s\mb}  ~.
\eeq
These are derivable from a superpotential functional~$W = \int_\X \O (H+\ii \dd \o)$  provided deformations are in holomorphic gauge \eqref{eq:holomorphicgauge} \cite{delaOssa:2015maa,McOrist:2016cfl,Ashmore:2018ybe,McOrist:2019mxh}. 

If there are F-terms there should be D-terms. In \citeE it is demonstrated there is an operator $\Dbar$ on an extension bundle who kernel amounts to F-terms and D-terms amounts to the adjoint of $\Dbar$ computed with respect to the moduli space metric in \citeM. They are
 \beq\label{eq:DtermsDefs}
\begin{split}
& \nabla^{{\textrm B} \,\mb} \D_{\mb\nb} - \half H^{\s\tb}{}_\nb \ccZ_{\s\tb}  + \frac{\ap}{4} \tr F_\nb{}^\lb \aa_\lb  - \frac{\ap}{4} R_\nb{}^{\mb\s\tb} \nabla_\mb \ccZ_{\s\tb} - \frac{\ap}{2} R_\nb{}^{\sb\t\mb} \nabla_\t \D_{\sb\mb} \= 0~,\\[5pt]
&\delb^\dag \ccZ_\n - \frac{\ap}{2} R^{\t\m\sb} \nabla_\m \ccZ_{\t\sb} \= 0~,\\[5pt]
&\delb_\A^\dag \aa  +  \half F^{\r\nb} \ccZ_{\r\nb} \= 0~.
\end{split}
 \eeq
These equations come from a combination of the  last three lines of \eqref{eq:ALLeomsgf} and \eqref{eq:spinmoduli}.

One can now use this to solve for the Hodge expansion of the physical degrees of freedom \citeSG. Dropping now the $\a$ subscripts and $(p,q)$ superscripts, we see that
$$
\D \= \D^{\rm harm} + \delb \k~, \qquad \ccZ^{1,1 } \= \ccZ^{\rm harm} + \delb^\dag \xi~, \qquad \fD \A \= \aa^{\rm harm} + \delb_\A \Psi + \delb_\A^\dag \psi~.
$$
By consideration of type it is also the case that
$
\ccZb^{0,2} = \del^\dag \wt \xi~.
$

\newpage

\section{Connections, covariant derivatives and the universal bundle}
\label{s:covDerivs}
In the the study of the string theory moduli space of Calabi-Yau manifolds special geometry could be phrased as the existence of a flat connection on the moduli space. Alternatively, parameter space derivatives were covariant with respect to a bundle and those derivatives commute - this is explicit in the seminal work of \cite{Candelas:1989bb,Strominger:1990pd} and later in the context of $tt^*$ e.g.~\cite{Bershadsky:1993cx} and in this section we study what happens when we do not necessarily make this assumption in the more general context of a heterotic theory.

Recall, that in the study of the moduli of a CY manifold (e.g. at the standard embedding in heterotic),  the top form $\O$ is a section of a line bundle $\cL$ over the moduli space $\M$; the corresponding covariant derivatives, at second order, have non-trivial mixed terms
$$
 \d y^\bb \,\d y^\a \, [\fD_\bb,\fD_\a]\O \=\!- \d y^\a \d y^\bb G_{\a\bb} \O
$$
where $G_{\a\bb}$ corresponds to the metric on deformations of complex structure of a CY manifold. This result follows from using the canonical choice of connection on the line bundle, that being a derivative of the \K potential $K_{,\a}$. We also use that $\fD_\ab \O = \del_\ab \O = 0$, a statement of holomorphy at first order.  As $\cL$ is a line bundle this is in fact the first Chern class  and the calculation here demonstrates it coincides with the metric on the moduli space of complex structures. It goes without saying, but this implies that $\d \O$ has mixed derivative terms at second order in the derivative expansion and the bundle $\cL$ is holomorphic but not flat. Both these are  lessons for the general case.

Consider a second example:  deformations of the connection for the gauge bundle. We write
\beq\notag
\begin{split}
 \d \A  &\= \d y^a \fD_a \A + \half \d y^a \d y^b \fD_{(a}\fD_{b)} \A + \cdots,\\
\end{split}
\eeq
where $\fD_a \A = \del_a A - \dd_A A\#_a$ and $A\#_a$ is a connection 1-form on the moduli space and as $\d y^a\d y^b = \d y^b \d y^a$ we get a projection onto the symmetric component. Taylor series for sections of bundles naturally involve symmetrised covariant derivatives and have a natural connection to the theory of jet bundles (the $k$th jet carrying information about the section, and its derivatives  up to the $k$th covariant derivative, symmetrised).  

The bundle $E$ being holomorphic means that $F^{0,2} = 0$. Differentiating this with respect to parameters:
\beq\label{eq:MCGauge}
\delb_\A \fD_a \A \= 0~,\qquad \delb_\A \left(\fD_a\fD_b \A \right) + \{ \fD_a \A, \fD_b \A\} \= 0~. 
\eeq
The first order term is solved by Hodge decomposition $\fD_a \A = \x_a^{\rm harm} + \delb_\A \m_a$ where $\x_a^{\rm harm}$ are zero modes of  $\Box_{\delb_\A}$ and $\m_a$ are sections of $H^0({\rm End}E)$. As this is a deformation of $\A=A^{0,1}$, there is no reality constraint on $\mu_a$.  We pick our complex structure on the moduli space so that 
$$
\fD_\a \A = \xi_\a^{\rm harm} + \delb_\A  \m_\a~,\qquad  \fD_\ab \A \= \delb_\A  \m_\ab~, 
$$
At second order,
$$
\fD_b \fD_\ab \A \= \delb_\A( \fD_b \m_\ab) + [\fD_b \A, \m_\ab]~,\qquad \fD_b \fD_\a \A \= \fD_b \xi_\a^{\rm harm} + \delb_\A( \fD_b \m_\a) + [\fD_b \A, \m_\a]~.
$$
We use the index $b$ as shorthand for both the holomorphic and antiholomorphic cases $b=\b,\bb$. 

The connection $A\# = A\#_a \dd y^a$ can be deformed $A\#_a \to A\#_a + \phi_a$ with  $\phi_a \dd y^a$ any endomorphism valued 1-form on $\M$ such that $(\phi_a)^\dag = -(\phi_a)$. Under this deformation,
$$
\fD_a \A \to \fD_a \A - \delb_\A \phi_a~,\qquad  \fD_a\fD_b \A \to \fD_a \fD_b \A - \delb_\A( \fD_a \phi_b) +  [\delb_\A \phi_b,\phi_a ]-  [\fD_a \A, \phi_b] - [ \fD_b \A,\phi_a ]~.
$$
As $\phi^\dag = -\phi$, in complex coordinates $\phi_\ab = (\phi_\a)^\dag$, a constraint not present for $(\mu_a)^\dag \ne \mu_a$. So we can choose $\phi_\ab = \mu_\ab$ to remove some but not all of the $\delb_\A$-exact terms
\beq\label{eq:SecOrdGauge}
\begin{split}
 \fD_\a \A &\= \xi_\a^{\rm harm} + \delb_\A \mu_\a \= \x_\a~,\quad  \fD_\ab \A \= 0~,  \\
  \fD_\a\fD_\b \A &\= \fD_\a\xi_\b~, \quad  \fD_\ab \fD_\b\A \= \fD_\ab \xi_\b~, \\
     \fD_\a\fD_\bb \A &\= \fD_\ab\fD_\bb \A \= 0~.
\end{split}
\eeq

Here $\x_\a$ is simply a representative of the cohomology group $H^1(\X,\End E)$. 
Using that $\delb_\A \xi_\a = 0$, we can check that  \eqref{eq:SecOrdGauge} satisfies the Maurier-Cartan equations:
$$
\delb_\A (\fD_\a\fD_\b \A) + \{ \fD_\a \A, \fD_\b \A\} \= 0~, \qquad \delb_\A (\fD_\ab\fD_\b \A)  \= 0~.
$$
As an aside, if in addition we require the hermitian Yang-Mills equation hold $\o^2 F= 0$ then at first order
$$
 \delb_\A^\dag \fD_\a \A \= 0~.
$$
This additional equation forces $\m_\a = 0$ and so $\x_\a = \x_\a^{\rm harm}$. This does not occur in the more general heterotic situation \citeSG. However, the lesson is important: imposing the D-term seems important to curvature of the connection on the moduli space, mirroring an observation of \citeHHP in the context of D-brane transport over moduli of CY manifolds.

  The choice of connection $A\#$ also determines the curvature $\IF_{ab}$ where
$$
[\fD_a, \fD_b] A \=\! -\dd_A \IF_{ab}~.
$$

The last equation of \eqref{eq:SecOrdGauge} together with stability of $E$ so that $H^0(X,\End E) = 0$ implies $\IF_{\ab\bb} = 0$. That $\IF$ is antihermitian implies $\IF_{\a\b} = 0$. Hence, going to holomorphic gauge implies $\IF_{\ab\mb} =0$ at first order and $\IF_{\ab\bb}=0$ at second order. In \citeUG we took $\IF$ to be  holomorphic as an assumption. Here  we have gone one better and  derived this.

Furthermore
  $$
 [\fD_\a,\fD_\b]\A \= \fD_\a \xi_\b - \fD_\a \x_\b \= 0~, \qquad  [\fD_\a,\fD_\bb]\A \= \fD_\bb \xi_\a \= \! - \delb_\A \IF_{\a\bb} ~,
 $$
We see that $\xi_\a$ is a $\fD$-closed when viewed as an endomorphism valued $1,0$-form on the moduli space. Furthermore, $\xi_\a$ is both $\fD$- and ${\ol \fD}$-closed if and only if $\IF_{ab} = 0$, a flat bundle on the space of deformations.  

In terms of small deformations, we find that $\IF_{ab} = 0$ if and only if $\d \A$ is
\beq\notag
\begin{split}
 \d \A  &\=  \d y^\a \fD_\a \A + \half \d y^\a \d y^\b \fD_{(\a}\fD_{\b)} \A + \cdots~.
\end{split}
\eeq
with $[\fD_\a,\fD_\b]\A=0$. 

On the other hand, if  $\IF_{\a\bb}\ne0$, we find an additional term
\beq\notag
\begin{split}
 \d \A  &\=  \d y^\a \fD_\a \A + \half \d y^\a \d y^\b \fD_{\a}\fD_{\b} \A + \half \d y^\ab \d y^\b \fD_\ab\fD_\b \A \cdots~,
\end{split}
\eeq
where because $[\fD_\a,\fD_\b]\A=0$, the symmetrisation above is on autopilot.

This example provides the first part of the paradigm for the full heterotic geometry.  The second part of the paradigm is to study deformations of the universal bundle, such that they preserve the holomorphic structure. In the baby example here it is simply $\IF^{0,2} = 0$ is invariant under perturbations. We pursue this in the next section.

 In Kodaira-Spencer theory of complex structure there is a deep connection between the existence of a flat connection and integrability of deformations. Similarly, in the $tt^*$ construction of Cecotti-Vafa, existence of a flat connection on the moduli space has a deep relationship with the theory being integrable. When the connection is flat it means we can find local coordinates in which we can eliminate the covariant derivatives and the study of our moduli space vastly simplified. Finally, in \citeHHP flatness of parallel D-brane transport over a CY moduli space is reliant on imposing D-terms. So with this circle of ideas in mind, we press on.

\newpage
\section{Deformations of the universal bundle for the Atiyah problem}
\label{s:Atiyah}

We now turn to the framework of universal geometry. A summary of results derived in \citeUG is provided in \sref{s:summaryUniv}. 

Universal geometry is a family of vector bundles which itself is a vector bundle and denoted $\cU \to \IX$, where $\IX$ is topologically $\IX = \M \times \X$.  It generalises the Kodaira-Spencer family of complex manifolds to families of heterotic supergravity theories. The base manifold $\IX$ is a complex manifold characterised by its complex structure $\IJ$, Hermitian form $\Iomega$, $\IH$, and a gauge field $\IA$. The real coordinates on $\IX$ are $(y^a, x^m)$, while the holomorphic coordinates are $(y^\a, x^\m)$. An Ehresmann connection, $c_a{}^m$, defines the product structure of $\IX$, implying that $\X$ is fibered over $\M$. The field strength $S_{ab}{}^m$ of this connection governs the integrability of the product structure. It is integrable if $S_{ab}{}^m = 0 $ and this is the same as the Nijenhuis tensor vanishes. We assume the product structure is integrable so that $S=0$.\footnote{This vanishing is an underlying assumption in the Kodaira-Spencer construction of a family of complex manifolds over their Kuranishi or moduli space.}

On each fibre of $\cU$ there is a heterotic structure defining an $N=1$ $d=4$ vacuum of the heterotic string. That is, for fixed $y_0\in \M$, we have $[X,E]$ with $\o,J,B,A$ satisfying the equations of Hull-Strominger \eqref{eq:Fterms}.  To realise this, the universal bundle $\cU$ is equipped with a connection $\IA = A + A^\sharp$, where $A$ is the connection on $E \to X$ and $A^\sharp$ is a connection on $\M$, as described in \sref{s:covDerivs}. We sometimes refer to $\IA$ as the extension of $A$ to the  universal bundle. Its field strength $\IF$ takes the conventional form: $\IF = \Id \IA + \IA^2$, and encodes both the structure of the bundle $E$ and its deformations. The structure group of the vector bundle in the heterotic theory, $E \to X$, is denoted by $\cG$, and under gauge transformations, the connection transforms as ${}^\Phi A = \Phi^{-1} A \Phi - \Phi^{-1} \dd \Phi$, where $\Phi$ is valued in the adjoint of $\cG$. Since $\Phi$ also depends on parameters, the connection $\IA$ extends to a connection on a $\cG$-bundle $\cU \to \IX$. In principle, we could allow $A^\sharp$ to transform within a larger structure group that contains $\cG$ as a subgroup, such as $\cG \times \cH$. However, for simplicity, we focus on $A^\sharp$ transforming in $\cG$, as the other components like $\cH$ decouple from the physical theory. Tangibility $[p,q]$ of a form means that $p$-legs lie along $\M$ and $q$-forms lie along $X$. 

While on each fibre of the universal bundle, the full supersymmetry equations and Bianchi identity hold, it is shown in \citeUS that the F-terms and Bianchi identity have an extension to the universal bundle:
\beq
N_\IJ = 0~, \quad \IF^{0,2} = 0~, \quad \IH \=  \Id^c \Iomega~, \quad   \Id \IH \=\! -\frac{\ap}{4} \tr \IF^2 + \frac{\ap}{4} \tr \IR^2~.
\eeq
The $[1,1]$ component of $\IF$ corresponds to  $\fD_a A$, the first-order deformation of $\A$, and the $[1,2]$ of $\IH$ to the first-order deformation $\ccB_a$. Similar relations hold for $\d \o$ and $\d J$. Notably, complex algebraic identities in deformation theory simplify elegantly as tensor identities of the universal bundle. A natural question arises: how does the universal bundle capture second and higher-order deformations? Are they linked to the $L_3$ algebra from \citeS? 

Given that \cite{Ashmore:2018ybe} analyses F-terms and the Bianchi identity, and key aspects of the the universal bundle is sensitive to this warrants further investigation. 

\subsection{Warm-up: deform \texorpdfstring{$\IF$}{F} with everything else fixed}
\label{s:deformF}
Suppose  $\{\cU_Y\}$ is a family of universal bundles, where  $Y$ is a label for family members. It is also a coordinate of an auxiliary moduli space $\wt \M$. As a warm-up we consider deformations of the universal bundle connection, given by $\IA \to \IA + \d \IA$, that preserve the condition $\IF^{0,2} = 0$, holding all other structures fixed. This corresponds to $Y_0 \to Y_0+\d Y$ where $\cU_{Y_0}$ is our basepoint: the universal bundle about which we are deforming. Within $\cU_{Y_0}$ we assume that for each $y\in \M$, the fibre $[X,E]$ satisfies heterotic supersymmetry equations: F-terms, D-terms and holomorphic gauge. 

Recall that $\IF^{0,2} = 0$ is equivalent to three statements \citeUGSG:
\begin{enumerate}
 \item the bundle $E\to \X$ is holomorphic $F^{0,2} = 0$;
 \item first order deformations being holomorphic $\fD_\ab \A = 0$;
 \item the connection $A^\sharp$ on the moduli space is holomorphic $\IF_{\ab\bb} = 0$.
\end{enumerate}
The first follows from supersymmetry. The second from holomorphic gauge. The third we assumed to be true in \citeUGSG but in fact what we have shown in \eqref{eq:SecOrdGauge} is that it follows from holomorphic gauge.  $\cU$ being holomorphic is equivalent to holomorphic gauge.

We demand  $\ccU_Y$ be holomorphic, which to first order in $\d$:
$$
\big(\Id_\IA \d \IA\big)^{0,2} \= 0~.
$$
As a form $\d \IA^{0,1} = \d \A \oplus \d \A\# $. Decomposing  according to tangibility:
\beq\label{eq:defUnivA}
\begin{split}
\delb_\A \d \A  &\=0~, \\
\fDb \d \A + \delb_\A \d \A\#  &\= 0~, \\[0pt]
\fDb \d \A\#     &\=0~,
\end{split}
\eeq
where we use that $\Idelb_\IA =\delb_\A \oplus   \fDb$. Small gauge transformations are
\beq\label{eq:UnivGauge}
\d \IA \sim \d \IA + \Idelb \Psi ~~\Rightarrow~~ \d \A \sim \d \A + \delb_\A \Psi~, ~~{\rm and}~~ \d \A\# \sim \d \A\# + \fDb \Psi~.
\eeq

The first equation in \eqref{eq:defUnivA} implies deformations of $\d \IA$ restricted to the fibre must satisfy the familiar equation for deformations of $\A$. To first order in $\d$, we can expand with respect to coordinates of $\ccU_{Y_0}$:
\beq
\d \A \in H^{0,1} (X, \EndE)~, \quad \d \A = \d y^\a (\fD_\a \A + \delb_\A \mu_\a)~,
\eeq
where we use \eqref{eq:UnivGauge} to impose holomorphic gauge $\fD_\ab \A = 0$ (see \citeSG for details on how to do this) and $\a$ runs over the dimension of the cohomology group, $\fD_\a \A_\mb = \IF_{\a\mb} $ and for this particular case is a harmonic representative of $H^{0,1} (X, \EndE)$ and $\mu_\a$ is an arbitrary section of $\EndE$. 

Looking ahead to the full moduli problem, we want deformations of the universal bundle, when restricted to the fiber, to solve the heterotic equations of motion. So far, we have imposed the F-term condition, but we must also satisfy the D-term, as seen in the last line of \eqref{eq:ALLeomsgf}. In this toy model, this is equivalent to requiring that the connections $\A$ and $\A + \d \A$ satisfy the Hermitian Yang-Mills equation. That being so,  $\mu_\a = 0$. However, we could also consider a deformation problem where we do not impose the same D-terms, which we expect to arise when considering how the universal bundle structure changes with $\ap$-corrections. This is because $\ap$-corrections are expected to modify the D-terms while leaving the F-terms unchanged \citeW. In that case, we must keep $\mu_\a$. For now however, we consider the simpler problem in which the fibres continue to solve the D-terms.\footnote{We thank Philip Candelas for this observation.}

That being so, we have
$$
\d \IA^{0,1} \= \d y^\a \fD_\a \A + \d Y^A \d_A \A\# + \cdots~,
$$
where the $\cdots$ are higher order terms in $\d Y^A$. Deformations of the connection $A\#$ on $\M$ are written  $\d \A\# = \d y^a \d_a \A\# + \d Y^{a'} \d_{a'} \A\#$ where $\d Y^A = (\d y^a, \d Y^{a'})$ and  $Y^{a'}$ are parameters that deform $A\#$ leaving $\A$ fixed. These are gauge transformations of $\A$. 

The second equation in \eqref{eq:defUnivA} is
\beq\label{eq:mixed1}
\fD_\ab \fD_\b \A - \delb_\A \d_\b \A\#_\ab \= 0~,\quad \fD_\ab \fD_\bb \A - \delb_\A \d_\bb \A\#_\ab \= 0~, \quad \delb_\A (\d_{a'} \A\#_\ab) \= 0~.
\eeq
Consider for a moment $\cU_{Y_0}$. On a neighbourhood of a point in the base $y\in \M$, we have $\IF^{0,2}=0$ on an open neighbourhood and so any hermitian covariant derivative also  vanishes: 
 \beq\label{eq:derivsF02}
\begin{split}
  \fD_\a \IF_{\bb\mb} &\= \del_\a\IF_{\bb\mb}  + [\A\#_\a,\IF_{\bb\mb}]   \=   \fD_\a \fD_\bb \A_\mb \= 0 ~, \\[2pt]
   \fD_\ab \IF_{\bb\mb} &\= \del_\ab\IF_{\bb\mb} -\G_\ab{}^\nb{}_\mb \IF_{\bb\nb}  + [\A\#_\ab,\IF_{\bb\nb}] \=  \fD_\ab \fD_\bb \A_\mb \= 0~,
\end{split}
\eeq
 where because $\IF_{\bb\nb}$ vanishes in a neighbourhood, all partial derivatives of it vanish, as do  any linear combinations. When this is combined with \eqref{eq:mixed1} and stability of $E$ (i.e. D-terms hold on the fibre) we have $\d_\bb \A\#_\ab = \d_\b \A\#_\a = 0$. 
Notably \eqref{eq:derivsF02} coincides with the result derived in \eqref{eq:SecOrdGauge}.  

We can learn something by considering $\wt \IF$ in $\ccU_Y$, which is related to the basepoint field strength by $\wt \IF = \IF + \Id_\IA \d \IA + \cdots$. The tangibility $[1,1]$ component of $\wt \IF$ is
$$
\wt \fD_a\wt \A \= \fD_a \A  + \fD_a \d \A  - \delb_\A \d \A\#_a~.
$$
We can interpret this as a simultaneous deformation of $\A$ and small gauge transformation. 

Decomposing into complex parameters of $\ccU_{Y_0}$, we see $\wt \fD_\ab \wt \A = 0$, leaving 
\beq\label{eq:DefF}
\begin{split}
 \wt \fD_\a \wt \A&\= \fD_\a \A     + \d y^\bb \,( \fD_\a \fD_\bb \A - \delb_\A \d_\bb \A\#_\a)+ \d y^\b\,  (\fD_\a \fD_\b \A  - \delb_\A \d_\b \A\#_\a ) ~,\\[5pt]
\end{split}
\eeq 
We use \eqref{eq:derivsF02} so that $\fD_\a\fD_\bb \A = 0$ and $\d_b A\#_a = 0$. We take deformations of the universal bundle to depend holomorphically on parameters  which forces  $\delb_\A (\d_\bb A\#_a) = 0$. Equivalently, fibres of $\ccU_Y$  satisfy the Hermitian Yang-Mills equation.  Putting it all together we get
$$
\delb_\A (\d_{\b} \A\#_\ab) \= 0, \quad
 \delb_\A (\d_{\bb} \A\#_\ab) \= 0, \quad \delb_\A (\d_{a'} \A\#_\ab) \= 0~.
$$
Using $H^0(X,\EndE)=0$ (fiberwise D-term), we see 
$$
\d_\b \A\# \=
 \d_\bb\A\#\=\d_{a'} \A\# \= 0~.
$$

Hence, deformations of $\A$ in $\ccU_Y$ are related to $\ccU_{Y_0}$ by
\beq\label{eq:DefF2}
\begin{split}
 \wt \fD_\a \wt \A &\= \fD_\a \A  + \d y^\b\,  \fD_\a \fD_\b \A  ~,\qquad \wt \fD_\ab\wt \A \= 0 ~,
\end{split}
\eeq 
Along the way we learn something about second order derivatives in $\ccU_{Y_0}$:
\beq
\fD_\ab \fD_\b \A \= 0~, \quad \fD_\a\fD_\bb \A \= 0~, \qquad \fD_\ab\fD_\bb \A \= 0~.
\eeq

We also need to take into account possible gauge transformations  $\d \IA^{0,1} \sim \d \IA^{0,1}  + \Idelb_\IA \Phi$. The leg along the manifold $\X$ is the usual gauge transformation law: $\d \A \sim \d \A + \delb_\A \Phi$. We want to preserve holomorphic gauge, and so $\delb_\A \Phi = 0$. Stability of the bundle implies then $\Phi = 0$. So in fact, there are no gauge transformations on $\IA$ permitted. 
 
One might worry that it was too fast to set $\d_a \A\#, \d_{a'}\A\#$ to zero; it could be a non-vanishing function of just parameters and the equation above implies it simply does not depend on $\X$. However, what we have shown is that for each  $y\in \M$, there is a section $\d_A\A\#$ of $H^0(X,\EndE)$. The cohomology vanishing tells us it must be the zero section. This did not depend on $y$, and so $\d_A \A\#$ is a zero section also over $\M$.

In conclusion, a deformation of $\A$ is holomorphic to second order in parameters:
\beq
 \d \A \= \d y^\a \fD_\a \A + \half \d y^\a \d y^\b \fD_\a \fD_\b \A + \cdots~,
 \eeq
and because of the relation
\beq
[\fD_a,\fD_b] \A \= -\delb_\A \IF_{ab} \= 0~.
\eeq
we find the bundle is flat
\beq
\IF_{ab} \= 0~.
\eeq

The main lesson is first order deformations of the universal bundle are connected to second order deformations of the original heterotic theory. The imposition of the F-terms on the universal bundle, equivalently holomorphy of the bundle, and D-terms (via $E$ being stable equivalently, solving $A$ solving the HYM) restricted to the fibres were restrictive.

It is important here that we took the connection $\IA$ to transform in the gauge group $\cG$, identified with that on fibres. We could have taken a more general situation in which $\IA$ transforms in a larger group such that it restricts to $\cG$ on fibres. This would naturally generate additional deformations of the universal bundle, but we do not study them as they do not obviously tell us about the heterotic theory, which is our primary interest.

\subsection{The Atiyah problem: deformations of  \texorpdfstring{$\IJ$}{J} and \texorpdfstring{$\IA$}{A}}
\label{s:UniAtiyah}
We now build on the previous subsection to introduce simultaneous deformations of complex structure $\IJ$ and the gauge connection $\IA$ preserving integrability $N_\IJ =0$  and holomorphy $\IF^{0,2} = 0$. In terms of tangibility, the complex structure is block diagonal $\IJ = J \oplus J^\sharp$, with $J^\sharp$ the complex structure on the moduli space $\M$. We demand deformations preserve $\IJ^2=  -1$ and block diagonal structure. We have two tools at the workbench: the paradigm of the previous subsection and the approach of Atiyah \cite{Atiyah:1955}. 

A deformation of complex structure $\IJ \to \IJ + \d \IJ$ preserving integrability can be phrased in terms of the Fr\"ohlicher-Nijenhuis bracket 
$$
[\IJ, \d \IJ]_{FN} \= 0~,  
$$

 Evaluate this on a complex structure in the usual way to find, to first order in $\d$,
\beq\label{eq:cpxdef}
\Idelb \d \IJ^{0,1\,\m} = 0~,\qquad  \Idelb \d \IJ^{0,1\,\a} \= 0~.
\eeq

We find
\beq\label{eq:IJDefns}
\delb \d J^\m \= 0~, \quad \fDb \d J^{0,1\,\m} \= 0~, \quad \delb \d J^{\sharp\,0,1\,\a} \= 0~,\quad  \fDb \d J^{\sharp\,0,1\,\a}\= 0~.
\eeq
The first equation implies
\beq\label{eq:dJ}
\d J^\m \= 2\ii \d y^\a \D_\a{}^\m~,
\eeq
where $\D_\a$ are a basis for $\delb$-closed $0,1$-forms on $X$ valued in $\ccT_X^{1,0}$. They can depend on parameters. The second equation implies $ \fD_\bb \fD_\a J^\m = 0$ and $\fD_\ab\fD_\bb J^\m = 0$. As we have $S_{ab}{}^m=0$, $\fD_\a\fD_\bb J^\m = 0$. As $\dim H^0_\delb(X,\IC) = 1$ the third equation implies $\d J^\sharp$ depends only on parameters. Together with the fourth equation this implies $\d J^{\sharp\,\a}$ are $\fDb$-closed $0,1$-forms on $\M$ valued in $\ccT_\cM^{1,0}$ and independent of $X$. This implies 
\beq\label{eq:dJsh}
\d J^{\sharp\,\a} \= 2\ii \d Y^{a'} \D_{a'}^\sharp{}^\a~,
\eeq
 where $\D_{a'}^\sharp$ are a basis for $H^1(\cM, \ccT_\cM^{1,0})$ and $a'$ denote parameters that leave $X$ fixed.

Hence, the universal family of complex structures has
$$
\d J \= \d y^\a \fD_\a J^\m + \half \d y^\a \d y^\b \fD_\a \fD_\b J^\m + \cdots~,
$$ 
where symmetrisation is automatic as $[\fD_\a,\fD_\b] J^\m = 0$.
 As a family of complex manifolds the universal geometry is complex and fibres depend holomorphically on parameters. This is what is typically said when studying Kodaira-Spencer theory.

Now we follow the paradigm of the previous subsection.  A first order deformation of  complex structure $\IJ$ preserving integrability \eqref{eq:cpxdef} together with holomorphic structure of the bundle $\IF^{0,2} = 0$ gives 
\beq
2\ii \left(\Id_\IA \d \IA   \right)^{0,2} \= \d \IJ^\m\w \IF_\m + \d \IJ^\a\w \IF_\a  ~.
\eeq
Decomposing into tangibility
\beq\label{eq:defUnivAJ}
\begin{split}
\delb_\A \d \A  &\=   \D{}^\m\w  F_{\m} ~, \\[2pt]
 \fDb \d \A + \delb_\A \d \A\#  &\= \D^\m\w \fDb \A^\dag_\m  + \D^{\sharp\,\a} \w \fD_\a \A    ~, \\[2pt]
\fDb \d \A\#  &\= \D^{\sharp\,\a} \w F^\sharp_\a ~.
\end{split}
\eeq
The first equation, together with the first equation of \eqref{eq:IJDefns}, gives the  Atiyah complex \cite{Atiyah:1955}. The solution to this is well-known, corresponding to a certain cohomology. We expand this in parameters $y^\a$ and basis for this cohomology e.g. to first order $\d \A = \d y^\a \fD_\a \A$ and $\d J^\m = 2\ii \d y^\a \D_\a{}^\m$. 

 The second equation has new informations. Using \eqref{eq:dJ}-\eqref{eq:dJsh}:\footnote{ We take $J^\sharp$ to be independent of $y^a$. A deformation of the parameter of the fibre doesn't modify the complex structure of the parameter space.}
\beq\label{eq:secondeqn}
\begin{split}
& \d y^\a \left( \fD_\bb \fD_\a \A - \delb_\A (\d_\a \A\#_\bb)  + \D_{\a}{}^\m \fD_\bb \A^\dag_\m 
\right)+  \d y^\ab \left( \fD_\bb \fD_\ab \A - \delb_\A (\d_\ab \A\#_\bb)  \right)\\[5pt]
 &\qquad  - \d Y^{a'}\left( \delb_\A( \d_{a'} \A\#_\bb) +\D_{a'\,\bb}^\sharp{}^\a \,\fD_\a \A \right) \= 0~.
\end{split}
\eeq
The three terms must independently vanish. The first term we will use momentarily; the second term is unchanged; the third term accounts for changes in complex structure of the parameter space and gives a condition for preserving holomorphy, in particular, holomorphic gauge $\fD_\ab \A = 0$. 
It also  an Atiyah equation for deformations of $A\#$. 

On a fixed $\ccU_{Y_0}$, we have $\IF^{0,2} = 0$ with $\IJ$ constant in complex coordinates. Using equation (2.16) of  
 \beq\label{eq:DerivIF}
\begin{split}
   \fD_\ab \IF_{\bb\mb} &\= \fD_\ab \fD_\bb \A_\mb  \= e_\ab\IF_{\bb\mb} -\G_\ab{}^\nb{}_\mb \IF_{\bb\nb}  + [A\#_\ab,\IF_{\bb\nb}]   \= 0~,\\
     \fD_\a \IF_{\bb\mb} &\= \fD_\a \fD_\bb \A \= e_\a\IF_{\bb\mb} -\G_\a{}^\n{}_\mb \IF_{\n\bb} + [A\#_\a,\IF_{\bb\mb}]   \=  -\D_\a{}^\m\fD_\bb \A_\m^\dag~,
\end{split}
\eeq
where $e_\a$ is the vielbein that diagonalises the product structure of the universal bundle; it is written in terms of the Ehresmann connection in for example \eqref{eq:ebasis}. 

We can derive  \eqref{eq:DerivIF} in two different ways: by evaluating $\fD_\a$ directly in terms of the connection $c_a{}^m$, and using that the curvature vanishes on a neighbourhood, meaning its partial derivatives all vanish and using a hermitian connection for $\fD_\a$ with respect to the moduli coordinates; or secondly by simply taking a perturbation of $\IF^{0,2} = 0$ with respect to $\IJ$ and $\IA$. 

We combine \eqref{eq:secondeqn} with \eqref{eq:DerivIF}. The first line of \eqref{eq:DerivIF} has $\fD_\ab\fD_\bb\A = 0$ and so with \eqref{eq:secondeqn} we find $\delb_\A \d_\ab \A\#_\bb \= 0$. This in turn implies $\d_\ab \A\#_\bb = 0$ as $H^0(X, {\rm End} E) = 0$. The second line of \eqref{eq:DerivIF} has mixed derivatives \beq\label{eq:mixedDerivA}
\begin{split}
  \fD_\bb \fD_\a \A &\=\!   - \D_{\a}{}^\m \fD_\bb \A^\dag_\m + \delb_\A (\d_\a \A^\sharp_\bb) ~, \qquad \fD_\a\fD_\bb \A \= \!- \D_\a{}^\m \fD_\bb \A^\dag_\m~.
\end{split}
\eeq
Taking commutators of \eqref{eq:mixedDerivA} gives field strengths
$$
[\fD_\a,\fD_\bb] \A \=\!-\delb_\A \IF_{\a\bb} \=\!-\delb_\A (\d_\a\A^\sharp_\bb)~, \qquad  [\fD_\ab\,\fD_\bb] \A \=\! -\delb_\A \IF_{\ab\bb} \=  0 ~.
$$
It follows that $\delb_\A (\IF_{\a\bb} - \d_\a \A^\sharp_\bb) = 0$ and from $H^0(X, {\rm End} E) = 0$ it follows that 
\beq\label{eq:IFab}
\IF_{\a\bb} \=  \d_\a \A^\sharp_\bb~, \qquad \IF_{\ab\bb} \= \d_\ab \A\#_\bb \= 0~.
\eeq
We can make additional progress by  following what was done in \eqref{eq:DefF}-\eqref{eq:DefF2} and consider $\wt \IF^{1,1} = \IF^{1,1} + \Id_\IA \d \IA$. This has a tangibility $[1,1]$ component given by
\beq\label{eq:Ftilde}
\begin{split}
 \wt \dd y^\a \wt \fD_\a\wt \A  &\= \dd y^\a\,  \fD_\a \A + \dd y^\a  \d y^\b\, \fD_\a \fD_\b \A  +\dd y^\a \d y^\bb (\delb_\A \IF_{\a\bb} -\D_\a{}^\m\fD_\bb \A_\m^\dag  ) \\[5pt]
 &\qquad\qquad -  \dd y^\ab  \d y^\b \left(\D_\b{}^\m \fD_\ab \A^\dag_\m\right)~. 
\end{split}
\eeq
We demand no antiholomorphic universal deformation parameters $\d y^\bb$ appear on the righthand side as a  generalisation of holomorphic gauge\footnote{Holomorphic gauge implies $\fD_\a A$ is a functional of $y^\a$ and not $y^\bb$. By extension, $\wt \fD_\a \wt \A$ is a functional of just $y^\a, \d y^\a$.} to the family $\{ \ccU_Y\}$. Then,  we wind up with $\delb_\A \IF_{\a\bb} = \D_\a{}^\m\fD_\bb\A_\m^\dag$ and 
\beq
\begin{split}
 \wt \dd y^\a \wt \fD_\a\wt \A  &\= \dd y^\a\,  \fD_\a \A  + \dd y^\a  \d y^\b\, \fD_\a \fD_\b \A -  \dd y^\ab  \d y^\b \left(\delb_\A \IF_{\a\bb}\right)~. 
\end{split}
\eeq

We require that $\wt \fD_\a \wt \A$ continue to satisfy the HYM equation. From this we find  $\Box_{\delb_\A} \IF_{\a\bb} = 0$ and so the connection is flat
\beq\label{eq:IFflat}
\IF_{\a\bb} \= \d_\a \A\#_\bb \= 0~, \quad \IF_{\ab\bb} \= \d_\ab \A\#_\bb \= 0~.
\eeq
 The HYM equation is thought of as a D-term condition and so we see a relationship between D-terms and flat connections. This is reminiscent of the relationship between the flat connection used to parallel transport D-branes over the CY moduli space and D-terms observed in \citeHHP. 
    
Finally, there is the last equation of \eqref{eq:defUnivAJ}. These are deformations of the complex structure on $\M$, preserving $F^\sharp$ remains holomorphic: it is the Atiyah problem for the moduli space itself. 

To summarise, deformations of the universal bundle preserving $\IF^{0,2}$ and integrable complex structure, D-term equations  on fibres gives
\beq\label{eq:derivJderivA}
\begin{split}
\fD_\a \fD_\bb J^\m &\=0~, \quad \fD_\ab \fD_\b J^\m \=0~, \quad \fD_\ab\fD_\bb J^\m \= 0~, \\[5pt]
  \fD_\bb \fD_\a \A &\=  0 ~, \quad \fD_\a\fD_\bb \A \=0 ~,\quad  \fD_\ab\fD_\bb \A \= 0~,
\end{split}
\eeq

\subsubsection{Comparison with direct differentiation}
As was the case in \citeUG, we get more information and much more quickly and easily than directly differentiating the equations of interest $N_J = 0$ and $F^{0,2} = 0$  with respect to parameters. To see this, note that with the result given by differentiating $F^{0,2} = 0$ with respect to parameters directly is given by: 
\beq\label{eq:SecondDerivA}
\delb_\A \,\fD_a \fD_b \A + \{ \fD_a \A, \fD_b \A \}  \= \frac{1}{2\ii}\, \fD_a \fD_b J^{\m 0,1} F_\m + \D_a{}^\m \fD_b F_\m + \D_b{}^\m \fD_a F_\m~,
\eeq
The components of this equation with at least one anti-holomorphic deformation are
\beq\notag
\begin{split}
\delb_\A (\fD_\ab \fD_\bb \A) &\= \frac{1}{2 \ii}\, \fD_\ab \fD_\bb J^{\m 0,1} F_\m ~,\\
\delb_\A (\fD_\a \fD_\bb \A + \D_\a{}^\m \fD_\bb \A_\m^\dagger) &\= \frac{1}{2 \ii}\,\fD_\a \fD_\bb J^{\m 0,1} F_\m ~,\\
\delb_\A (\fD_\bb \fD_\a \A + \D_\a{}^\m \fD_\bb \A_\m^\dagger) &\= \frac{1}{2 \ii}\,\fD_\bb \fD_\a J^{\m 0,1} F_\m  ~.
\end{split}
\eeq	 
Without more information inputted into the problem, we cannot kill the right hand side. So while these equations are consistent with what we already derived in  universal geometry, the universal geometry gives us additional information and more powerful results. It is analogous to demanding that $\IF^{0,2} = 0$ required the fields be holomorphic $\fD_\ab\A=0$ in parameters. We learn that $\D_\a{}^\m\fD_\bb \A^\dag_\m$ is $\delb_\A$-exact and $\fD_a \fD_b \A$ with any of the indices antiholomorphic vanishes. 

The purely holomorphic component of \eqref{eq:SecondDerivA} matches the Maurier-Cartan equation type equation in \eqref{eq:defUnivAJ}:
\beq\notag
\delb_\A \,(\fD_\a \fD_\b \A) + \{ \fD_\a \A, \fD_\b \A \} \= \frac{1}{2\ii}\, \fD_\a \fD_\b J^{\m 0,1} F_\m + \D_\a{}^\m \fD_\b F_\m{}+ \D_\b{}^\m \fD_\a F_\m{}~,
\eeq

\subsubsection{Atiyah complex for the universal bundle}
\label{s:AtiyahUniv}
If we define a bundle 
 \beq
 \IQ \= {\rm End}\, \IE \oplus \ccT_\IX{}^{1,0}~.
\eeq
with sections
$$
 \Iq \= 
\begin{pmatrix}
 \d \IA^{0,1} \\
 \d \IJ^{0,1\,M}
\end{pmatrix}
$$
where $M = (\m,\a)$ is holomorphic along $\IX$. We define an operator to act on these sections
 \beq
{\overline{  \ID}} \= 
\begin{pmatrix}
    \Idelb_\IA & \hat \IF \\
 0 & \Idelb
\end{pmatrix}~,
 \eeq
 where $\hat \IF(\d \IJ) =  \d \IJ^M \IF_M$. The operator is nilpotent ${\overline{  \ID}}^2 = 0$ in virtue of the Bianchi identity $\Id_\IA \IF= 0$ \citeUG. Its kernel, by construction, captures the deformations  studied in \eqref{eq:defUnivAJ} but as it is identical to the situation in \cite{Atiyah:1955} the space of deformations can be understood using the same techniques:
 \beq
 \ker  \IDb \= H^{0,1}_{\Idelb_\IA} ( \IX, \End \IE) \oplus \ker \hat \IF~.
 \eeq
 The first term we studied section \sref{s:deformF}. We found there that $(\Id_{\IA} \d \IA)^{0,2}= 0$ decomposes according to tangibility
 \beq
 \delb_\A \d \A \,\oplus\, \left( \delb_\A \d \A^{\sharp\,0,1} + \fDb \d \A \right) \,\oplus\, \left( \fDb \d \A\# \right)~.
 \eeq
Requiring D-terms, in this case the HYM equation, be preserved on fibres from \eqref{eq:IFab}, \eqref{eq:IFflat} implies $\d \A\# = 0$ and $\d \A$ are elements of $ H^{0,1}(X, {\rm End E})$:
 \beq\label{eq:HendIE}
  H^{0,1}_{\Idelb_\IA} ( \IX, \End \IE) \= H_\delb^{0,1}(X, {\rm End E}) ~.
 \eeq
 The second term $\ker \hat \IF$ is understood in cohomology and so is identical to the analysis in \sref{s:UniAtiyah}:
 \beq
 \hat \IF (\d \IJ) \= e^{mp} \d J_m{}^n F_{np} \oplus \dd y^a e^m \left( \d J_m{}^n \fD_a A_n + \d J^\sharp_a{}^b \fD_b A_m    \right) \oplus \dd y^{ac} \d J^\sharp_a{}^b \IF_{bc}~.
 \eeq
 The first term corresponds to the equation on the fibre and we write as $\ker F$. The middle term gives conditions on second order derivatives. The last term are  deformations of $A\#$ preserving $\IF^{0,2} = 0$ and these we write as $\ker F^\sharp$. Hence, 
\beq\label{eq:KerIF}
\ker \IF \= \ker F \oplus \ker F^\sharp~.
\eeq

\newpage
\section{Deformations of the universal heterotic geometry}
\label{s:deformationuniv}

\subsection{Review of universal geometry}

We build on the lessons of the previous section for the full universal geometry. For this, the final ingredient is the complexified hermitian form $\ccZ$:
\beq\notag
\ccZ_\a \= \ccB_\a + \ii \fD_\a \o~, \quad \text{and} \quad \ccZb_\a \=   \ccB_\a - \ii \fD_\a \o~,
\eeq
which is the generalisation to heterotic geometry of the variation of the complexified \K class familiar in special geometry $\d B + \ii \d \o$. The hermitian form on the universal geometry includes $\o$ and the \K form on $\M$:
\beq
\Iomega \= \o + \o^\sharp~,
\eeq
where there are no off-diagonal terms \citeUG. There is a three-form on the universal geometry $\IH$ which includes $H$ on the fiber $\X$, and is defined analogously
\beq
\IH ~=~ \Id \IB - \frac{\ap}{4} \Big(\CS[\IA] - \CS[\ITheta]\Big)~,~~~\text{where}~~~
\CS[\IA]~=~\tr\!\left(\IA\,\Id \IA +\smallfrac23 \IA^3\right)~,
\label{eq:IHdef3}\eeq
where $\IB$ is the extension of the Kalb--Ramond field
\beq\notag
\IB \= \frac12 B_{mn}\,e^m e^n +\IB_{am}\, \dd y^a e^m  +
\frac12 \IB_{ab}\,\dd y^a\dd y^b \= B + \IB_a\, \dd y^a + {B}^\sharp~ .
\eeq
$\IH$ decomposes as\footnote{The term with three legs along the moduli space vanishes due to $\M$ having a \K metric. The term with two legs along the moduli space $\IH_{ab}$ needs to be clarified.}
\beq
 \IH~=~\frac{1}{3!}\,\dd y^{abc}\,\IH_{abc}+\frac{1}{2}\,\dd y^{ab}\,\IH_{ab}+\dd y^a\,\IH_a+H\ ,
\notag\eeq
where the term with one leg along the moduli space $\M$ will be relevant in what follows:
\beq\label{eq:ccBdef3}
 \IH_a \=  \ccB_a \= \fD_a B + \frac{\ap}{4}\tr{(A\,\fD_a A)}-\dd\IB_a ~, \quad \fD_a B \= \del_a B  -\frac{\ap}{4} \tr (A\#_a\, \dd A) ~.
\eeq
This gives an interpretation to the one--form  $\IB_a$ as the mixed component $\IB_{am}$ of the universal $\IB$--field \citeUG. Under the analogue of a background gauge transformation $\IB \sim \IB + \frac{\ap}{4} \tr \IY \IA + \frac{\ap}{4} \IU$ applied to $\IB_a$ and $\ccB_a$ we find the transformation law for the covariant derivative of $B$:
\beq\label{eq:ccBmixedBackground}
 {}^\Phi \fD_a B \= \fD_a B - \frac{\ap}{4} \Big( \tr{(\fD_a A \, Y)} - \mathfrak{U}_a \Big) ~, \quad \dd \mathfrak{U}_a \= 0 ~.
\eeq

Small gerbe transformations $\ccB_a \to \ccB_a + \dd \mfb_a$ are realised by a deformation 
\beq\label{eq:smallGerbeIB}
\IB_a \to \IB_a - \mfb_a~.
\eeq
We view $\IB_a$ in an analogous fashion to $A\#_a$ for the gauge transformations: it is a `gerbe--connection' on the moduli space $\M$. 

The hermitian form $\Iomega$  on the universal geometry has a quantity
$$
\Id^c\Iomega \=\frac{1}{3!} \IJ^P \IJ^Q \IJ^R (\Id \Iomega)_{PQR}~.
$$
It being hermitian means $\Iomega$ is $1,1 $ and so
$$
\Id^c\Iomega \= \ii (\Id \Iomega)^{(2,1)} - \ii (\Id\Iomega)^{(1,2)}~.
$$
The term $\Id^c\Iomega$ has vanishing component with all legs along $\M$ due to the metric  $g\#_{\a\bb}$ being K\"ahler. The remaining components are given by
\beq\label{Idcom}
\begin{split}
(\Id^c\Iomega)_{\a\phantom{\b}} \=&~ \ii\,\fD_\a\o^{1,1 }-\ii \fD_\a \o^{0,2} ~,\\[0.3cm]
(\Id^c\Iomega)_{\a\b} \=& -\ii\,S{}_{\a\b}{}^\m\,\o_\m \= 0~,
\hskip40pt (\Id^c\Iomega)_{\ab\bb}\= \ii\,S{}_{\ab\bb}{}^\mb\,\o_\mb \= 0~ ,\\[0.3cm]
(\Id^c\Iomega)_{\a\overline\b} \=& -\ii\,S{}_{\a\bb}{}^{\mb}\,\o_{\mb}+
\ii\,S{}_{\a\overline\b}{}^\m\,\o_\m \= 0~.
\end{split}\eeq
In order for the Bianchi identity below to hold we require $S=0$. A related reason is for the complex structure deformations being integrable to form a universal family of complex manifolds $\X$ over the moduli space $\M$ in the sense of Kodaira--Spencer (the total space being a complex manifold requires $S=0$). We can rewrite this as
\beq\label{eq:dcomega}
\Id^c\Iomega \= \ii (\deth - \dethb) \omega + \ii \dd y^\a (\fD_\a \omega^{1,1 } - \fD_\a \omega^{0,2}) + \ii \dd y^\bb (\fD_\bb \omega^{2,0 } - \fD_\bb \omega^{1,1 }) ~.
\eeq
While $\omega$ is type $1,1 $, its derivative  $\fD_\a$ is type $(2,1)\oplus (1,2)$: $\fD_\a \omega {\=} \fD_\a\omega^{1,1 } + \fD_\a\omega^{0,2}$, and this expresses the type changing property of variations with respect to complex structure.

As shown in \citeUG, on a universal bundle $\ccU_{Y_0}$ the F-term 
\beq
\IH = \Id^c \Iomega~.
\label{eq:extendedsusy}
\eeq
holds given certain F-terms and their deformations. By first considering the tangibility with one leg along $\M$,  $\IH_a = (\dd^c\Iomega)_a$, we find
\beq\begin{split}\label{susyoncF}
& \ccZ_\a^{2,0 }~=~ \ccZb_\a^{2,0 } ~=~  0 ~,\\[0.2cm]
 &\ccZb_\a^{1,1 } ~=~ 0~,\\[0.2cm]
 &\ccZ_\a^{0,2}  ~=~ 0~.
\end{split}\eeq
This captures three of the F-terms, all in holomorphic gauge -- not surprising considering $\IF^{0,2} = 0$ captured the Atiyah equation in holomorphic gauge \citeUG. There is one remaining F-term and for this we turn the eye of Sauron to the Bianchi identity for $\Id\IH$ on $\IX$: 
\beq
\Id\IH~=-\frac{\ap}{4}\Big(\tr{\IF^2}-\tr{\IR^2}\Big) \=  \Id( \Id^c\Iomega)\ .
\label{eq:Bianchi2}\eeq
The curvatures $\IF$ and $\IR$ are of type $1,1 $ to first order in $\ap$ and so only the type $(2,2)$ part of this relation is non-vanishing.

We start with the component with one leg along $\M$ focusing on holomorphic variation with index $\a$. The first equality of the previous equation is 
\beq\label{Bid13}
 (\Id\IH)_\a~=-\frac{\ap}{2}\Big(\tr{(\fD_\a\cA\ F)}-\tr{(\fD_\a\th\ R)}\Big)\ .
\notag\eeq
Meanwhile $(\Id \Id^c \Iomega)_\a$  is simplified using
\beq
\begin{split}
 \dd (\Id^c \Iomega)_\a ~&=~ \ii \dd (\fD_\a \o^{1,1} - \fD_\a \o^{0,2})~, \\
   \fD_\a(\dd^c\o)~&=~2\ii\,\D_\a{}^\m\,(\del\o)_\m-2\ii\,\delb(\D_\a{}^\m\,\o_\m) +
    \ii (\del - \delb) \fD_\a \o\ ,
\end{split}\notag\eeq
and so\beq\begin{split}
 \delb( \ccZ_\a^{1,1 })&\= 2\ii\,\D_\a{}^\m\,(\del\o)_\m+\frac{\ap}{2}\Big(\tr{(\fD_\a\cA\ F)}-\tr{(\fD_\a\th\ R)}\Big)~. 
\end{split}\label{susyoncF2}\eeq
Note: we get additional non-physical terms if $S\ne 0$. 

Now consider tangibility $[2,2$], with two legs along $\M$
\beq\label{Bianchitang22}
\begin{split}
\fD_\a(\Id^c\Iomega)_\b-\fD_\b(\Id^c\Iomega)_\a~&=\!-\frac{\ap}{2}\,\Big(\tr{(\fD_\a\cA\,\fD_\b\cA)}-\tr{(\fD_\a\th\,\fD_\b\th)}\Big)\ ,\\[0.2cm]
\fD_\a(\Id^c\Iomega)_{\bb}-\fD_{\bb}(\Id^c\Iomega)_\a~&=\!-\frac{\ap}{2}\,\Big(\tr{(\fD_\a\cA\,\fD_{\bb}\cA^\dagger + \IF_{\a\bb} F)}-\tr{(\fD_\a\th\,\fD_{\bb}\th^\dagger + \IR_{\a\bb} R)}\Big)~.
\end{split}
 \raisetag{1.8cm}
\eeq
We have used $\IF_{\a\b} = \IR_{\a\b} = 0$. 
With three legs, or four legs along the moduli space the equation identically vanishes due to $S=0$ and $\IF_{ab} = \IR_{ab}= 0$ and \K property respectively. Using \eqref{eq:dcomega} we find
\beq\label{Bianchitang22b}\notag
\begin{split}
 - \ii [\fD_\a, \fD_\b] \omega^{0,2}
~&=\!-\frac{\ap}{2}\,\Big(\tr{(\fD_\a\cA\,\fD_\b\cA)}-\tr{(\fD_\a\th\,\fD_\b\th)}\Big)\ ,\\[0.2cm]
\fD_\a(\Id^c\Iomega)_{\bb}-\fD_{\bb}(\Id^c\Iomega)_\a~&=\!-\frac{\ap}{2}\,\Big(\tr{(\fD_\a\cA\,\fD_{\bb}\cA^\dagger + \IF_{\a\bb} F)}-\tr{(\fD_\a\th\,\fD_{\bb}\th^\dagger + \IR_{\a\bb} R)}\Big)~.
\end{split}
 \raisetag{0cm}
\eeq
The first equation is an identity. To see this, recall $[\fD_\a,\fD_\b]\o^{1,1 } = ([\del_\a,\del_\b]\o)^{1,1} = 0$ and $[\fD_\a,\fD_\b]\o^{0,2} = ([\del_\a,\del_\b]\o)^{0,2} = 0$ while the first term on the right hand side vanishes as we want a hermitian metric on deformations of the holomorphic bundle, the second term on the right hand side vanishes for the same reason, or can be computed explicitly by using $R_{\m\n} = 0$, the curvature of the connection $\th$. Hence, the only non--trivial equation is the second one and it is crucial to the construction of the \K potential. Simplifying it somewhat it reads:
\beq\begin{split}\label{secondorderom}
  &\fD_\a\fD_{\bb}\o^{1,1 } \=\!-\frac{\ii\ap}{4}\,\Big(\tr{(\fD_\a\cA\,\fD_{\bb}\cA^\dagger + \IF_{\a\bb} F)}-\tr{(\fD_\a\th\,\fD_{\bb}\th^\dagger + \IR_{\a\bb} R)}\Big)\\[0.2cm]
  &  +\D_\a{}^\m\,\D_{\bb\,\m}{}^\rb\o_\rb  +\D_{\bb}{}^{\nb} \D_{\a\,\nb}{}^\r \o_\r + 2 \D_\a{}^\m\,\D_{\bb}{}^\nb\o_{\m\nb} ~.
\end{split}\eeq
with last line coming from
$\D_\a{}^\m\,(\fD_{\bb}\o^{2,0 })_\m+\D_{\bb}{}^{\nb}\,(\fD_\a\o^{0,2})_{\nb}$~. We utilise this relation in \sref{s:KahlerPotRedone}. 

\subsection{Deformations of the universal geometry}
\label{s:DefUniv}
We now study deformation of the universal bundle so that the F-terms along with the Green-Schwarz Bianchi identity continue to be satisfied
\beq
N_\IJ \= 0~, \quad \IF^{0,2} \= 0~, \quad \IH \=  \Id^c \Iomega~, \quad   \Id \IH \=\! -\frac{\ap}{4} \tr \IF^2 + \frac{\ap}{4} \tr \IR^2~.
\eeq
 Furthermore, we demand holomorphy be satisfied and when restricted to fibres the fields satisfy the D-terms of heterotic: HYM and a balanced metric. 

As for previous sections we take our family of universal bundles to be denoted $\ccU_Y$ with $Y$ coordinate for the parameter space of universal bundles. We now consider deformations $\Iomega \to \Iomega + \d \Iomega$ and $\IB \to \IB + \d \IB$, in addition to $\IJ \to \IJ + \d \IJ$ and $\IA \to \IA + \d \IA$. In the previous section we studied how $N_\IJ=0$ and $\IF^{0,2}=0$ were modified under perturbations. Part of the data we required was that $\IJ$ remained block diagonal and here we require $\Iomega$ remain block diagonal in the e-basis of \citeUG.  Now we study how $\IH= \Id^c\Iomega$ subject to the Bianchi identity \eqref{eq:Bianchi2} behaves under perturbations. 

As for previous sections, we work to first order in $\d$ in the universal geometry, and this will give us equations for second order in deformations of the heterotic geometry. 

A first order deformation of $\IH$ satisfying the Bianchi identity is 
\beq\label{eq:dIH}
\d \IH \= \Id  \left( \d \IB +\frac{\ap}{4}\tr{(\IA\,\d \IA)} \right)  - \frac{\ap}{2} \tr\left( \IF \d \IA\right) +  \frac{\ap}{2} \tr\left( \IR \d \ITheta\right)~,
\eeq
while a first order deformation of $\Id^c\Iomega$ 
\beq\label{eq:defUnivHdc}
\d( \Id^c\Iomega) \= \half \d \IJ^M \IJ^N \IJ^P (\Id \Iomega)_{MNP}  + \frac{1}{3!} \IJ^M \IJ^N \IJ^P (\Id \d \Iomega)_{MNP}~.  
\eeq

Formally, these equations are identical to \eqref{eq:Fterms} except with a change of font and so we can read off deformations evaluated in the complex structure of $\ccU_{Y_0}$:
\beq\label{eq:moduliEqnReal}
\begin{split}
 \Idelb \d \IZ^{0,2} &\= 0~, \\
\Idel \d \IZ^{0,2} + \Idelb \d \IZ^{1,1 } &\= \d \IJ{}^M (\Idel\Iomega)_M + \frac{\ap}{2} \tr \big( \d \IA^{0,1} \IF\big) -  \frac{\ap}{2} \tr \big( \d \ITheta^{0,1} \IR\big)~,
\end{split}
\eeq
where $M=(\a,\m)$ is a holomorphic index and 
\beq
\d \IZ \=  \d \IB +\frac{\ap}{4}\tr{(\IA\,\d \IA)}  - \frac{\ap}{4}\tr{(\IR\,\d \ITheta)} +  \ii \d \Iomega~.
\eeq
As discussed in \citeUG $\ITheta$ is the Hull connection on $\ccT_\IX$ and $\IR$ its curvature. 
 
This decomposes according to tangibility. The tangibility $[0,3]$ component is
\beq\label{eq:FibermoduliEqnReal}
\begin{split}
 \delb \d \ccZ^{0,2} &\= 0~, \\
\del\d \ccZ^{0,2} + \delb \d \ccZ^{1,1 } &\= \d J{}^\m (\del\o)_\m + \frac{\ap}{2} \tr \big( \d \A F\big) -  \frac{\ap}{2} \tr \big( \d \Th R\big)~.
\end{split}
\eeq
We recognise this as a first order deformation of one of the F-terms. Once we include deformations of $N_J =0$ and $F^{0,2}=0$ studied in  section \sref{s:Atiyah}, together with holomorphic gauge \citeSG, we get $(\d \ccZ, \d J^\m, \d \A)$ form a vector  which are sections of a $Q$-bundle in the previous subsection and furthermore, in the kernel of a $\Dbar$-operator defined in \citeE. The Hermitian Yang-Mills equation and balanced equation, force $\ccY$ to be the harmonic representative with respect to the $\Dbar$-operator but not the $\delb$-operator. From this we conclude $\d \ccZ  = \d y^a \ccZ_a$, with $\d y^a$ the usual heterotic parameters spanning the $\Dbar$-cohomology. We return to this cohomology later.

The tangibility $[1,2]$ component has new information. The bit proportional to $\d_\b$ is
\beq\label{eq:Zb}
\begin{split}
 \fD_\ab  \ccZ_\b^{0,2} &\= \delb \left(\d_\b \IZ_{\ab}^{0,1} \right)~, \\
 \fD_\a  \ccZ_\b^{0,2}  &\=\D_\b{}^\m\, \ccZ^{0,1}_{\a\,\m} 
 +  \delb \d_\b \IZ_\a^{0,1}\\
\fD_\ab \Z_\b^{1,1}&\= \frac{\ap}{2} \tr \big( \fD_\b \A \fD_\ab\A^\dag + \IF_{\b\ab} F\big) -  \frac{\ap}{2} \tr \big( \fD_\b \th \fD_\ab\th^\dag + \IR_{\b\ab} R\big) \\[4pt]
&\qquad\qquad + \D_\b{}^\m \ccZ_{\ab\,\m}^{1,0} +\del \d_\b \IZ_\ab^{0,1}  + \delb \d_\b \IZ_\ab^{1,0} ~,
\raisetag{2cm}
\end{split}
\eeq
where from the previous subsection $\tr \fD_\a \A \fD_\b \A = 0$ and $\IF_{\b\ab}=\d_\b \A\#_\ab$. We also have $\th = \Th^{0,1}$ and the term $\tr \big( \fD_\b \th \fD_\ab\th^\dag \big)$ is to be written in terms of the underlying moduli $\d J^\m = 2\ii \D^\m$ and $\d \ccZ^{1,1}$.

The bit proportional to $\d_\bb$ is
\beq\label{eq:Zbb}
\begin{split}
 \fD_\ab  \ccZ_\bb^{0,2} &\= \delb\left( \d_\bb \IZ_{\ab}^{0,1}\right) ~,
 \\
 \fD_\a  \ccZ_\bb^{0,2}  &\=  \delb\,( \d_\bb \IZ_\a^{0,1}\,)~,\\
 \fD_\ab  \ccZ_\bb^{1,1} &\=  \delb (\d_\bb \IZ_\ab^{1,0}) +  \del (\d_\bb \IZ_\ab^{0,1} )~.
\end{split}
\eeq

By analogous reasoning to \eqref{eq:DerivIF}, for a fixed universal bundle we have $\IH = \Id^c \Iomega$ and so derivatives vanish. The tangibility $[1,2]$ component is
\beq
\fD_a (\IH - \Id^c \Iomega)_b  \= 0~, 
\eeq

 Projecting onto complex type using that $\fD_\b \fD_\a \o^{p,q} = (\del_\b\del_\a\o)^{p,q}$\\[3pt]
\beq
\begin{split}
  \fD_\a \ccZ_\b^{2,0} + \fD_\a \ccZb_\b^{1,1} +\left( \fD_\a \ccZ_\b^{0,2} -   \D_\a{}^\m   \ccZ_{\b\,\m}^{0,1}\right) &\=  0 \\[5pt]  
  \fD_\a \ccZb^{2,0}_\bb +\left( \fD_\a \ccZ_\bb^{1,1}  -  \D_\a{}^\m \ccZ^{1,0}_{\bb\,\m}\right) 
  + \left(\fD_\a \ccZb_\bb^{0,2}  -\D_\a{}^\m \ccZb_{\bb\,\m}^{0,1}   \right) &\= 0~,
\end{split}
\eeq
 and so
  \beq
\begin{split}
  \fD_\a \ccZ_\b^{2,0 }  &\= 0~, \qquad\qquad~~   \fD_\ab \ccZ_\bb^{1,1}  \= 0~, \qquad \qquad  ~~\fD_\a \ccZ_\b^{0,2} \=   \D_\a{}^\m \ccZ_{\b\,\m}^{0,1}  ~,\\[3pt]
\fD_\ab \ccZ_\b^{2,0 } &\=    \D_\ab{}^\mb \ccZ_{\b\,\mb}^{1,0}~, \quad   \fD_\a \ccZ_\bb^{1,1}  \=   \D_\a{}^\m \ccZ_{\bb\,\m}^{1,0}~, \quad      \fD_\ab \ccZ_\b^{0,2}  \= 0 ~.\\
\end{split}
 \eeq
 where \eqref{susyoncF} is used and repeated below
 \beq\begin{split}\notag
& \ccZ_\a^{2,0 }~=~ \ccZb_\a^{2,0 } ~=~  0 ~,\\[0.2cm]
 &\ccZb_\a^{1,1 } ~=~ 0~,\\[0.2cm]
 &\ccZ_\a^{0,2}  ~=~ 0~.
\end{split}\eeq
Comparing with \eqref{eq:Zb} we see that $\d_\b \IZ_\a^{0,1} = \d_\b \IZ_\ab^{0,1}= 0$.  The last line is exactly \eqref{secondorderom} and so we determine $\d_\b \IZ_\ab^{1,0} =0$. Repeating with \eqref{eq:Zbb}, we see $ \d_\b \IZ_\a^{1,0} = 0$. The rest follow by complex conjugation. Hence, $\d_b \IZ_{am} = 0$. Equations \eqref{eq:Zb}-\eqref{eq:Zbb} simplify:
\beq\label{eq:Zb2}
\begin{split}
 \fD_\ab  \ccZ_\b^{0,2} &\= 0~, ~\qquad\qquad\qquad\qquad   \fD_\ab  \ccZ_\bb^{0,2} \= 0~,
 \\
 \fD_\a  \ccZ_\b^{0,2}  &\=\D_\b{}^\m\, \ccZ^{0,1}_{\a\,\m}~, \quad\qquad\qquad  \fD_\a  \ccZ_\bb^{0,2}  \=  0~, \\
  \fD_\a \ccZ_\b^{2,0} &\= 0~, ~\qquad\qquad\qquad\qquad \fD_\ab\ccZ_\b^{2,0} \= \D_\ab{}^\m \ccZ_{\b\,\mb}^{1,0}~,\\
   \fD_\a \ccZ_\bb^{1,1}  &\=   \D_\a{}^\m \ccZ_{\bb\,\m}^{1,0}~, ~~\qquad\qquad   \fD_\ab  \ccZ_\bb^{1,1} \= 0~,\\
\fD_\bb \Z_\a^{1,1}&\= \frac{\ap}{2} \tr \big( \fD_\a \A \fD_\bb\A^\dag  + \IF_{\a\bb}F\big) - \frac{\ap}{2} \tr \big(\fD_\a \th \fD_\bb \th^\dag + \IR_{\a\bb} R \big) + \D_\a{}^\m \ccZ_{\bb\,\m}^{1,0}~,
\end{split}
\eeq
where we have included the contribution of the Hull connection to the tangent bundle. The simplicity of these equations is striking. 

The expressions for $\fD_\a \ccZ_\bb^{1,1}$ and $\fD_\bb \ccZ_\a^{1,1}$ after solving for $\fD_\a\fD_\bb \o^{1,1}$ give \eqref{secondorderom}, repeated here
\beq\begin{split}
  &\fD_\a\fD_{\bb}\o^{1,1 } \=\!-\frac{\ii\ap}{4}\,\Big(\tr{(\fD_\a\cA\,\fD_{\bb}\cA^\dagger + \IF_{\a\bb} F)}-\tr{(\fD_\a\th\,\fD_{\bb}\th^\dagger + \IR_{\a\bb} R)}\Big)\\[0.2cm]
  &  +\D_\a{}^\m\,\D_{\bb\,\m}{}^\rb\o_\rb  +\D_{\bb}{}^{\nb} \D_{\a\,\nb}{}^\r \o_\r + 2 \D_\a{}^\m\,\D_{\bb}{}^\nb\o_{\m\nb} ~.
\end{split}\eeq

This collection of results is partially reproduced by direct differentiation of the Hull-Strominger system, a calculation undertaken in sections to follow. However, the universal geometry calculation gives a stronger set of results. 

The F-terms have a nice upper triangular structure, so that the analysis of the Atiyah case in \sref{s:UniAtiyah}  can be embedded in the full heterotic problem. Note that the D-terms do not obey this structure as can be seen from the last two lines of \eqref{eq:ALLeomsgf}. 
 
That being so, the analysis leading to \eqref{eq:mixedDerivA}-\eqref{eq:Ftilde} is unchanged and these augment \eqref{eq:Zb2}.

\subsection{Other tangibliities}

Recall from section 5 of \citeSG that $\IH_{ab}$ can be viewed as a curvature for the mixed term $\IB_a$ analogous to the curvature $\IF_{ab}$ for $A\#_a$ for the gauge bundle. However, as the curvature $S_{ab}=0$ it follows from \eqref{Idcom} that have $(\Id^c\Iomega)_{ab} = 0$. The F-term $\IH=\Id^c\Iomega$ implies the gerbe `curvature'  $\IH_{ab} = 0$ is also flat. 

Consider the tangibility $[2,1]$ component of \eqref{eq:moduliEqnReal}. Using $\d \IZ_{m\bb} = 0$, $\o^\sharp$ is \K and independent of $X$, $\IF_{ab}=0$, both equations simplify
\beq\label{eq:fDbZa}
\begin{split}
\delb \d \ccZ^{\sharp\,0,2}  &\= 0~, \qquad  \del \d \ccZ^{\sharp\,0,2} + \delb \d \ccZ^{\sharp\,1,1}  \= 0~.
\end{split}
\eeq
Together with $H^0(X,\IZ)=1$ these imply $\d \ccZ^{\sharp\,0,2}$ and $\d \ccZ^{\sharp\,1,1}$ are constants over $X$, depending just on parameters. This is consistent with, for example, $\o^\sharp$ depending solely on parameters: deformations are respecting this. Hence, deformations of $\ccZ^\sharp$ decouple from the fiber and are confined to the base manifold, the moduli space. 

Furthermore, as the moduli space metric is \K the tangibility $[3,0]$ component of the three-form vanishes $\IH_{abc} = 0$ and this is to be preserved  under deformations. This means 
\beq\label{eq:fDbZb}
\begin{split}
 \fDb \d \ccZ^{\sharp\,0,2} &\= 0~, \qquad \fD \d \ccZ^{\sharp\,0,2} + \fDb \d \ccZ^{\sharp\,1,1} \= 0 ~,
\end{split}
\eeq
where we use the moduli space is \K in that $\fDb \o^\sharp = 0$ and the result from the previous section, $F^\sharp=0$. The first equation derives from deformations of the complex structure of the moduli space. If $H^{0,2}(\M,\IC)=0$ then  $ \d \ccZ^{\sharp\,0,2}$ is $\fDb$--exact and so using the second equation can be reabsorbed into $\ccZ^{\sharp\,1,1}$ as a field redefinition. Even if this cohomology is non-vanishing, as we work locally we use the Poincare lemma to write $\d \ccZ^{\sharp\,0,2}$ as a $\fDb$-exact term, and absorb it into $\d \ccZ^{\sharp\,1,1}$ by a field redefinition. The space of deformations would then be the space of deformations of $\fDb$-closed $(1,1)$-forms.\footnote{If we work locally around a point in the moduli space, then  $\d \ccZ^{\sharp\,0,2}$ id exact by the Poincaré lemma, and so this field redefinition may be applied.}

\subsection{Summary of results}
Its worthwhile to bring our results together. 

Derivatives:
\beq\label{eq:AJ2}
\begin{split}
 \fD_\a\fD_\bb J^\m &\= \fD_\ab\fD_\b J^\m \= \fD_\ab\fD_\bb J^\m \= 0~, \\[6pt]
  \fD_\bb \fD_\a \A& \= 0  ~, \quad \fD_\a\fD_\bb \A \= \! \delb_\A \IF_{\a\bb}~,\quad  \fD_\ab\fD_\bb \A \= 0~.
\end{split}
\eeq
As described in \sref{s:UniAtiyah}, we have imposed  $\wt \fD \wt \A$ is a functional of holomorphic parameters, which  implies $\delb_\A \IF_{\a\bb} =-\D_\a{}^\m\fD_\bb \A_\m^\dag$. We have used this to simplify the  above. 
\beq\label{eq:Zb2b}
\begin{split}
 \fD_\ab  \ccZ_\b^{0,2} &\= 0~, ~\qquad\qquad\qquad\qquad   \fD_\ab  \ccZ_\bb^{0,2} \= 0~,
 \\
 \fD_\a  \ccZ_\b^{0,2}  &\=\D_\b{}^\m\, \ccZ^{0,1}_{\a\,\m}~, \quad\qquad\qquad  \fD_\a  \ccZ_\bb^{0,2}  \=  0~, \\
  \fD_\a \ccZ_\b^{2,0} &\= 0~, ~\qquad\qquad\qquad\qquad \fD_\ab\ccZ_\b^{2,0} \= \D_\ab{}^\m \ccZ_{\b\,\mb}^{1,0}~,\\
   \fD_\a \ccZ_\bb^{1,1}  &\=   \D_\a{}^\m \ccZ_{\bb\,\m}^{1,0}~, ~~\qquad\qquad   \fD_\ab  \ccZ_\bb^{1,1} \= 0~,\\
\fD_\bb \Z_\a^{1,1}&\= \frac{\ap}{2} \tr \big( \fD_\a \A \fD_\bb\A^\dag  + \IF_{\a\bb}F\big) - \frac{\ap}{2} \tr \big(\fD_\a \th \fD_\bb \th^\dag + \IR_{\a\bb} R \big) + \D_\a{}^\m \ccZ_{\bb\,\m}^{1,0}~,
\end{split}
\eeq

Curvatures:
\beq
\begin{split}
 \qquad  S_{ab}{}^m &\= 0~,\qquad \IF_{\a\b} \=\IF_{\ab\bb}\= 0~,\\
 \IH_{abc}&\= 0~,\qquad \IH_{ab} \= 0~.
\end{split}
\eeq
A quick comment on the connection $\ITheta$ on the tangent bundle. This is the extension of the Hull connection to the universal geometry. As calculated in \citeUG, this connection has a curvature $\IR$ and it has
$$
\IR_{\a\b} \= \IR_{\ab\bb} \= 0~,
$$
where $\IR_{\a\bb}$ has non-vanishing components. We can divide these up according to the endmorophism indices: $\IR_{\a\bb}{}^c{}_d$ corresponds to the curvature of the Chern connection on the moduli space\footnote{As the moduli space is \K, the Chern connection, Hull connection and Levi-Civita connection are all the same.}; $\IR_{\a\bb}^\m{}_\r$ and $\IR_{\a\bb}{}^\m{}_\rb$ are both non-vanishing a contain information about derivatives of the complex structure moduli of the fibre $\D{}^\m$ and hermitian moduli of the fibre $\ccZ^{1,1}$. It is interesting that curvatures of the connections $\IA$ and $\ITheta$ seems to take distinct roles in the universal geometry; more so that in the original heterotic theory. 
.

 \newpage
\section{\texorpdfstring{$\IDb$}{Db}  operator for a \texorpdfstring{$\IQ$}{Q}-bundle: the pullback of \texorpdfstring{$Q$}{} to the universal geometry} 

\subsection{\texorpdfstring{$\Db$}{Db} operator for \texorpdfstring{$Q$}{Q}-bundle: review}
We review \citeE, \citeES. It is helpful in the first instance to include deformations of the the connection on the tangent bundle, despite the fact such deformations are not part of the heterotic string moduli space. The reason being this seems a more natural way to embed the construction in the universal bundle.

Recall from \citeE that on $\X$ we require a complex manifold, hermitian form $\o$, a $J$ and a $H$ and a gauge field satisfying F-terms. Deformations of these, supplemented by spurious deformations of the connection $\th$, are captured by a $Q$ bundle topologically \citeOS,\citeE:
\beq
Q\=(\ccT_\X{}^{1,0})^* \oplus {\rm End} E \oplus \ccT_\X^{1,0}~.
\eeq
Denote sections $q$ of $\Omega^{0,1}(Q)$ by
\beq
q \= \begin{bmatrix} ~\ccZ^{1,1} ~ \\ \varphi \\ \aa \\ \D \end{bmatrix} \in \begin{bmatrix}~ \Omega^{0,1}((T^{1,0}X)^*)~ \\ \Omega^{0,1}({\rm End} \, TX)\\ \Omega^{0,1}({\rm End} \, E) \\ \Omega^{0,1}(T^{1,0} X) \end{bmatrix}~.
\eeq
The F-terms are \eqref{eq:moduliEqnHol} repeated here
\beq
\begin{split}
&\delb\, \D^\m \= 0~, \\[0.15cm]
&\delb_\A \aa +  F_\m \w \D^\m\=0~, \\
& \delb \ccZ^{1,1 }  - 2\ii \, \D{}^\m (\del\o)_\m  -\frac{\ap}{2} \tr \big( F \aa \big) + \frac{\ap}{2} \big( R \varphi \big)  \= 0~,
 \end{split}
\eeq
where from \citeUG 
\beq\label{eq:dthp}
\varphi \=  \nabla^\LC_\m  \D{}^\n + \half \, \nabla^{\LC\,\n} \, \ccZ_{\m\rb} \dd x^\rb ~.
\eeq
It will be helpful to suppress this knowledge from our brain for the moment. 

The F-terms with the spurious degrees of freedom are in the kernel of an operator denoted by $\Dbar: \Omega^{(0,p)}(Q) \rightarrow \Omega^{(0,p+1)}(Q)$ given by
\beq \label{Dbar-defn}
\Dbar \= \begin{bmatrix}
\delb & -\ap \cR^* & \ap \cF^* & \cH \\
0 & \delb_\th & 0 & \hat R \\
0 & 0 & \delb_\A & \hat F \\
0 & 0 & 0 &  \delb
\end{bmatrix}~,
\eeq
where
\beq\label{eq:cFdef0}
\hat F: \Omega^{(0,p)}(T^{1,0}X) \to \Omega^{(0,p+1)}({\rm End} \, E)~, \qquad \cF^*: \Omega^{(0,p)}({\rm End} \, E) \to \Omega^{(0,p)}((T^{1,0}X)^*)~,
\eeq
is defined by
\beq\label{eq:cFdef}
\hat F (\D) \= F_{\m \nb} \, \dd x^\nb \w \D^\m~, \qquad \cF^* (\aa) \= {\rm Tr} \, F_{\m \nb} \, \dd x^\m \otimes \dd x^\nb \wedge \aa~,
\eeq
with analogous definitions for $\cR, \cR^*$, 
while 
\beq\label{eq:cHdef}
\cH: \Omega^{(0,p)}(T^{1,0}X) \to \Omega^{(0,p)}((T^{1,0}X)^*)~, \qquad \cH (\D) \= H_{\r \nb\m} \,\dd x^\r \otimes \dd x^{\nb} \wedge \D^\m~.
\eeq

\subsection{\texorpdfstring{$\IDb$}{Db}-operator for \texorpdfstring{$\IQ$}{Q} bundle on universal geometry}
On the universal manifold $\IX$, where topologically $\IX = \X\oplus \M$, we have a complex manifold with hermitian form $\Iomega$, complex structure $\IJ$, a $\IH$ and a gauge field $\IA$ satisfying F-terms and the Bianchi identity (provided curvature for the Ehreshmann connection $c_a{}^m$ vanishes $S_{ab}{}^m=0$):
\beq
N_\IJ = 0~, \quad \IF^{0,2} = 0~, \quad \IH \=  \Id^c \Iomega~, \quad   \Id \IH \=\! -\frac{\ap}{4} \tr \IF^2 + \frac{\ap}{4} \tr \IR^2~.
\eeq
 So the form of $\IDb$ operator on the universal space is the same as that on the manifold $\X$. Define a bundle $\IQ$, which is topologically
 \beq
\IQ \= (\ccT_\IX^{1,0})^* \oplus {\rm End}\, \ccT_\IX \oplus {\rm End} \IE \oplus \ccT_\IX{}^{1,0}~.
\eeq
Sections of $\IQ$  the case of interest being $(0,1)$-forms, denoted
$$
\IY \= 
\begin{pmatrix}
 \d \IZ^{1,1} \\
 \d\ITheta^{0,1}\\
 \d \IA^{0,1} \\
 \d \IJ^M
\end{pmatrix},
$$
with $M=(\a,\m)$ a holomorphic index. 
On sections of $\IQ$ we define an operator inspired by the beautiful work of \citeOS
 \beq \label{IDbar-defn}
\IDb \= \begin{bmatrix}
\Idelb & -\ap \IR^* & \ap \IF^* & \IH \\
0 & \Idelb_\Th & 0 & \hat\IR \\
0 & 0 & \delb_\IA & \hat\IF \\
0 & 0 & 0 &  \Idelb
\end{bmatrix}~,
\eeq
 where $\hat \IF(\d \IJ) =  \d \IJ^M \IF_M$, $\hat \IR(\d \IJ) =  \d \IJ^M \IR_M$, while $\IF^*(\d \IA^{0,1}) = \half \tr (\IF \w \d \IA^{0,1})$, $\IR^*(\d \ITheta^{0,1}) = \half \tr (\IR \w \d \ITheta^{0,1})$, $\IH(\d \IJ) = -\ii \IH^{1,1}_M \d \IJ^M$. Just as for \citeOS, this defines a double extension structure. They  naturally generalise \eqref{eq:cFdef0}-\eqref{eq:cHdef} to the universal space. It is automatically the case that $\IDb^2 = 0$ as formally the operator satisfies the same equations as the original heterotic problem.  The new information is in the decomposition into tangibility. We have included deformations of $\ITheta$ -- just as for the heterotic case these deformations are spurious modes and to be understood as being determined in terms of the underlying geometric structures such as $\IJ$, $\IB$ and $\Iomega$. Just as for \citeE, we need to correctly eliminate these from the cohomology calculation, however doing so is beyond the scope of the current work.

Universal bundle moduli plus the universal spurious modes are elements of the kernel $\IDb \IY = 0$. The first line of this equation corresponds to the discussion in \sref{s:DefUniv}; the second through fourth lines to the discussion in \sref{s:UniAtiyah}.  

Following \citeOS, there is a long-exact sequence calculation in cohomology,  that determines the space of universal bundle deformations as
\beq
H^1_\IDb(\IX,\IQ) \= H^1(\IX,\ccT_\IX^*)\oplus \IS~, 
\eeq
where 
\beq
\IS \subset H_{\Idelb}^1(\IX,{\rm End}\ccT_\IX)\oplus H_\Idelb^1(\IX,{\rm End}E) \oplus (\ker \hat\IF \cap\ker \hat\IR)
\eeq
and $\IS$ corresponds to the first line of \eqref{IDbar-defn} understood in $\Idelb$-cohomology, e.g.
\beq
-\ii \IH_M^{1,1} \d \IJ^M + \frac{\ap}{2} \tr \left( \IF \w \d \IA^{0,1} \right) -  \frac{\ap}{2} \tr \left( \IR \w \d \ITheta^{0,1} \right) \= \Idelb(\cdots)~,
\eeq
and the decomposition of this equation into its underlying components is described in the previous sections. In addition to eliminating the spurious modes, it would be interesting to understand if there is an adjoint  $\IDb^\dag$ and its relation to the D-terms following \citeE. It would also be interesting to extend the work of \citeES and explore whether $\IDb$ has a field strength and its relation to the equations of motion.

\newpage

\section{Second order deformations of heterotic moduli}
\label{s:DirectDiff}
In this section we work just with the Hull--Strominger system. We re-derive   the equations studied previously via the universal bundle by direct differentiation of the structures. 

\subsection{Complex structure}
Complex structure satisfies an algebraic equation $J^2 = -\Ione$ and the Nijenhuis equation $N_J=0$. A deformation of complex structure we expand in parameter derivatives $\d J = \d y^a \fD_a J + \half \d y^a \d y^b \fD_a \fD_b J + \cdots$. 

Recall, a deformation of the algebraic equation is
$$
\d J_m{}^p J_p{}^n + J_m{}^p \d J_p{}^n \= 0~,
$$
where $\d J = \d y^a \del_a J$. 
Evaluating in complex coordinates
$$
\d J_\m{}^\n \= 0~, \qquad \d J_\mb{}^\n \cong \d y^a (\del_a J)_\mb{}^\n \cong \d y^a \fD_a J_\mb{}^\n~,
$$
where recall \citeM that $(\del_a W)^{(p,q)} = \fD_a W^{(p,q)} = \del_\a W^{(p,q)} - \D_\a{}^\m W_\m^{(p-1,q)} + \D_\a{}^\m W_\m^{(p,q-1)}$.

A deformation of the algebraic equation, correct to second order evaluated in complex coordinates is then 
\beq\notag
\fD_a J_\n{}^\m  \= 0~, \qquad \fD_a \fD_b J_\m{}^\n \=  \frac{\ii}{2} (\fD_a J_{\m}{}^\sb\, \fD_b J_{ \sb}{}^\n + \fD_b J_{ \m}{}^\sb\, \fD_a J_{ \sb}{}^\n)~, 
\eeq
The second order equation can also be obtained by differentiating  the first order equation. 
A  deformation of Nijenhuis tensor $N_J = \frac{1}{4} [J,J]_{FN}$ to second order gives
\beq\label{eq:cpxstFN}
[\fD_a J,J]_{FN} \=0, \qquad [\fD_a\fD_b J, J]_{FN} + [\fD_a J, \fD_b J]_{FN} \=0~,\\
\eeq
where $FN$ denotes the Frochlicher--Nijenhuis bracket. The advantage of this form is that by a variation of complex structure we obtain the second equation from the first. When evaluated on a complex structure the deformation equations reduce to the conventional equations as written in say \cite{Tian:1987,todorov1989}.\beq\label{eq:cpxDefEOM}
\begin{split}
&\delb \left(\fD_a J{}^\m  \right) \=0~, \\[2pt]
 &2\ii\, \delb\big( \fD_a \fD_b J_\nb{}^\m \dd x^\nb\big) \= \fD_a J{}^\r\, \del_\r \fD_b J{}^\m + \fD_b J{}^\r\, \del_\r \fD_a J{}^\m ~,
\end{split}
\eeq
where we understand $\fD_a J^\m = \fD_a J_\nb{}^\m \dd x^\nb$. No ambiguity should arise as the other components vanish. The second equation is of Maurier-Cartan type and its important to note that it is not obtainable from the first by a variation of complex structure, unlike \eqref{eq:cpxstFN}.

The conventional approach to solving these goes as follows. At first order, the solution to \eqref{eq:cpxDefEOM} is
$$
\fD_\a{} J^\m\= 2\ii  \D_\a{}^\m~, \qquad \fD_\ab{} J^\m \=  \delb \k_\ab{}^\m~,
$$
where $\k_a{}^m$ is a moduli space 1-form valued in $T_\X$ and $\D_\a{}^\m$ is a $\delb$-closed holomorphic vector valued form. Sadly, as the Maurier Cartan equation is not a derivative of the $\delb$-closed equation, we cannot obtain $\fD_a \fD_b J$ by differentiating $\fD_a J$. A small diffeomorphism is parameterised by a vector $\ve_a{}^m$. We have the freedom to choose $\ve_\ab{}^\m = -\k_\ab{}^\m$, which sets $\fD_\ab J^\m = 0$.  The Maurier-Cartan equation simplifies for three components:
\beq\label{eq:delbJzeroso}
\delb\big( \fD_a \fD_b J_\nb{}^\m \dd x^\nb\big) \=  0~, \quad {\rm for} \quad (a,b) = (\a,\bb); (\ab,\b); (\ab,\bb)~.
\eeq
Therefore, 
\beq\label{eq:secondJ}
\fD_\ab \fD_\b J^\m \= \delb \k_{\ab\b}{}^\m~, \quad \fD_\a \fD_\bb J^\m \= \delb \k_{\a\bb}{}^\m~, \quad \fD_\ab \fD_\bb J^\m \= \delb \k_{\ab\bb}{}^\m~.
\eeq
The small diffeomorphism acts on 
$$
\fD_a \fD_b J^\m \to \fD_a \fD_b J^\m + \delb (\del_a \ve_b{})^\m + \cdots~,
$$
and so provided we assume $\k_{\a\bb} = - \del_\a\ve_\bb$, $\k_{\ab\b} = - \del_\ab\ve_\b$, $\k_{\ab\bb} = - \del_\ab\ve_\bb$, we can use the gauge symmetries to set the right hand side of these second order derivatives in \eqref{eq:secondJ}  to zero. The universal bundle calculation above does this automatically.

A second order variation of  $\dd\O=0$ 
\beq\label{eq:ddOparts}
\begin{split}
\delb \,\fD_a \fD_b \O^{(3,0)} + \del \fD_a \fD_b \O^{(2,1)} &\= 0~,\quad \delb \fD_a \fD_b \O^{(2,1)} + \del \fD_a \fD_b \O^{(1,2)} \= 0~,\\
 \delb \fD_a \fD_b \O^{(1,2)} &\= 0~.
\end{split}
\eeq
These derivatives are covariant with respect to the line bundle symmetry $\O \to \l \O$, with $\l\in\IC^*$. Using $\O$ is a holomorphic $(3,0)$-form, the possible non-zero components of $\fD_a \fD_b \O$ are of type $(3,0)$, $(2,1)$ and $(1,2)$. Using the gauge in which the terms in \eqref{eq:secondJ} all vanish, it follows that $\fD_\ab \fD_\b \O = \fD_\b \fD_\ab \O = \fD_\ab \fD_\bb \O = 0$. So the non-zero terms are  $\fD_\a \fD_\b \O^{(3,0)}$, $\fD_\a \fD_\b \O^{(2,1)}$ and $\fD_\a \fD_\b \O^{(1,2)}$. Combining this  with the solutions to \eqref{eq:ddOparts}, we can expand the second order deformation of $\O$ as 
\beq\label{eq:deltasquaredOmega}
\fD_\a\fD_\b \O \=  \del (\eta_{\a \b}{}^{(2,0)} + \x_{\a \b}{}^{(1,1)}) + \delb (\eta_{\a \b}{}^{(2,0)} + \x_{\a \b}{}^{(1,1)})~.
\eeq

\subsubsection{Holomorphy of \texorpdfstring{$F$}{}}
Differentiating $F^{0,2}=0$ twice gives 
\beq\notag
2\ii\left( \delb_\A \,\fD_a \fD_b \A + \{ \fD_a \A, \fD_b \A \}\right) \= 2 \ii \fD_a \fD_b J^{\m 0,1} F_\m + \D_a{}^\m \fD_b F_\m + \D_b{}^\m \fD_a F_\m~,
\eeq
The components of this equation with at least one anti-holomorphic deformation are
\beq\notag
\begin{split}
\delb_\A (\fD_\ab \fD_\bb \A) &\= 2 \ii \fD_\ab \fD_\bb J^{\m 0,1} F_\m~,\\
\delb_\A (\fD_\a \fD_\bb \A + \D_\a{}^\m \fD_\bb \A_\m^\dagger) &\= 2 \ii \fD_\a \fD_\bb J^{\m 0,1} F_\m ~,\\
\delb_\A (\fD_\ab \fD_\b \A + \D_\b{}^\m \fD_\ab \A_\m^\dagger) &\= 2 \ii \fD_\ab \fD_\b J^{\m 0,1} F_\m ~,
\end{split}
\eeq	 
while the purely holomorphic component has a part that resembles Maurier-Cartan equations together with a 2nd order deformation of the Atiyah equation
\beq\notag
2\ii \left( \delb_\A \,(\fD_\a \fD_\b \A) + \{ \fD_\a \A, \fD_\b \A \} \right) \= \fD_\a \fD_\b J^{\m 0,1} F_\m + \D_\a{}^\m \fD_\b F_\m{}+ \D_\b{}^\m \fD_\a F_\m{}~,
\eeq

\subsubsection{\texorpdfstring{$H=d^c\o$}{} and the Bianchi identity}
\label{s:BianchiSecondDeriv}
The second order deformations of $H=\dd^c\o$ and the Bianchi identity \eqref{eq:Anomaly0} yield
\beq\notag
\begin{split}
\dd \fD_a \B_b - J^s \del_s \fD_a \fD_b \o \= &2i \Big( \fD_a \fD_b J^s (\dd \o)_s - \dd (\fD_a \fD_b J^s \o_s)\\
& + \D_a{}^s (\dd \fD_b \o)_s + \D_b{}^s (\dd \fD_a \o)_s - \dd \left(\D_a{}^s  \fD_b \o_s + \D_b{}^s  \fD_a \o_s\right) \Big)\\
&+ \frac{\ap}{2} \tr \Big( (\fD_a \fD_b A) F +(\fD_b A) \fD_a F \Big)~.
\end{split}
\eeq
For the moment we omit the corresponding terms with $\fD_a\fD_b \Th$ in the interest of space. Projecting this equation onto its $(0,3)$ and $(1,2)$ parts on $\ccX$ gives
\beq\notag
\begin{split}
\delb \fD_a \Z_b^{0,2} \= &2 \ii \Big( \fD_a \fD_b J^{\r 0,1} (\delb \o)_\r -\delb (\fD_a \fD_b J^{\r 0,1} \o_\r)\\
&\ \ \ \ + \D_a{}^\r (\del \fD_b \o^{0,2})_\r + \D_b{}^\r (\del \fD_a \o^{0,2})_\r  \Big)\\
&+ \frac{\ap}{2} \tr \Big( (\fD_b \A) \fD_a F^{0,2} \Big)~,
\end{split}
\eeq
and
\beq\notag
\begin{split}
\delb \fD_a \Z_b^{1,1 } + \del \fD_a \Z_b^{0,2} \= &2 \ii \Big( \fD_a \fD_b J^{\r 0,1} (\del \o)_\r - \del (\fD_a \fD_b J^{\r 0,1} \o_\r)\\
&\ \ \ \ + \fD_a \fD_b J^{\sb 0,1} (\delb \o)_\sb - \delb (\fD_a \fD_b J^{ \sb 0,1} \o_\sb)\\
&\ \ \ \ + \fD_a \fD_b J^{\r 1,0} (\delb \o)_\r - \delb (\fD_a \fD_b J^{ \r 1,0} \o_\r)\\
&\ \ \ \ + \D_a{}^\r (\del \fD_b \o^{1,1 })_\r  + \D_a{}^\sb (\delb \fD_b \o^{0,2})_\sb\\
&\ \ \ \ - \del (\D_a{}^\r \fD_b \o_\r^{0,1}) - \delb (\D_a{}^\sb \fD_b \o_\sb^{0,1})\\
&\ \ \ \ + \D_b{}^\r (\del \fD_a \o^{1,1 })_\r + \D_b{}^\sb (\delb \fD_a \o^{0,2})_\sb\\
&\ \ \ \ - \del (\D_b{}^\r \fD_a \o_\r^{0,1}) - \delb (\D_b{}^\sb \fD_a \o_\sb^{0,1}) \Big)\\
&+ \frac{\ap}{2} \tr \Big( (\fD_a \fD_b \A) F + (\fD_b \A) \fD_a F^{1,1 } - (\fD_b \A^\dagger) \fD_a F^{0,2} \Big)~.
\end{split}
\eeq
The $(2,1)$ and $(3,0)$ are determined by complex conjugation.

\subsubsection{HYM}
A second order deformation of the Hermitian Yang Mills equation $\o^2 F = 0$ tells us that
\beq\notag
\begin{split}
&(\fD_a \fD_b \o) \o F + (\fD_b \o) (\fD_a \o) F + (\fD_b \o) \o \fD_a F\\
&+ (\fD_a \o) (\fD_b \o) F + \o (\fD_a \fD_b \o) F + \o (\fD_b \o) \fD_a F\\
&+ (\fD_a \o) \o \fD_b F + \o (\fD_a \o) \fD_b F + \o^2 \fD_a \fD_b F \= 0~. 
\end{split}
\eeq
All but the last term is symmetric under $a \leftrightarrow b$, meaning that
\beq\notag
\o^2 [\fD_a, \fD_b] F \= -\o^2 [F, \IF_{ab}] \= 0~,
\eeq
where we have used $[\fD_a, \fD_b] A = -\dd_A \IF_{ab}$.

\subsection{Small Gauge Transformations and deformations of connections}
The point here is that there is a relation between deformations of connections on the moduli space and gauge symmetries of deformations of heterotic structures. A toy example is to consider a first order deformation of the gauge field $\d A = \d y^a \fD_a A$, where the covariant derivative is necessary as gauge transformations can depend on parameters:
$$
\fD_a A \= \del_a A+ [A\#_a,A]~, 
$$
Under $A\#_a \to A\#_a + \phi_a$ we have $\fD_a A \to \fD_a A+ \dd_A \phi_a$, in other words  $\d A \to \d A +\dd_A \d \phi$. Taking the $0,1$-component, this we see a small gauge transformation corresponds precisely to a deformation of the connection $A\#$. How does this relationship generalise to higher orders?

First consider the perspective of the gauge transformation taken to second order in deformations. We write
\beq\notag
\begin{split}
A &\rightarrow A' \= A + \d A + \tfrac{1}{2} \d^2 A~,\\
\Phi &\rightarrow \Phi' \= \Phi + \d \Phi + \tfrac{1}{2} \d^2 \Phi~,
\end{split}
\eeq	
where $\Phi$ is an element of the structure group of the bundle $V$. 
The action of $\Phi$ on $A$ is  
\beq\notag
\begin{split}
^{\Phi '} \! A' &\= \bigg( \Phi + \d \Phi + \tfrac{1}{2} \d^2 \Phi \bigg) \bigg( A + \d A + \frac{1}{2} \d^2 A \bigg) \bigg(1 - \Phi^{-1} \d \Phi  + \Phi^{-1} \d \Phi \Phi^{-1} \d \Phi  - \frac{1}{2}  \Phi^{-1} \d^2 \Phi  \bigg) \Phi^{-1}\\[3pt]
&\hspace{9mm} - \dd \bigg( \Phi + \d \Phi + \frac{1}{2} \d^2 \Phi \bigg) \bigg( 1- \Phi^{-1} \d \Phi  + \Phi^{-1} \d \Phi \Phi^{-1} \d \Phi  - \frac{1}{2}  \Phi^{-1} \d^2 \Phi \bigg) \Phi^{-1}~.
\end{split}
\eeq
We can isolate small gauge transformations by setting background transformations to identity, yielding
\beq\notag
\begin{split}
^{\Phi '} \! A' \big|_{\Phi = \Ione} 
&= \hspace{5mm} A\\
&\hspace{5mm} + \d A - \dd_A \d \Phi \big|_{\Phi=\Ione}\\
&\hspace{5mm} + \tfrac{1}{2} \Big( \d^2 A - \dd_A \big( \d^2 \Phi \big|_{\Phi=\Ione} - \d \Phi \big|_{\Phi=\Ione} \d \Phi \big|_{\Phi=\Ione} \big)\\
&\hspace{15mm} + [\dd_A \d \Phi \big|_{\Phi=\Ione}, \d \Phi \big|_{\Phi=\Ione}] - 2[\d A, \d \Phi \big|_{\Phi=\Ione}] \Big) \nonumber\\
&= \hspace{5mm} A\\
&\hspace{5mm} + \d A - \dd_A \d \phi\\
&\hspace{5mm} + \tfrac{1}{2} \Big( \d^2 A - \dd_A  \d^2 \phi  + [\dd_A \d \phi, \d \phi ] - 2[\d A, \d \phi ] \Big)~,\\
\end{split}
\eeq
where $\phi$ is an element of the Lie algebra associated to $\Phi$, that is $\Phi = e^\phi$. In the last line we have used
$$
\d\phi = \d\Phi\big|_{\Phi=\Ione}~,\qquad\qquad \d^2 \phi \= \d^2 \Phi \big|_{\Phi=\Ione} - \d \Phi \big|_{\Phi=\Ione} \d \Phi \big|_{\Phi=\Ione}~,
$$
which follows from $\Phi = e^\phi$.
We identify the action of small gauge transformations order-by-order 
\beq\label{eq:Agauge2ndorder}
\d A \sim \d A - \dd_A \d \phi~, \qquad \d^2 A \sim \d^2 A - \dd_A \, \d^2 \phi  + [\dd_A \d \phi, \d \phi ] - 2[\d A, \d \phi ]~.\\[2pt]
\eeq
What is interesting is that the second order gauge transformation is captured by the same deformation of $A\#$ used at first order. That is, under $A\#_a \to A\#_a + \phi_a$, 
\beq
\begin{split}
 \fD_a \fD_b A &\to 
 \fD_a \fD_b A - (\dd_A \fD_a \phi_b) +  [\dd_A \phi_b,\phi_a ]-  [\fD_a A, \phi_b] - [ \fD_b A,\phi_a ]~.   \\[2pt]
\end{split}
\eeq
We get precisely \eqref{eq:Agauge2ndorder} with the identifications $\d \phi = \d y^a \phi_a$ and $\d^2 \phi =  \d y^a \d y^b \fD_a \phi_b$. The field strength for the connection also deforms
$$
\IF_{ab} \to \IF_{ab} + \fD_a \phi_b - \fD_b \phi_a~.
$$
So because small gauge transformations are not physical, the curvature of the connection on the moduli space is also not physical. Instead the choice of gauge on $\X$ is connected to the choice of certain connections on the moduli space.

We find a similar situation with other gauge symmetries: diffeomorphisms and gerbe transformations. For diffeomorphisms on $\X$: $x \rightarrow f(x)$,  we perform the deformations
\beq\notag
\begin{split}
f &\rightarrow f + \d f +\frac{1}{2} \d^2 f~,\\
\x &\rightarrow \x + \d \x +\frac{1}{2} \d^2 \x~,\\
V &\rightarrow V + \d V +\frac{1}{2} \d^2 V~,
\end{split}
\eeq
for any one-form $\x$, and vector $V$. After setting $f$ to identity in order to isolate the small diffeomorphisms, this leads us to identify
\beq\label{eq:smalldiff}
\begin{split}
\d^2 V &\sim \d^2 V + [(\d^2 f - \d f^s \del_s \d f), V] + [\d f, \d V] + [\d f, \d V + [\d f, V]]~,\\
\d^2 \x &\sim \d^2 \x + \dd \left( (\d^2 f - \d f^s \del_s \d f)^s \x_s \right) + (\d^2 f - \d f^s \del_s \d f)^s (\dd \x)_s\\
&\ \ \ + \dd (\d f^s \d \x_s) + \d f^s (\dd \d \x)_s\\
&\ \ \ + \dd \left( \d f^s \left( \d \x + \dd (\d f^t \x_t) + \d f^t (\dd \x)_t \right)_s \right)\\
&\ \ \ + \d f^s \left( \dd \left( \d \x  + \d f^t (\dd \x)_t \right) \right)_s~.
\end{split}
\eeq
From here it is straightforward -- but not necessarily easy -- to calculate the action of a small diffeomorphism on any  tensor. The action is captured by deformations of the pre-connection (Ehreshman connection) $c_a{}^m$.\footnote{In \citeUG it is demonstrated that $c$ defines an (almost) product structure for the universal bundle: the fibering of the heterotic structures over their moduli space. It is necessitated by the simple observation that diffeomorphisms on $\X$ should, in general, depend on parameters and so deformations are described by covariant derivatives. That being so, identification of coordinate basis for forms and vectors is gauge dependent; an invariant formulation comes via the Ehreshman connection. } In effect, this is a connection for diffeomorphisms of $\X$ depend on parameters. If $c \to c - \ve$ then the corresponding connection $\G_a = \dd c_a$ transforms as $\G_a{}^m{}_n \to \G_a{}^m{}_n - \del_n \ve_a{}^m$. A first order a deformation of a 1-form $\d\xi = \d y^a \nabla_a \xi$ transforms
$$
\nabla_a \xi \to \nabla_a \xi + \ve^s (\dd \x)_s + \dd (\ve^s\x_s)~, \qquad \nabla_a V \to \nabla_a V + [\ve_a, V]~,
$$
where the first equation is precisely the action of a Lie derivative, the second the action of a Lie bracket. At second order, we repeat the process. It is algorithmic but tedious. 
\beq
\begin{split}
 \nabla_a \nabla_b \x &\to \nabla_a \nabla_b \xi + \dd ((\nabla_a \ve_b{}^s)\xi_s) + \nabla_a \ve_b{}^s (\dd \x)_s \\
 &\quad  + \dd (\ve_b{}^s\nabla_a \x_s) + \ve_b{}^s (\dd \nabla_a \x)_s+ \\
 &\quad + \dd \left(\ve_a{}^m (\nabla_b \x  + \ve_b{}^s (\dd \xi)_s + \dd (\ve_b{}^s \x_s))_m\right)+\\
  &\quad + \ve_a{}^m \left( \dd (\nabla_b \xi + \ve_b{}^s (\dd \x)_s ) \right)_m ~, \\[3pt]
   \nabla_a\nabla_b V &\to \nabla_a \nabla_b V + [\nabla_a\ve_b, V] + [\ve_b , \nabla_a V] + [\ve_a, \nabla_b V+ [\ve_b,V]]~,
\end{split}
\eeq
where we identify, in a slight generalisation of the gauge field case,
\beq\notag
\d^2 f - \d f^m \del_m \d f \= \d y^a \d y^b \nabla_a \ve_b~. 
\eeq

Putting these observations together, along with a corresponding second order action for small gerbe transformations, we find an action on the second order deformations of heterotic structure:
\beq\label{sosg}
\begin{split}
2 \ii \, \fD_a \fD_b J^{\m0,1} &\sim 2 \ii \fD_a \fD_b J^{\m0,1} + \delb \nabla_a \ve_b{}^\m\\
\fD_a \fD_b \A &\sim \fD_a \fD_b \A + \nabla_a \ve_b{}^\r F_\r - \delb_\A \nabla_a \phi_b\\
\fD_a \Z_b &\sim \fD_a \Z_b + \nabla_a \ve_b{}^m (H + \ii \dd \o)_m - \frac{\ap}{2} \tr \{ F \nabla_a \phi_b \} + \dd (\ccb_{ab} + \ii \nabla_a \ve_b{}^m \o_m)\\
\fD_a \Zb_b &\sim \fD_a \Zb_b + \nabla_a \ve_b{}^m (H - \ii \dd \o)_m - \frac{\ap}{2} \tr \{ F \nabla_a \phi_b \} + \dd (\ccb_{ab} - \ii \nabla_a \ve_b{}^m \o_m)\\
\fD_a \fD_b \O &\sim \fD_a \fD_b \O + \dd (\nabla_a \ve_b{}^m \O_m)~,
\end{split}
\eeq

\subsection{Second Order Gauge Fixing}
Of particular interest are the second order equations of motion with an anti-holomorphic first derivative; these are
\beq\label{eq:soeombb}
\begin{split}
\delb \,\fD_\ab \fD_\bb J^{\m 0,1} &\= 0~, \quad\quad \delb \,\fD_\a \fD_\bb J^{\m 0,1} \= 0\\
\delb_\A (\fD_\ab \fD_\bb \A) &\= 2 \ii \, \fD_\ab \fD_\bb J^{\m 0,1} F_\m~,\\
\delb_\A (\fD_\a \fD_\bb \A) &\= 2 \ii \, \fD_\a \fD_\bb J^{\m 0,1} F_\m - \delb_\A (\D_\a{}^\m \fD_\bb \A_\m^\dagger)~,\\
\delb \fD_\ab \Z_\bb^{0,2} &\= 2 \ii \Big( \fD_\ab \fD_\bb J^{\r 0,1} (\delb \o)_\r - \delb (\fD_\ab \fD_\bb J^{\r 0,1} \o_\r) \Big),\\
\delb \fD_\a \Z_\bb^{0,2} &\= 2 \ii \Big( \fD_\a \fD_\bb J^{\r 0,1} (\delb \o)_\r - \delb (\fD_\a \fD_\bb J^{\r 0,1} \o_\r) \Big)~,\\
\delb \fD_\ab \Z_\bb^{1,1 } + \del \fD_\ab \Z_\bb^{0,2} &\= 2 \ii \, \fD_\ab \fD_\bb J^{\r 0,1} (\del \o)_\r + \frac{\ap}{2} \tr \Big( (\fD_\ab \fD_\bb \A) F \Big)~,\\[5pt]
\delb \fD_\a \Z_\bb^{1,1 } + \del \fD_\a \Z_\bb^{0,2} &\= 2 \ii \, \fD_\a \fD_\bb J^{\r 0,1} (\del \o)_\r - \delb (\D_\a{}^\r \Z_{\bb \ \r}^{1,0})\\[5pt]
&+ \frac{\ap}{2} \tr \Big( (\fD_\a \fD_\bb \A) F + (\D_\a{}^\r \fD_\bb \A^\dagger_\r) F \Big)~.
\end{split}
\eeq
With the non-trivial elements of the relevant cohomologies assigned to the first order deformations, we use the freedoms parametrised by $\ve_{\ab \bb}{}^\m$ and $\ve_{\a \bb}{}^\m$, to remove the remaining $\delb$-exact parts of $\fD_\ab \fD_\bb J^{\m 0,1}$ and $\fD_\a \fD_\bb J^{\m 0,1}$, yielding
\beq\notag
\begin{split}
\fD_\ab \fD_\bb J^{\m 0,1} &\= 0~,\quad \fD_\a \fD_\bb J^{\m 0,1} \= 0~.
\end{split}
\eeq
This fixes our freedom for $\ve_{ab}{}^\m$ in the same way as it does for $\ve_{b}{}^\m$ at first order, and drastically simplifies the right hand sides of \eqref{eq:soeombb}.\\
We can then use the parameters $\phi_{\ab \bb}$ and $\phi_{\a \bb}$ to set
\beq\notag
\begin{split}
\fD_\ab \fD_\bb \A &\= 0\\
\fD_\a \fD_\bb \A &\= -\D_\a{}^\m \fD_\bb \A^\dagger_\m~;
\end{split}
\eeq
in a corresponding manner, exhausting the gauge freedom associated with $\phi$.\\
For the deformations of $\Z$, we use part of the freedom of $\ccb_{\ab \bb}^{0,1}$ and $\ccb_{\a \bb}^{0,1}$ to set 
\beq\notag
\begin{split}
\fD_\ab \Z_\bb^{0,2} &\= 0~,\quad \fD_\a \Z_\bb^{0,2} \= 0~,\\
\ccb_{\ab \bb}^{0,1} + \ii \fD_\ab \ve_\bb{}^\m \o_\m &\= \delb \psi_{\ab \bb}~,\quad
\ccb_{\a \bb}^{0,1} + \ii \fD_\a \ve_\bb{}^\m \o_\m \= \delb \psi_{\a \bb}~,
\end{split}
\eeq
from which we also get $\fD_\ab \Zb_\bb^{0,2} = 0$, on account of the fact that $\fD_\ab \fD_\bb \o^{0,2} = 0$.\\ 
We use some of the freedom of $\ccb_{\ab \bb}^{1,0}$, $\ccb_{\a \bb}^{1,0}$, $\fD_\ab \ve_\bb{}^\sb$, and $\fD_\a \ve_\bb{}^\sb$, to set 
\beq\notag
\begin{split}
\fD_\ab \Z_\bb^{1,1 } &\= 0~,\qquad \fD_\a \Z_\bb^{1,1 } \= \D_\a{}^\r \Z_{\bb \, \r}^{1,0}~,\\[5pt]
\ccb_{\ab \bb}^{1,0} + \ii \fD_\ab \ve_\bb{}^\sb \o_\sb &\= \delb \psi_{\ab \bb}~,\quad \ccb_{\a \bb}^{1,0} + \ii \fD_\a \ve_\bb{}^\sb \o_\sb \= \delb \psi_{\a \bb}~,
\end{split}
\eeq
and, finally, we use $\ccb_{\a \bb}^{1,0} - \ii \fD_\a \ve_\bb{}^\sb \o_\sb$ to set
\beq\notag
\begin{split}
\fD_\a \Zb_\bb^{2,0 } &\= 0~,\quad \ccb_{\a \bb}^{1,0} - \ii \fD_\a \ve_\bb{}^\sb \o_\sb \= \del \tilde{\psi}_{\a \bb}~.
\end{split}
\eeq
This is made possible by using \eqref{eq:delbJzeroso} to simplify the complex conjugate of the sixth line of \eqref{eq:soeombb}.

We can then use $\varepsilon_{\ab \bb}{}^\sb$ to remove the $\eta^{2,0 }$ terms of \eqref{eq:deltasquaredOmega}, meaning that gauge parameters that remain unfixed are just those parametrised by $\psi$ and $\tilde{\psi}$. This leaves us with
\beq\label{eq:sogaugefixing}
\begin{split}
\fD_\ab \fD_\bb J^{\m 0,1} &\= 0~,\quad \fD_\a \fD_\bb J^{\m 0,1} \= 0~,\\
\fD_\ab \fD_\bb \A &\= 0~,\quad \fD_\a \fD_\bb \A \=\! -\D_\a{}^\m \fD_\bb \A^\dagger_\m~,\\
\fD_\ab \Z_\bb^{0,2} &\= 0~,\quad \fD_\a \Z_\bb^{0,2} \= 0~,\\
\fD_\ab \Z_\bb^{1,1 } &\= 0~,\quad \fD_\a \Z_\bb^{1,1 } \= \D_\a{}^\r \Z_{\bb \, \r}^{1,0}~,\\
\fD_\ab \Zb_\bb^{0,2} &= 0~,\quad \fD_\a \Zb_\bb^{2,0 } = 0~.
\end{split}
\eeq
These results are to be compared with \eqref{eq:Zb2}-\eqref{eq:AJ2}. It is spectacular that after all this hard work that the universal bundle calculation captures these results. In fact, it gives more information in particular regarding the curvatures of connections on the moduli space and some additional information about derivatives. It says for example that the contraction of $\D_\a$ and $\fD_\bb\A_\m^\dag$ are orthogonal. The lesson learned in \citeUG for first order calculations has reappeared: the universal bundle gives a shortcut to otherwise difficult algebraic calculations.

\newpage
\section{\texorpdfstring{$\d \o^{0,2}$}{}, the \K potential and the moduli space metric}
\label{s:KahlerPotRedone}
In \sref{s:deformationuniv} we clarified that the tangability $[2,2]$ component of the universal Bianchi identity, \eqref{Bianchitang22}, naively contains several non-trivial equations, however only one of them is in fact non-trivial. It relates $\fD_\a \fD_\bb \o^{1,1}$ to a product of first order deformations of the gauge connection. This clarifies an open question in \citeUG. Here we clarify an open question in \citeM. 

The dimensional reduction of heterotic supergravity gives a moduli space metric
\beq\begin{split}
 g\#_{\a\bb} &\= \frac{1}{V}\int_X \Big( \D_\a{}^\m\star\D_{\bb}{}^{\nb}\,g_{\m\nb} + \fD_\a\o^{1,1 }\star\fD_{\bb}\o^{1,1 } + \frac{\ap}{4}\tr{ ( \fD_\a A \star \fD_{\bb}A ) } \Big)\\[0.1cm]
 &\hspace{3cm}~+~ \frac{\ap}{2V}\int_X \vol ~ \Big( \D_{\a\mb\nb}\,\D_{\bb\r\s} + \fD_\a\o_{\r\mb}\,\fD_{\bb}\o_{\s\nb} \Big) \, R^{\mb\r\nb\s} +  \cO(\ap^2) ~.\label{eq:modulimetric}
\end{split}\raisetag{2cm}\eeq
When $H=\cO(\ap)$, which is always the case for string theory ameniable to a dimensional reduction, the term 
$$
\fD_\a \o^{0,2} \= (\del_\a\o)^{0,2} \= \D_\a{}^\m \o_\m = \ii\D_{\a\,[\mb\nb]} \dd x^{\mb\nb} \=  \cO(\ap)~.
$$
This is straightforward to see by first checking that if $H=0$ then $\o$ is \K and the \K condition kills this term. However, at first order it is non-zero. It does not explicitly appear in the metric above and so its role in the moduli problem is obscure. It is there implicitly, being the first term in 
$$
\D_{\a\,\mb\nb} \= \D_{\a\,[\mb\nb]}  + \D_{\a\,(\mb\nb)}~, 
$$
while in which in the dimensional reduction $\d g_{\mb\nb} =2 \D_{\a\,(\mb\nb)}$. It is straightforward to check that the term $\D_{\a\,[\mb\nb]}$ derives from a variation of the real B-field both the symmetric and antisymmetric component combine precisely to give an inner product on the complete tensor $\D_\a$ and not its symmetric component. 

Part of its mystery came from the \K potential proposed in \citeM:
\beq\notag
K \= K_1 + K_2 \=\! - \log \left( \frac{4}{3} \int  \o^3 \right) - \log \left( \int  \ii \O \Ob \right)~.
\eeq
After differentiating this twice with respect to parameters, in \citeM the inner product of $(\del_\a \o)^{0,2}$ was discarded as its contribution to the moduli space metric was $\ap^2$. While this is true, what we briefly clarify here is that if we don't discard this term, then the \K potential gives exactly the metric \eqref{eq:modulimetric} that comes from the dimensional reduction, even without discarding this term. 

Starting with $\del_\a \del_\bb K_1$, we have
\beq\notag
\begin{split}
\del_\a \del_\bb K_1 &= - \del_\a \left( \frac{1}{\int \o^3} \int 3 (\del_\bb \o) \o^2 \right)\\
&= - \frac{1}{V} \int \fD_\a \o^{0,2} \star \fD_\bb \o^{2,0 } + \frac{1}{V} \int \fD_\a \o^{1,1 } \star \fD_\bb \o^{1,1 } - \frac{1}{2V} \int \o^2 \del_\a \del_\bb \o~.
\end{split}
\eeq
The third term is precisely the one that is dealt with in \eqref{secondorderom} and gives
\beq\notag
\begin{split}
- \frac{1}{2V} \int \o^2 \del_\a \del_\bb \o &= \frac{\ii}{4V} \int \ii \o^2 \{ \del_\a, \del_\bb \} \o^{1,1 }\\
&= \frac{\ii \ap}{8 V} \int \tr \left( \fD_\a \A\, \fD_\bb \A^\dagger \right) + \frac{2}{V} \int \del_\a \o^{0,2} \star \del_\bb \o^{2,0 }~,
\end{split}
\eeq
where we have used $\dd (\o^2) = 0$ and $\o^2 F = 0$, as well as the identity
$$
- \frac{1}{2V} \D_\a{}^\r (\fD_\bb \o^{2,0})_\r \o^2 = \frac{1}{V} \fD_\a \o^{0,2} \star \fD_\bb \o^{2,0}~,
$$
and thus we arrive at
\beq\label{eq:ddK1}
\del_\a \del_\bb K_1 \= \frac{\ii \ap}{8 V} \int \o^2 \tr \left( \fD_\a \A\,\fD_\bb \A^\dagger \right) + \frac{1}{V} \int \del_\a \o \star \del_\bb \o~.
\eeq
The deformation $\del_\a \del_\bb K_2$ gives
\beq\notag
\begin{split}
\del_\a \del_\bb K_2 &\= - \del_\a \left( \frac{1}{\int \ii \O \Ob} \int \ii \O \del_\bb \Ob \right) \= \frac{1}{V} \int \D_{\a (\mb \nb)} \D_\bb{}^{(\mb \nb)} \vol - \frac{1}{V} \int \D_{\a [\mb \nb]} \D_\bb{}^{[\mb \nb]} \vol~.
\end{split}
\eeq
Using
$
\fD_\a \o^{0,2} \star \fD_\bb \o^{2,0} = 2 \, \D_{\a [\mb \nb]} \D_\bb{}^{[\mb \nb]} \vol
$
and combining with \eqref{eq:ddK1}, we arrive at \eqref{eq:modulimetric}.

\newpage
\section{Conclusions}
In this work, we investigate first-order deformations of the universal bundle that preserve its defining conditions, which formally mirror the F-terms in heterotic string theory along with D-terms on the fibers. Our motivation is that first-order deformations of the universal bundle serve as a streamlined approach to second-order deformations within the original heterotic framework. Indeed, we uncover additional insights. By enforcing the structure of the universal bundle, we obtain vanishing conditions for certain second-order derivatives within the heterotic theory. We demonstrate the circumstances under which field strengths for connections on the moduli space are holomorphic, and, in specific cases, reduce to flatness. Extending this to the study of D-terms on the universal bundle, as inspired by prior work such as \cite{Herbst:2008jq}, could reveal whether connections on the moduli space achieve flatness.

There are numerous potential avenues for further investigation. One could, for instance, relax the preservation of D-terms on fibres to explore a broader deformation space. As we maintain F-term constraints that remain unrenormalized in $\ap$, this approach could impose conditions on the nature of $\ap$-corrections within the original heterotic theory. Our current findings include initial observations regarding the kernel of the $\IDb$-operator on the universal bundle, raising many questions, motivated by analogous results in heterotic theories. For example, with the explicit construction of the connection on the tangent bundle available in \citeUG, we might eliminate certain terms, in the spirit of \citeE. Key questions include: Does this operator manifest a non-trivial curvature? Is it Hermitian-Yang-Mills? Can an adjoint operator be consistently defined, and what would its kernel represent?

Finally, it would be intriguing to explore the potential of universal deformation theory to impose constraints on the algebraic structures satisfied by deformations in the original heterotic theory. Prior research, such as that in \cite{Ashmore:2018ybe}, approached this via superpotential constructions in heterotic theories, realising the F-terms we examine here. The universal bundle structure could provide a pathway to identifying the algebraic properties of deformations, especially once spurious modes are appropriately removed.

\vskip2cm
\subsection*{Acknowledgements}
We would like to thank Philip Candelas, Xenia de la Ossa, Johanna Knapp for as always enjoyable, enlightening and helpful conversations. JM is in part supported by an ARC Discovery Project Grant DP240101409. JM, ES and MS would like to thank the MATRIX Research Institute where part of this work was completed.

\newpage
\appendix

\section{Hodge dual relations}
It is useful to recount from the appendix of \citeSG adjoint differential operators and some Hodge dual relations for forms on $\X$.

The Hodge $\star$ operator acts on type
\beq\notag
 \star : \O^{(p,q)}(X) \to \O^{(N-q,N-p)}(X) ~.
\eeq

Given $k$-forms $\eta$, $\x$ and a metric $\dd s^2 = g_{mn} \dd x^m \otimes \dd x^n$ on $\X$, the Hodge dual defines an inner product
\beq\notag
 (~\cdot~,~\cdot~) ~~: ~~ \O^k(X) \times \O^k(X) \to \mathbb{R} ~,
\eeq
with
\beq\label{eq:HodgeInner}
 (\eta,\xi)     \=  \frac{1}{V\, k!} \int_X \vol \, \eta^{m_1 ... m_k} \, \xi_{m_1 ... m_k}  ~.
\eeq
Now consider two forms $\eta_k$ and $\xi_l$, where the subscript denotes their degree and $k \leq l$,. Contraction is 
\beq\notag
 \lrcorner : \O^k(X) \times \O^l(X) \to \O^{l-k}(X) ~,
\eeq
and acts as follows
\beq\notag
 \eta_k \, \lrcorner \, \xi_l  \=  \frac{1}{k!(l-k)!} \, \eta^{m_1 ... m_k} \, \xi_{m_1 ... m_k \, n_1 ... n_{l-k}} \, \dd x^{n_1 ... n_{l-k}}  \=  \frac{1}{k!} \,\eta^{m_1 ... m_k} \, \xi_{m_1 ... m_k} ~.
\eeq
An interesting feature of this operator is that it is the adjoint of the wedge product
\beq\label{contradjwedg}
 (\s_{l-k} \, \lrcorner \, \xi_l , \eta_k)  \=  (\xi_l , \s_{l-k} \w \eta_k) ~.
\eeq
Recall, the de Rahm operator $\dd$ can be written
\beq\notag
 \dd  \=  \dd x^m \, \nabla^{\LC}_m ~.
\eeq
Using  \eqref{contradjwedg} and  integration by parts
\beq\notag
 ( \dd \eta , \xi )  \=  ( \dd x^m \, \nabla^{\LC}_m \eta , \xi )  \=  ( \nabla^{\LC}_m \eta , \xi^m )  \=  ( \eta , - \nabla^{\LC}_m \xi^m ) ~.
\eeq
It follows that
\beq\label{eq:AdjointdLC}
 \dd^\dag \xi_k  \=\!  - \nabla^{\LC}_m \xi^m  \=\!  - \frac{1}{(k-1)!} \, \nabla^{\LC}_n \xi^n{}_{m_1 ... m_{k-1}} \, \dd x^{m_1 ... m_{k-1}} ~.
\eeq

The de Rham differential splits into the sum of Dolbeault operators $\dd = \del + \delb$. Analogously, the codifferential also splits $\dd^\dag = \del^\dag + \delb^\dag$ where
\beq\label{eq:deladjoint}
\begin{split}
 &\del^\dag : \O^{(p,q)}(X) \to \O^{(p-1,q)}(X) \qquad , \qquad \del^\dag  \=\!  - \star \delb \, \star ~,\\[0.1cm]
 &\delb^\dag : \O^{(p,q)}(X) \to \O^{(p,q-1)}(X) \qquad , \qquad \delb^\dag  \=\!  - \star \del \, \star ~.
\end{split}\eeq

\vskip0.5cm
 One-forms, type $1,0$:
\beq\label{eq:Hodge1form}
 \star\, \eta^{1,0} \=\! -\ii \, \eta^{1,0} ~ \frac{\o^2}{2} ~.
\eeq
 Two-forms, types $2,0 $ and $1,1 $:
\beq\begin{split}\label{eq:Hodge2form}
 &\star\, \eta^{2,0 }  \=  \eta^{2,0 } \, \o ~,\\[0.1cm]
 &\star\, \eta^{1,1 }  \=\! -\ii \, \eta_{\m}{}^\m ~ \frac{\o^2}{2} - \eta^{1,1 } \, \o \= (\o \, \lrcorner \, \eta^{1,1 }) ~ \frac{\o^2}{2} - \eta^{1,1 } \, \o ~,
\end{split}\eeq
 Three-forms, types $(3,0)$ and $(2,1)$:
\beq\begin{split}\label{eq:Hodge3form}
 &\star\, \eta^{(3,0)}  \=\!  - \ii \, \eta^{(3,0)} ~,\\[0.1cm]
 &\star\, \eta^{(2,1)}  \= \ii \, \eta^{(2,1)} - \eta_{\m}{}^\m{}\,^{1,0} \, \o \= \ii \, \eta^{(2,1)} - \ii \, (\o \, \lrcorner \, \eta^{(2,1)}) \, \o ~,
\end{split}\eeq
Four-forms, types $(3,1)$, and $(2,2)$
\beq\label{eq:Hodge4form}
\begin{split}
&\star \, \eta^{(3,1)}  \= \!  - \ii \, \eta_\m{}^{\m 2,0 } ~,\\[0.1cm]
&\star \, \eta^{(2,2)}  \= \frac{1}{2} \, \eta_{\m \n}{}^{\m \n} \, \o + \ii \, \eta_{\m}{}^\m{}^{1,1 } ~,
\end{split}
\eeq
The five-form, type $(3,2)$:
\beq\label{eq:Hodge32}
\star \eta^{(3,2)} \= - \frac{\ii}{2} \, \eta_{\m \n}{}^{\m \n 1,0} ~,
\eeq
%
%
%
The six-form $(3,3)$:
\beq\label{eq:Hodge33}
\star \eta^{(3,3)} \= - \frac{\ii}{3!} ~ \eta_{\m \n \r}{}^{\m \n \r} ~.
\eeq
From these relations, it follows that the hodge star squares to
\beq\label{hodgesquared}
\star^2 \, \eta^{(p,q)} \= (-1)^{p+q} \, \eta~.
\eeq

\section{Connection symbols on \texorpdfstring{$\IX$}{X}}
We enumerate some commonly used connection symbols in heterotic theories. We list the components in complex coordinates. We also give expressions for various divergences which are useful for calculations in the paper. 

$\X$ is a complex manifold with complex structure $J$ and hermitian metric $g$. A vector bundle $V\to X$ is hermitian if there is a hermitian inner product on sections of the bundle. For example $\ccT_X$ has a hermitian structure facilitated by the hermitian metric $\dd s^2 = 2 g_{\m\nb} \dd x^\m \otimes \dd x^\nb$. The bundle is holomorphic (or has a holomorphic structure) if the total space $V$ is a complex manifold with complex structure $\IJ$ and the projection map $\pi: V \to X$ is holomorphic. That is, fibres consist only of holomorphic sections according to $\IJ$. This is equivalent to the transition functions being purely holomorphic. For example, the bundle $\ccT_X$ is not a holomorphic bundle while   $\ccT_X^{1,0}$ is a holomorphic bundle.  

A connection $\nabla$ on $\ccT_X$ is metric compatible if $\nabla g = 0$. A connection $\nabla$ is hermitian if it preserves the hermitian structure. That is, it is metric compatible and $\nabla J = 0$. In terms of components, a hermitian connection has $\G_\m{}^\n{}_\rb = \G_\m{}^\nb{}_\r = 0$. There may be more than one Hermitian connection. 

If $V$ is a holomorphic bundle then a connection $\nabla$ is compatible  with its holomorphic structure if $\nabla^{0,1} = \delb$. In terms of components $ \G_\mb{}^\n{}_\r = 0$. For example,  if $V$ is a section of $\ccT_X^{1,0}$ then $\nabla$ is compatible with the holomorphic structure if $\nabla^{0,1} V = \dd x^\mb \left(\del_\mb V^\n + \G_\mb{}^\n{}_\r V^\r \right) \del_\n = \delb V$. 

\subsubsection*{Levi-Civita}
Levi--Civita is the unique metric compatible connection with no torsion (symmetric in lower indices). It is hermitian if the manifold is \K but not in general. 
\beq\label{eq:LC}
\begin{split}
 &\G^{\LC}{}_{\m}{}^{\n}{}_\r  \=  \frac{1}{2} \, g^{\n\sb}(\del_\m g_{\r\sb} + \del_\r g_{\m\sb})  \=  g^{\n\sb} \, \del_\m g_{\r\sb} - \frac{1}{2} H_{\m}{}^{\n}{}_{\r}  \=  g^{\n\sb} \, \del_\r g_{\m\sb} + \frac{1}{2} H_{\m}{}^{\n}{}_{\r} ~,\\[0.1cm]
 &\G^{\LC}{}_{\m}{}^{\nb}{}_{\r}  \=  0 ~,\\[0.1cm]
 &\G^{\LC}{}_{\m}{}^{\n}{}_{\rb}  \=  \frac{1}{2} \, g^{\n\sb}(\del_{\rb} g_{\m\sb} - \del_{\sb} g_{\m\rb})  \=  \frac{1}{2} H_{\m}{}^\n{}_{\rb} ~,\\[0.1cm]
 &\G^{\LC}{}_{\m}{}^{\nb}{}_{\rb}  \=  \frac{1}{2} \, g^{\nb\s}(\del_\m g_{\rb\s} - \del_{\s} g_{\m\rb})  \=\!  -\frac{1}{2} H_{\m}{}^{\nb}{}_{\rb} ~.
\end{split}
\eeq

\subsubsection*{Bismut}

The supersymmetry Killing spinor of heterotic supergravity (to first order in $\ap$) is covariantly constant with respect to the connection
$\G_m^\Bi = \G_m^{\LC} - \half H_m$. Writing $J$ as a spinor bilinear it follows that  $\o$ and $J$ are covariantly constant with respect to this connection $\nabla^\Bi J = \nabla^\Bi \o = 0$ and so it follows $\G^\Bi$ is metric compatible and hermitian. The torsion of $\G^\Bi$ is  completely antisymetric and equal to~$H=\dd^c \o$ i.e. $T^m{}_{np} {\=} H^m{}_{np}$. A geometric statement is that there is a unique connection on $\ccT_X$ that is hermitian with completely antisymmetric torsion. This is the Bismut connection.
\medskip
\beq\label{eq:Bismut}
\begin{split}
 &\G^{\Bi}{}_{\m}{}^{\n}{}_\r  \=  g^{\n\sb} \, \del_\r g_{\m\sb}  \=  g^{\n\sb} \, \del_\m g_{\r\sb} - H_{\m}{}^{\n}{}_{\r} ~,\\[0.1cm]
 &\G^{\Bi}{}_{\m}{}^{\nb}{}_{\r}  \=  0 ~,\\[0.1cm]
 &\G^{\Bi}{}_{\m}{}^{\n}{}_{\rb}  \=  0 ~,\\[0.1cm]
 &\G^{\Bi}{}_{\m}{}^{\nb}{}_{\rb}  \=  g^{\nb\s} (\del_\m g_{\s\rb} - \del_\s g_{\m\rb})  \=\!  - H_{\m}{}^{\nb}{}_{\rb} ~.
\end{split}\eeq

\subsubsection*{Hull}

While the spinor in heterotic is covariantly constant with respect to $\G^\Bi$,  a different connection 
$\G^\Hu$ appears in the heterotic action. It  is not hermitian but has completely antisymmetric torsion with opposite sign $-H$. Hence,  $\G_m^\Hu = \G_m^{\LC} + \half H_m$. This we call the Hull connection.
\medskip

\beq\begin{split}
 &\G^{\Hu}{}_{\m}{}^{\n}{}_\r  \=  g^{\n\sb} \, \del_\m g_{\r\sb}  \=  g^{\n\sb} \, \del_\r g_{\m\sb} + H_{\m}{}^{\n}{}_{\r} ~,\\[0.1cm]
 &\G^{\Hu}{}_{\m}{}^{\nb}{}_{\r}  \=  0 ~,\\[0.1cm]
 &\G^{\Hu}{}_{\m}{}^{\n}{}_{\rb}  \=  g^{\n\sb}(\del_{\rb} g_{\m\sb} - \del_{\sb} g_{\m\rb})  \=  H_\m{}^\n{}_{\rb} ~,\\[0.1cm]
 &\G^{\Hu}{}_{\m}{}^{\nb}{}_{\rb}  \=  0 ~.
\end{split}\eeq

\subsubsection*{Chern}
The Chern connection is the unique connection which is hermitian ($\nabla g = \nabla J = 0$) and compatible with the holomorphic structure of $\ccT_X^{1,0}$. The connection has no mixed indices. If the manifold is non--\K then it has torsion.
\beq\label{eq:Chern}
\begin{split}
 &\G^{\Ch}{}_{\m}{}^{\n}{}_\r  \=  g^{\n\sb} \, \del_\m g_{\r\sb}  \=  g^{\n\sb} \, \del_\r g_{\m\sb} + H_{\m}{}^{\n}{}_{\r} ~,\\[0.1cm]
 &\G^{\Ch}{}_{\m}{}^{\nb}{}_{\r}  \=  0 ~,\\[0.1cm]
 &\G^{\Ch}{}_{\m}{}^{\n}{}_{\rb}  \=  0 ~,\\[0.1cm]
 &\G^{\Ch}{}_{\m}{}^{\nb}{}_{\rb}  \=  0 ~.
\end{split}\eeq

\subsection*{Divergences}

The divergence of a vector $\ve^\m$ taken with respect to a generic connection
\beq
 \nabla_\m \ve^\m  \=  \del_\m\ve^\m + \ve^\m \, \G_\n{}^{\n}{}_{\m} ~.
\eeq
To compute this we need the following contraction
\beq\begin{split}
 \G^{\LC}{}_\n{}^\n{}_\m & \=  \del_\m\log{\sqrt g} + \frac{1}{2} H_{\m\n}{}^{\n} ~,\\[0.1cm]
 \G^{\Bi}{}_\n{}^\n{}_\m & \=  \del_\m\log{\sqrt g} ~,\\[0.2cm]
 \G^{\Hu}{}_\n{}^\n{}_\m  \=  \G^{\Ch}{}_\n{}^\n{}_\m & \=  \del_\m\log{\sqrt g} + H_{\m\n}{}^\n ~.
\end{split}\eeq
The four choices above give
\beq\begin{split}\label{divve}
 \nabla_\m^{\LC}\ve^\m & \=  \del_\m\ve^\m + \ve^\m \, \del_\m\log{\sqrt g} + \frac{1}{2} \ve^\m \, H_{\m\n}{}^\n ~,\\[0.1cm]
 \nabla_\m^{\Bi}\ve^\m & \=  \del_\m\ve^\m + \ve^\m \, \del_\m\log{\sqrt g} ~,\\[0.2cm]
 \nabla_\m^{\Hu}\ve^\m  \=  \nabla_\m^{\Ch}\ve^\m & \=  \del_\m\ve^\m + \ve^\m \, \del_\m\log{\sqrt g} + \ve^\m \, H_{\m\n}{}^\n
\end{split}\eeq
As for the vector-valued form $\D_{\nb}{}^\m$ we have, for a generic connection
\beq
 \nabla_\m\D_{\nb}{}^\m  \=  \del_\m\D_{\nb}{}^\m + \D_{\nb}{}^\m \, \G_\r{}^\r{}_\m - \G_\m{}^{\rb}{}_{\nb} \, \D_{\rb}{}^\m ~,
\eeq
and the four choices above give
\beq\begin{split}\label{divD}
 \nabla_\m^{\LC}\D_{\nb}{}^\m & \=  \del_\m\D_{\nb}{}^\m + \D_{\nb}{}^\m \, \del_\m \log{\sqrt g} + \frac{1}{2} \D_{\nb}{}^\m \, H_{\m\r}{}^\r - \frac{1}{2} \D^{\m\r} \, H_{\m\r\nb} ~,\\[0.1cm]
 \nabla_\m^{\Bi}\D_{\nb}{}^\m & \=  \del_\m\D_{\nb}{}^\m + \D_{\nb}{}^\m \, \del_\m \log{\sqrt g} - \D^{\m\r} \, H_{\m\r\nb} ~,\\[0.2cm]
 \nabla_\m^{\Hu}\D_{\nb}{}^\m  \=  \nabla_\m^{\Ch}\D_{\nb}{}^\m & \=  \del_\m\D_{\nb}{}^\m + \D_{\nb}{}^\m \, \del_\m \log{\sqrt g} + \D_{\nb}{}^\m \, H_{\m\r}{}^\r ~.
\end{split}\eeq

\subsection{Adjoint of the \texorpdfstring{$\del$}{}-operator}

The $\del^\dag$-operator is defined in \eqref{eq:deladjoint} and acts on a $(p,q)$--form. We want to derive an expression for it in terms of covariant derivatives analogus to that of $\dd^\dag$ in \eqref{eq:AdjointdLC}. It easiest to do this, at least initially, with a hermitian operator as these preserve type. For that reason we focus on Chern \eqref{eq:Chern} and Bismut \eqref{eq:Bismut}. 

Given a $(p+1,q)$--form $\eta$ and a $(p,q)$--form $\x$, the adjoint is 
$$
(\del \xi\, , \eta) \= (\xi \, , \del^\dag \eta)~,
$$
where the inner product is the usual one in \eqref{eq:HodgeInner} extended to act on complex forms. Note that 
$$
\del \xi \= \frac{1}{p!q!} \left\{ \nabla^\Ch_\m \xi_{\m_1 \cdots \m_p \nb_1 \cdots \nb_q}  + \frac{p}{2} \, H_\m{}^\r{}_{\m_1} \x_{\r\m_2\cdots \nb_q} \right\} \dd x^{\m\m_1\cdots\nb_q}~.
$$
Now, 
\beq
(\del \xi\, , \eta) \=  \frac{1}{p!\, q!}  \int \left\{ \nabla^\Ch_\m \xi_{\m_1 \cdots \m_p \nb_1 \cdots \nb_q}  + \frac{p}{2} \, H_\m{}^\r{}_{\m_1} \x_{\r\m_2\cdots \nb_q} \right\} \eb^{\m\m_1\cdots\nb_q}\vol\\[3pt]
\eeq
To integrate by parts, we need to take care of the fact that the divergance theorem involves the Levi-Civita connection: $\int \nabla_m^\LC V^m \vol = 0$ for any well-defined vector $V^m$. This involves converting the Chern connection to the Levi-Civita connection
$$
\int \nabla^\Ch_\m V^\m \vol \= \int \left\{ \nabla^\LC_m V^m \, + H_\r{}^\r{}_\m V^\m \right\}\vol \= \int \left(H_\r{}^\r{}_\m V^\m\right) \vol ~.
$$
Thence
\beq\notag
(\del \xi\, , \eta) \= \frac{1}{p!q!} \int \xi_{\m_1\cdots\nb_q} \left\{ - \nabla_\m^\Ch \eb^{\m\m_1\cdots\nb_q} + H_\r{}^\r{}_\m \eb^{\m\m_1\cdots\nb_q} -\frac{p}{2} H^{\m_1}{}_{\m\r}\eb^{\m\r\m_2\cdots\nb_q} \right\} \= (\x,\del^\dag \eta)~;
\eeq
so we infer
\beq
\begin{split}
 \del^\dag \eta^{(p+1,q)} &\=\frac{1}{p!q!} \left\{ - \nabla^{\Ch\,\m} \eta_{\m\m_1\cdots\nb_q} - H_\r{}^\r{}^\m \eta_{\m\m_1\cdots\nb_q} -\frac{p}{2} H_{\m_1}{}^{\m\r}\eta_{\m\r\m_2\cdots\nb_q}\right\} \dd x^{\m_1\cdots\nb_q}\\
 &\=  - \nabla^{\Ch\,\m} \eta^{(p,q)}_{\m} - H_\r{}^\r{}^\m \eta^{(p,q)}_{\m} -\frac{1}{2} H^{\m\r\,1,0}\eta^{(p-1,q)}_{\m\r}~.
\end{split}
\eeq
Using \eqref{eq:Bismut}, \eqref{eq:Chern} we can rewrite this straightforwardly in terms of Bismut
\beq
 \del^\dag \eta^{(p+1,q)} \=  - \nabla^{\Bi\,\m} \eta^{(p,q)}_{\m} + H^{\m\nb\,0,1} \eta_{\m\nb}^{(p,q-1)} +\frac{1}{2} H^{\m\r\,1,0}\eta^{(p-1,q)}_{\m\r}~.
\eeq

We can write this in terms of the Levi-Civita connection however it is more complicated as it is not a hermitian connection.  From \eqref{eq:AdjointdLC}, recall  $\dd^\dag \eta = - \nabla^{\LC\,m} \eta_m$ and that $\dd^\dag=\del^\dag + \delb^\dag$. As $\eta$ is a $(p+1,q)$--form, we can project onto type:
$$
(\dd^\dag \eta)^{(p,q)} = \del^\dag \eta^{(p+1,q)} = - \nabla^{\LC\,\m} \eta_\m^{(p,q)} - \nabla^{\LC\,\mb} \eta_\mb^{(p,q)} 
$$
Note the last term is really due to it being non--hermitian.  
Using \eqref{eq:LC}, \eqref{eq:Bismut}, \eqref{eq:Chern},   we can check this matches the expressions for the Chern and Bismut connections.

\section{Calabi-Yau}
In the case of the standard embedding, we have the manifold is \K $\dd \o = 0$. In that case, together with it being balanced, we show $\Z_\a^{0,2}$ vanishes. 

The Kahler condition  $\dd \o = 0$ has a first order deformation
\beq\label{calabiDeltaomega}
\begin{split}
\del \,\fD_\a \o{}^{1,1 } &\= 0\\
\del \,\fD_\a \o{}^{0,2} + \delb \,\fD_\a \o{}^{1,1 } &\= 0\\
\delb \,\fD_\a \o{}^{0,2} &\= 0~.
\end{split}
\eeq
This tells us that
\beq\label{eq:kahlersol}
\begin{split}
\fD_\a \o^{1,1 } &\= \del \psi^{0,1} + \x_\a^\textrm{harm}\\
\fD_\a \o^{0,2} &\= \delb \psi^{0,1}~,
\end{split}
\eeq
where $\x_\a^\textrm{harm}$ is a harmonic representative of $H_\del^{1,1 }(X,\IR)$. We use that  $h^{2,0 } = 0$.\\
On the other hand, the metric being balanced $\dd \o^2 = 0$ implies 
\beq\label{eq:balanced}
\begin{split}
\del^\dagger \,\fD_\a \o{}^{1,1 } - \delb^\dagger \,\fD_\a \o{}^{0,2} &\= 0\\
\delb^\dagger \,\fD_\a \o{}^{1,1 } &\= 0~.
\end{split}
\eeq
Substituting \eqref{eq:kahlersol} into the first line of \eqref{eq:balanced}, and taking the inner product with $\psi^{0,1}$ yields
\beq
\Braket{\del \psi^{0,1} + i \delb \psi^{0,1} | \del \psi^{0,1} + i \delb \psi^{0,1} } = 0~,
\eeq
and thus 
\beq
\begin{split}
\fD_\a \o^{1,1 } &\= \x_\a^\textrm{harm}\\
\fD_\a \o^{0,2} &\= 0~.
\end{split}
\eeq
It follows that $\Zb_\a^{0,2} = -2i \fD_\a \o{}^{0,2} = 0$, meaning that on a Calabi-Yau manifold, the only surviving degrees of freedom are
\beq
\begin{split}
&\D_\a{}^\n~,\quad \fD_\a \A~, \quad \Z_\a^{1,1 }.
\end{split}
\eeq
	
	\newpage
	
	\section{Summary of universal geometry}
	\label{s:summaryUniv}
We summarise the extension of the remaining fields in the heterotic theory to the universal geometry, following closely \citeUG. 

The field strength $H$ is related to $B$ by the relation
\beq
H \= \dd B - \frac{\ap}{4}\Big(\CS[A] - \CS[\Th]\Big)~,
\label{Hdef}\eeq
where $\CS$ denotes the Chern--Simons three-form
\beq
\CS[A] \= \tr\!\left(A\dd A +\frac{2}{3}\, A^3\right) ~.
\notag\eeq
The Chern--Simons forms transform under gauge transformation and so does $B$, with the transformation law for $B$ chosen to ensure that $H$ is gauge invariant. This transformation law is 
\beq
{}^{\Phi,\Psi} B \= B + \frac{\ap}{4}\Big( \tr \big(Y\! A - Z \Th\big) +  U\! - W \Big)~,
\label{eq:BTransf2} \eeq
with 
\beq
Y\=\dd\Phi \Phi^{-1}~,\qquad Z\=\dd\Psi \Psi^{-1}~,
\notag\eeq
and $U$ and $W$ are such that $\dd U {\=} \frac{1}{3} \tr (Y^3)$ and $\dd W {\=} \frac{1}{3} \tr (Z^3)$ and $\Psi$ denotes gauge transformations of the connection $\Th$ on $\ccT_\X$. 

As $H$ is gauge invariant its variation with respect to the parameters can simply be given as a partial derivative. In this way we arrive at a relation of the form
\beq
\del_a H \= \dd \ccB_a - \frac{\ap}{2} \tr \big(D_a A\, F\big) + \frac{\ap}{2} \tr \big( D_a \Th R \big)~.
\label{eq:Hderiv}\eeq
This relation identifies a gauge invariant quantity $\ccB_a$ that is defined up to the addition of a $\dd$-closed form.

Let us define extended forms of $B$ and $H$ that are related by
\beq
\IH \= \Id \IB - \frac{\ap}{4} \Big(\CS[\IA] - \CS[\ITheta]\Big)~,~~~\text{where}~~~
\CS[\IA] \= \tr\!\left(\IA\,\Id \IA +\frac{2}{3}  \IA^3\right)~.
\label{eq:IHdef2}\eeq
It is pleasing that the important quantity $\ccB_a$ turns out to be a mixed component of the gauge invariant tensor $\IH$
\beq
\ccB_a \= \frac12 \IH_{amn} \dd x^m \dd x^n~.
\notag\eeq
We take $\IH$ to satisfy an extended supersymmetry relation  and a Bianchi identity
\begin{equation}\notag
 \IH \= \Id^c\Iomega ~,\quad \Id\IH~=-\frac{\ap}{4}\Big(\tr{\IF^2}-\tr{\IR^2}\Big)\ ,
\end{equation}  
whose mixed components give important relations \eqref{susyoncF}, \eqref{susyoncF2} and \eqref{secondorderom} among the heterotic moduli. 

So far, we have discussed the consequence of allowing the gauge functions $\Ph$ and $\Ps$ to be functions of both $x$ and $y$. One further consequence, is the introduction of a covariant basis of forms and a corresponding dual basis of vectors
\beq
\label{eq:ebasis}
\begin{split}
 e^m &\= \dd x^m + c_a{}^m\dd y^a~,  e^a \= \dd y^a~, \\[5pt]
e_m &\= \del_m~,  e_a\= \del_a - c_a{}^m \del_m~.
\end{split}
\eeq
The quantity $c^m{\=}c_a{}^m \dd y^a$ is a connection which transforms, under $x\to \tilde{x}(x,y)$, in the form
\beq
c^{\widetilde{m}} \= \pd{x^{\widetilde{m}}}{x^n}\, c^n - \pd{x^{\widetilde{m}}}{y^b}\, \dd y^b~,
\label{eq:ctransf}\eeq
and this ensures that the forms $e^m$ and vectors $e_a$ transform as expected
\beq
e^{\widetilde{m}} \= \pd{x^{\widetilde{m}}}{x^n}\, e^n~,~~~
e_{\widetilde{a}}  \= \pd{y^b}{y^{\widetilde{a}}}\, e_b~.
\notag\eeq
With these basis forms, we write the extended vector potential, for example, as
\beq
\IA \= A_m e^m + A^\sharp_a\dd y^a~,~~~\text{with}~~~A^\sharp_a \= \L_a - A_m c_a{}^m~. 
\notag\eeq
The connection $c_a{}^m$ also plays an important geometrical role in relation to the fibration $\IX$, in that it determines an almost product structure $\IL$ that provides a splitting of the tangent space of the fibration $\cT_\IX$ into vertical and horizontal subspaces. Being a fibration, $\IX$ naturally encodes a vertical projection $\IX\longrightarrow \M$. A horizontal structure, equivalent to a local choice of $c_a{}^m$, is not invariantly defined. The freedom inherent in choice of $c_a{}^m$ corresponds precisely to the freedom to make coordinate transformations.  This splitting of the tangent space into a vertical and horizontal component is a product structure \cite{YanoBook}. We take it to be integrable, which as described in \citeUG does not mean we can set $c_a{}^m = 0$; instead it simply means the curvature associated to the Ehreshman connection vanishes. In fact, the product structure and complex structure of $\IX$ have a non-vanishing Fr\"olicher--Nijenhuis bracket and our convention will be to utilise the Frobenius theorem to find complex coordinates the diagonalise the complex structure, but not the product structure. Hence, the connection $c_a{}^m$ remains part of our lives.

It is necessary to covariantise the de Rham operator $\dd$ along the manifold $\X$ so that it defines a fiber-wise cohomology. This is described in detail in \citeUG, when it is not ambigious we simply denote it by $\dd$. There is a complentary operator $\fD_a$ which describes how tensors on $\X$ evolve under a change in parameters. These are actually components of the de Rham operator $\Id= \dd  \oplus \fD$ on $\IX$. There is a third possible component which we have not written which is to do with the integrability of the product structure defined by $c_a{}^m$. As is described in \citeUG, the structure is required to be integrable for the universal geometry to capture some second order deformations of the hermitian form $\o$ and so we take as a starting assumption to be integrable. That being the case, the decomposition of the de Rham operator is as written.

The connection $c^m$ can be identified with the cross term in the `minimal' extended metric. We have two metrics that arise naturally:
the metric $g_{mn}$ on the manifold $X$ and the metric $g^\sharp_{ab}$ on the moduli space $\M$. If we combine these into a `minimal' extended metric
\beq
\Id\Is^2 \= g_{mn} e^m e^n + g^\sharp_{ab} \dd y^a \dd y^b~,
\notag\eeq
and write this out in terms of the basis forms $\dd x^m$ and $\dd y^a$ we find the cross term
\beq
\Ig_{ma} \= g_{mn} c_a{}^n~.
\notag\eeq

The connection $c^m$ appears also in an interesting way in relation to the variation of the complex structure of $X$. The first order variation of the complex structure is recorded in a form $\D_\a{}^\m{\=}\D_{\a\nb}{}^\m \dd x^\nb$ which is defined by
$$
\D_\a{}^\m \= \del_\a \dd x^\m \Big|^{0,1}
$$
It will also be shown later that $\D_\a{}^\m$ is also related to $c^\m$ by
\beq
\D_{\a\,\nb}{}^\m \= -\del_\nb c_\a{}^\m~.
\label{eq:extrinsic1}\eeq
This non-vanishing reflects the observation made above that we cannot set $c_a{}^m = 0$.

\newpage
	
	\section{Dictionary to convert conventions}

\makebox[\linewidth]{    
\resizebox{1.1\columnwidth}{!}{    
\renewcommand{\arraystretch}{1.5}    
    \begin{tabular}{r|c|c|c|c|}    
    \multicolumn{1}{c}{} & \multicolumn{1}{c}{Strominger} & \multicolumn{1}{c}{AQS} & \multicolumn{1}{c}{MarSpar} & \multicolumn{1}{c}{OS} \\
    \cline{2-5}    
    $\dd H = \frac{\ap}{4} \tr$ & 
    $-(\frac{4}{30 \ap} F^2 - \frac{4}{\ap} R^2)$ & 
    $- (F^2 - R^2)$ &  
    $(F^2 - R^2)$ & 
      $(F^2 - R^2)$ \\
    \cline{2-5}
    SUSY $\d \Psi_m =$& 
    $(\nabla^{\LC}_m - \frac{1}{4} H_m)\ve$ &
    $(\nabla^\LC_m - \frac{1}{8}  H_{m}{}^{ab} \g_{ab})\ve$  &
    $ \nabla^\LC_m \ve + \frac{1}{8}\hat H_{mab} \g^{ab} \ve~^{\$} $ &
   $ \nabla_m^\LC + \frac{1}{8} H_{mab} \g^{ab} $ \\
    \cline{2-5}
            Connections & 
             & 
            $\nabla_m e_a = \Th_m{}^b{}_a e_b $  & 
            $ $ & 
                      $\nabla_m^\LC:  \G^\LC{}_{mn}{}^p$ \\
    \cline{2-5}
     SUSY connection &
       & 
      $ \Th_m^\LC{\,}^a{}_b - \half H_m{}^a{}_b{}^{\$} $& 
      $$ & 
            $\G_{mn}^\LC{}^p +\half H_{mn}{}^p $\\
    \cline{2-5}
    $H =$ & 
    $\frac{\ii}{2} (\delb - \del) \o$* & 
    $\ii(\del - \delb) \o $ & 
    $\ii (\delb - \del) \o$ & 
    $\ii (\del - \delb) \o$**  \\
    \cline{2-5}
    Bianchi connection & 
    $ $ & 
    $\Th^+$& 
  $\Th^-$ & 
    {\rm instanton}\\
    \cline{2-5}
    Torsion (Bianchi) & 
    $ $ & 
    $$ & 
    $$ & 
    $$  \\
    \cline{2-5}
    Torsion (SUSY) & 
    $-H_{mnp} $ & 
    $ T=H$ & 
    $T^k{}_{mn} = -\hat H^k{}_{mn}{~}^{\$\$}$  & 
    $ T^k{}_{mn} = H^k{}_{mn}$  \\
    \cline{2-5}
    \end{tabular}
}}

\bigskip
{\bf Table Notes}:
$\$$: Given the gravitino variation there appears to be a sign discrepancy in writing the action of $\nabla^+$ on a vector in MS.  

$\$\$$: The torsion   computed using Cartan structure equations gives opposite sign in MarSpar. What's written here is computed via the structure equations.

$*$ There is a small error in the Strominger paper,  (2.12): the coefficient in front of both $H$ terms is out by a factor of $-1/2$ and so replace $H\to -\half H$ to get correct equations. Afterwards, (2.17) gives $H = \ii (\del-\delb)\o$. This is now compatible with the form of the Bianchi identity.

$**$ We use $\o(X,Y) = g(JX,Y)$ for any pair of vectors $X,Y$, as a comptability relation. In Bismut's original paper uses the opposite convention. To map conventions in OS to AQS requires sending $H\to -H$, $B \to -B$ and $\o \to -\o$. This changes the sign of the Ricci scalar in the supergravity action, see appendix of OS.

The $\tr$ and generators $T^a$ of the gauge groups are defined so that $tr (T^a T^b)$ is positive definite. This leads to an action in \cite{Martelli:2010jx}:
$$
S = \int \vol \left( R + 4(\del\Phi)^2 - \frac{1}{2} |H|^2 - 2\ap \left[\tr |F|^2  - \tr |R^-|^2 \right] \right) + \cO(\ap^2)~.
$$

{\bf Relation between $H$ and $\o$}\\

We now check the relation between $\o$ and $H$ in  MarSpar notation. As $J = -\ii \ve^\dag \g_m{}^n \ve \,\dd x^m\otimes \del_n$ we have $\nabla^+_m J = 0$. Expanding this out using the MarSpar action on a vector:
\beq\label{eq:BismutonJ}
\begin{split}
 \nabla_m^+ J &\=  \left\{ \nabla_m^\LC J_n{}^p +\half H_{mq}{}^p J_n{}^q - \half H_{mn}{}^q J_q{}^p \right\} \dd x^n \otimes \del_p \= 0~.
\end{split}
\eeq
Contracting with $g_{pt}$ on the components, using $J_n{}^p g_{pt} = \o_{nt}$:
\beq
 \nabla_m^\LC \o_{nt} \= -\half H_{mqt} J_n{}^q + \half H_{mn}{}^q J_q{}^p g_{pt}~.
 \eeq
 Contracting with $J^mJ^nJ^p$ where $J^m = J_t{}^m \dd x^t$:
 \beq\label{eq:BismutonJ2}
\begin{split}
J^m J^n J^t (\nabla_m^\LC \o_{nt}) &\= -\half J^m J^n J^t H_{mqt} J_n{}^q + \half J^m J^n J^t H_{mn}{}^q J_q{}^p g_{pt}\\
&\= -\half J^m  J^n  H_{mnq} J^t J_t{}^q + \half J^m J^n H_{mn}{}^q J_q{}^p g_{pt} J^t \\
&\= \half J^m  J^n  H_{mnq} \dd x^q + \half J^m J^n H_{mnq}\dd x^q \\
\frac{1}{3} (\dd \o)_{mnp}J^m J^n J^p&\= J^m J^n H_{mnq} \dd x^q~.
\end{split}
 \eeq
 where $J^2 = -1$ and $ J_q{}^p g_{pt} J_s{}^t \dd x^s  = g_{qs} \dd x^s$. We now use the Nijenhueis tensor vanishing to show $J^m J^n H_{mnq} \dd x^q = H$. This is most easily derived by evaluating the LHS in complex coordinates and using that the dilatino equation implies $H^{3,0} = 0$ . We therefore find 
  \beq
 \dd^c \o \= H~, \quad \dd^c \o \= \frac{1}{3!} J^m J^n J^p (\dd \o)_{mnp}~.
 \eeq

 \subsection*{Map from AQS to MarSpar}
To map from AQS to MarSpar we have
\beq\label{eq:AQSmap}
 H \= - \hat H~, \qquad B\=\! - \hat B~, \quad \ap \to \ap/8~.
\eeq
Note that this will send
$$
\ccZ  = \ccB  + \ii \d \o \to - \ccZb = -(\ccB - \ii \d \o)~.
$$
In holomorphic gauge in AQS conventions $\ccZ^{1,1}$ is associated to holomorphic parameters as is the case in CY manifolds, say \cite{Candelas:1990pi}. It solves an equation
$$
\delb \ccZ_\a^{1,1 } \= 2\ii \, \D_\a{}^\m (\del\o)_\m +\frac{\ap}{2} \tr \big( \fD_\a \A \, F\big)  ~.
$$
where $y^\a$ are holomorphic parameters. The map \eqref{eq:AQSmap} results in
$$
\delb \ccZb_\a^{1,1 } \=\! -2\ii \, \D_\a{}^\m (\del\o)_\m -\frac{\ap}{2} \tr \big( \fD_\a \A \, F\big)  ~.
$$
and so holomorphic parameters become associated to $\ccZb_\a^{1,1}$. To remedy this, the field redefinition could be augmented by one on the moduli space, e.g. complex conjugation of parameters.

\newpage

\providecommand{\href}[2]{#2}\begingroup\raggedright\endgroup

\end{document}